\documentclass[journal]{IEEEtran}
\usepackage[deletedmarkup=xout,commentmarkup=footnote]{changes}
\definechangesauthor{pc}
\definechangesauthor{jxp}
\definechangesauthor{dr}
\definechangesauthor{eg}

\usepackage[printonlyused,smaller]{acronym}

\makeatletter
\def\endthebibliography{%
  \def\@noitemerr{\@latex@warning{Empty `thebibliography' environment}}%
  \endlist
}
\makeatother

\usepackage{cite}
\ifCLASSINFOpdf
\else
\fi
\usepackage{float}
\usepackage{graphicx}
\usepackage{caption}
\usepackage{subcaption}

\usepackage[cmex10]{amsmath}
\usepackage{hyperref}
\usepackage{url}
\usepackage{amsthm}
\usepackage{xcolor}

\usepackage{xpatch}
\xpatchcmd{\sout}
  {\bgroup}
  {\bgroup}
  {}{}

\newcommand{\mname}{{\sc Nenya}} % Name for the new algorithm
\newcommand{\ulmo}{{\sc Ulmo}}
\newcommand{\ssta}{\ac{SSTa}}
\newcommand{\sst}{\ac{SST}}
\newcommand{\ssl}{\ac{SSL}}

\newcommand{\LL}{\ac{LL}}
\newcommand{\zslope}{\ensuremath{\alpha_{\rm AS}}} % Slope of the power spectrum along-scan [CONFIRM!!]
\newcommand{\mslope}{\ensuremath{\alpha_{\rm AT}}} % Slope of the power spectrum in the along-track direction
\newcommand{\slope}{\ensuremath{\alpha_{\rm min}}} % Minmium of the two 
\newcommand{\ablat}{\ensuremath{|{\rm latitude}|}}

\newcommand{\DT}{\ensuremath{\Delta T}} % T90-T10
\newcommand{\DTone}{\ensuremath{\Delta T = 1-1.5\,{\rm K}}} % DT = 1-1.5 K
\newcommand{\DThalf}{\ensuremath{\Delta T = 0.5-1\,{\rm K}}} % DT = 0.5-1 K
\newcommand{\DTzero}{\ensuremath{\Delta T = 0-0.5\,{\rm K}}} % DT = 0-0.5 K
\newcommand{\DTtwo}{\ensuremath{\Delta T = 1.5-2.5\,{\rm K}}} % DT = 1.5-2.5 K
\newcommand{\DTfour}{\ensuremath{\Delta T = 2.5-4\,{\rm K}}} % DT = 2.5-4 K
\newcommand{\DTfive}{\ensuremath{\Delta T > 4\,{\rm K}}} % DT > 4 K

\newcommand{\weak}{\ensuremath{U_0=[0, 2] \, {\rm and} \, U_1=[-2.5, -0.3]}} % DT1
\newcommand{\strong}{\ensuremath{U_0=[4,7.3] \, {\rm and} \, U_1=[2.4,4]}} % DT1

\newcommand{\ncutouts}{$\approx 8,000,000$} % 96% clear; v4
\newcommand{\cutout}{cutout} 

\newcommand{\ndim}{256}  % Number of dimensions
\newcommand{\ntrain}{600,000}  % Number of training cutouts
\newcommand{\nvalid}{150,000}  % Number of validation cutouts
\newcommand{\nbatch}{256}  % Number of images in a batch
\newcommand{\mlatentv}{\tilde{ \mathbf{z}}^{(i)}}  % Latent vector
\newcommand{\latentv}{$\mlatentv$}

\newcommand{\relf}{\ensuremath{f_r}} % Relative fraction f_r
\newcommand{\relb}{\ensuremath{f_b}} % Relative frequency f_b
\newcommand{\fracc}{\ensuremath{f_c}} % Relative frequency f_b

\begin{document}
\title{The Fundamental Patterns of Sea Surface Temperature}

\author{J. Xavier Prochaska,
    Erdong Guo,
    Peter~C.~Cornillon,
    Christian E.~Buckingham%
    %    John~Doe,~\IEEEmembership{Fellow,~OSA,}
    %    and~Jane~Doe,~\IEEEmembership{Life~Fellow,~IEEE}% <-this % stops a space
\thanks{J. Xavier Prochaska is an Affiliate of
the Ocean Sciences Department at the 
University of California, Santa Cruz and 
a Simons Pivot Fellow.}%
\thanks{E.~Guo was a member of the 
Department of Computer Sciences at the University of California, Santa Cruz during the formation of this work
and is an Affiliate 
of the Institute of Theoretical Physics, Chinese Academy of Sciences.}% <-this % stops a space
\thanks{P.~C.~Cornillon is a Professor Emeritus of Oceanography at the Graduate School of Oceanography, University of Rhode Island.}%
\thanks{C.~E.~Buckingham is a Research Assistant Professor in the Department of Estuarine and Ocean Sciences at the University of Massachusetts, Dartmouth.}
\thanks{Manuscript received XXX; revised XXX.}}


\markboth{Journal of \LaTeX\ Class Files,~Vol.~13, No.~9, September~2014}%
{Shell \MakeLowercase{\textit{et al.}}: Bare Demo of IEEEtran.cls for Journals}



\maketitle

\begin{abstract}
For over 40 years, remote sensing observations of the Earth’s oceans have yielded global measurements of \ac{SST}. 
With a resolution of approximately 1\,km, these data trace physical processes like western boundary currents, cool upwelling at eastern boundary currents, and the formation of mesoscale and sub-mesoscale eddies. To discover the fundamental patterns of SST on scales smaller than 10\,km, we developed an unsupervised, deep contrastive learning model named \mname. We trained \mname\ on a subset of 8 million cloud-free cutout images ($\sim 80 \times 80$\,km$^2$) from the \ac{MODIS} sensor, with image augmentations to impose invariance to rotation, reflection, and translation. 
The 256-dimension latent space of \mname\ defines a vocabulary to describe the complexity of SST and associates images with like patterns and features. We used a dimensionality reduction algorithm to explore cutouts with a temperature interval of $\DT = 0.5-1$\,K, identifying a diverse set of patterns with temperature variance on a wide range of scales. We then demonstrated that SST data with large-scale features arise preferentially in the Pacific and Atlantic Equatorial Cold Tongues and exhibit a strong seasonal variation, while data with predominantly sub-mesoscale structure preferentially manifest in western boundary currents, select regions with strong upwelling, and along the Antarctic Circumpolar Current. We provide a web-based user interface to facilitate the geographical and temporal exploration of the full \ac{MODIS} dataset. Future efforts will link specific SST patterns to select dynamics (e.g., frontogenesis) to examine their distribution in time and space on the globe.
\end{abstract}

\begin{IEEEkeywords}
IEEEtran, journal, \LaTeX, paper, template.
\end{IEEEkeywords}

% \fi
\IEEEpeerreviewmaketitle

%\acused{SSTa}
\acresetall
%\acused{SSTa}

\section{Introduction}
\IEEEPARstart{D}{ecades} of \ac{IR} observations made from satellite-borne sensors 
have been used to generate twice-daily (clouds permitting), global \ac{SST} fields. 
While dependent upon the specific satellite sensor examined, as well as the level of processing applied, \ac{IR}   measurements nominally have spatial resolutions of $1$~km and temperature resolutions of several tenths Kelvin. Moreover, the swath widths of these sensors typically range from $1000$ to $2500$~km, permitting resolution of both mesoscale (horizontal scales of $30$-$200$~km) 
and sub-mesoscale ocean phenomena 
(horizontal scales of $1$-$50$~km)\footnote{Including wave-driven currents in the category of sub-mesoscale phenomena, the corresponding horizontal scales must be modified: $0.1$-$50$~km.}.
One notes the two length-scales overlap, in part because
of the varying radius of deformation with latitude leading
to different ranges for each throughout the literature.
Examples of mesoscale phenomena include strong western boundary current systems such as the Gulf Stream, Kuroshio, and Aghulhas Currents, upwelling systems along eastern boundaries such as the California and Benguela Current Systems, tropical instability waves, 
large-scale eddies and/or 
baroclinic Rossby waves. 
Examples of sub-mesoscale processes include 
finer-scale fronts, filaments, and eddies that often result from stirring and straining by the mesoscale currents, and consequent instabilities that take place at these locations \cite{thomas2008submesoscale,mcwilliams2016review}. 
In summary, \ac{IR} satellite-derived measurements yield powerful datasets for exploring the physical properties and processes of our ocean over a broad range of spatial and temporal scales.
% and scientists have
% leveraged them for a wide range of oceanographic study,
% e.g.,  xxxx, marine heat wave 
% phenomena \cite{oliver}, xx.

Impressively, one notes that 
data from even a single sensor, such as the \ac{MODIS} dataset,
satisfies most characteristics of so-called ``big data.'' 
For example, the \ac{L2} \sst\ archive of the processed Aqua \ac{MODIS} data stream ($2003$-$2021$, inclusive) 
comprises over 
2 million individual 5-minute granules 
totalling over 20~TB. While much of these
data have been probed by human investigations, 
the dataset far exceeds
one's capacity for an analysis by visual inspection alone.
Moreover, the scope of this dataset poses computational 
challenges, both in terms of storage and processing resources. Yet, these hurdles are surmountable with modern approaches
tailored to big data including cloud storage
and modern-day \ac{ML} techniques.
%In short, to perform studies of \ac{SST} across the globe and
%over long time-periods, one must xxx

With this as background and armed with the \ac{MODIS} dataset and \ac{ML} tools, we are motivated to study \sst\ patterns across
the global ocean and over long time-periods to 
address the following questions. 
  What are the fundamental patterns and characteristics of \sst\ imagery?
  When and where do these patterns and characteristics manifest?
  And, ultimately, what do these patterns reflect about the underlying dynamics?

Our approach is to develop and apply
\ac{ML} techniques originally introduced for 
natural images (e.g., cats, dogs). By modifying and extending
their application to \ac{SST} imagery (in itself
a class of natural images), we may perform novel
investigations of the global ocean.

In an earlier study \cite{ulmo}, we constructed and applied a \ac{PAE} \cite{pae} 
nicknamed \ulmo\ to nighttime \ac{MODIS} \ac{L2} 
\sst\ data.
The primary goals of this effort were to construct
a massive dataset for exploration 
($\sim 12,000,000$ thumbnails or ``{\cutout}s'' of
dimension $\approx 125 \times 125$\,km$^2$)
and to identify outliers 
from the full dataset
(i.e., unusual phenomena at the ocean surface).
The \ac{PAE} is an unsupervised \ac{ML} model which
first encodes each \cutout\ into a 
reduced-dimension latent 
space (the autoencoder).  
One then generates probabilistic statements on the cutout distribution, specifically the likelihood of a given cutout occurring within the full distribution reported as a 
\ac{LL} metric.
%One then transforms this complex latent
%space into a Gaussian manifold 
%using a normalized flow.
%Finally, one generates 
%probabilistic statements on the 
%\cutout\ distribution.  
%Specifically, one calculates the likelihood of 
%a given \cutout\ occurring within the 
%full distribution reported
%as a \ac{LL} metric.

After calculating the \ac{LL} metric for $\sim 12,000,000$ cutouts,
\cite{ulmo} showed that the majority of 
outlier patterns  occur within previously known mesoscale
eddying regions of the ocean, such as western boundary currents.  While this result may have been anticipated
by those familiar with physical oceanography, 
it verified the efficacy of unsupervised \ac{ML} techniques
applied to these massive \sst\ datasets.  
On the other hand, the \ulmo\
\ac{LL} metric fails to capture the full complexity of 
\ac{SST} imagery--i.e., data with nearly identical 
\ac{LL} may arise from distinctly different \ac{SST} patterns
associated with distinct physical processes.
%Furthermore, the \ac{PAE} results have limited capacity for
%identifying trends--both  geographic and
%temporal--within the dataset.  
This limitation motivates the introduction of a new
\ac{ML} algorithm in this manuscript. 
%\textcolor{red}{I wonder if we could come up with a reason to name this particular ML algorithm as ELMO instead of ULMO. That would be fun. Maybe Extreme Learning Method for Ocean-identification (ELMO).}
%By the way, I searched for the reason for the acronym ULMO but could not find its meaning. Oh, I just found it--character in Tolkien's novels.

To further address our motivating questions,
we build a new model -- nicknamed
\mname\footnote{The ring of power from the
Lord of the Rings associated
with water and \ulmo.} -- built on the
unsupervised \ac{ML} algorithm termed 
self-supervised (or contrastive) learning 
\cite{hadsell2006dimensionality, bachman2019learning, chen2020simple, chen2020big, chen2020improved, he2020momentum, tian2020makes}.
This deep learning approach was developed to generate associations
amongst imagery with similar characteristics while 
separating them from data with contrasting nature.
In the process, one may explore the diversity of 
patterns/scenes/etc.\ that occur within a massive imagery
dataset (e.g., \cite{hayat2021}).
Regarding the ocean, this algorithm offers the opportunity
to describe the fundamental patterns of \ac{SST} fields
without introducing rigid rules-based or statistical metrics
that may predispose the outcome. 
Furthermore, it is straightforward afterward
to apply any such metric to gain physical insight
into the results.

%\begin{itemize}
% \item Importance of \ac{SST} patterns
% \item Introduce \ulmo\ and \ac{PAE} \cite{pae}
% \item Introduce \ac{LL}, \DT
% \item Introduce \ac{SSL}
%\end{itemize}

% An example of a floating figure using the graphicx package.
% Note that \label must occur AFTER (or within) \caption.
% For figures, \caption should occur after the \includegraphics.
% Note that IEEEtran v1.7 and later has special internal code that
% is designed to preserve the operation of \label within \caption
% even when the captionsoff option is in effect. However, because
% of issues like this, it may be the safest practice to put all your
% \label just after \caption rather than within \caption{}.
%
% Reminder: the "draftcls" or "draftclsnofoot", not "draft", class
% option should be used if it is desired that the figures are to be
% displayed while in draft mode.
%
%\begin{figure}[!t]
%\centering
%\includegraphics[width=2.5in]{myfigure}
% where an .eps filename suffix will be assumed under latex, 
% and a .pdf suffix will be assumed for pdflatex; or what has been declared
% via \DeclareGraphicsExtensions.
%\caption{Simulation results for the network.}
%\label{fig_sim}
%\end{figure}

% Note that IEEE typically puts floats only at the top, even when this
% results in a large percentage of a column being occupied by floats.

% %%%%%%%%%%%%%%%%%%%%%%%%%%%%%%%%%%%%%%%%%%%%%%%%%%%%%%%%%%%%%%%%%%%

\section{Data and Basic Metrics}
\label{sec:data}

\subsection{The Cutout Sample}
\label{sec:cutouts}

In this manuscript we analyze the 
nighttime \ac{L2} \ac{SST} Aqua  \ac{MODIS}
dataset (https://oceancolor.gsfc.nasa.gov/data/aqua/) 
for the period 2003-2021 inclusive.
%\deleted[id=pc]{[Remove clouds with PC algorithm?]} 
Our approach to data extraction and pre-processing
follow the procedures described in
\cite{ulmo}.  Specifically, we first generate \textit{patches} with a size of approximately ${128 \times 128}$\,pixel$^2$ that are at least
$96\%$ clear, land-free 
and of good quality.
This is a stricter clear criterion
than the $95\%$~clear criterion
adopted in \cite{ulmo} as subsequent work has identified
spurious results from improperly masked clouds
or poor in-painting applied to images
with $> \sim 5$\%\ cloud coverage.
We consider this to be the best trade-off between
sample size and quality for the analysis
presented here.

The patches were drawn randomly from the \ac{MODIS}
granules, restricted to lie within 480~pixels of nadir
(to maintain $\sim 1$-km resolution)
and required to have less than 50\%\ overlap.
By allowing for partial overlap, we generate more
`views' of a given field.

We then pre-process each patch by: 
 (i) in-painting the bad (flagged) pixels;
 (ii) median filtering with a 3-pixel window in the 
 along-track direction;
 (iii) resizing the patch to a ${64 \times 64}$\,pixel$^2$ \textit{cutout} 
 using the local mean of each $2 \times 2$ region; and
 (iv) subtracting the mean temperature to produce an
\sst\ anomaly (i.e.,\ \acs{SSTa}) cutout.
Table~\ref{tab:cutouts} lists a small
set of the \ncutouts\ cutouts
resulting from the parent sample;
the full table is available online.

%We note that the 99\%\ constraint deviates from \cite{ulmo}
%who used a 95\%\ clear criterion.
%Our investigations since that publication revealed that
%when the in-painting algorithm is applied to cutouts
%with $\sim 5\%$ clouds can introduce spurious results.
%These spurious effects are largely mitigated in 99\%\ clear
%data, but there are two knock-on effects:
% (i) the dataset is reduced to $\approx 1/3$ of the parent sample;
% (ii) sharp \ac{SST} gradients can be erroneously flagged as
% clouds and thereby not satisfy our stricter cloud-free criterion.
%[Comment on frequently this may occur]

% %%%%%%%%%%%%%%%%%%%%%%%%%%%%%%%%%%%%
\subsection{\texorpdfstring{\LL}{}~and~\texorpdfstring{\DT}{}} %\label{sec:cutouts}

To build physical intuition for the results that follow, 
we introduce several metrics to describe the 
individual cutouts.  Two of these metrics detailed in
this sub-section are the \ulmo\ \ac{LL} metric briefly
described in the introduction
and the temperature
range \DT\, defined as the difference in $90$th and $10$th percentiles of the \ac{SSTa} distribution:
$\DT \equiv T_{90} - T_{10}$ where $T_{N}$ denotes the
$N^{th}$-percentile. 
%\textcolor{red}{Is the SSTa distribution calculated for a single cutout?} %--i.e.,\ the value of temperature anomaly T_{N} below which $N$ percent of the data can be found.

For each cutout we calculate the \ac{LL} metric 
using the \ulmo\ \ac{PAE} of \cite{ulmo}.  
Conceptually, the \ac{LL} metric describes the relative
occurrence of a given image within the
ensemble distribution.
%This distribution, however, is more complex that
%one generated by a simple statistic
%(e.g., the mean temperature).  
Here, the distribution
is constructed from the set of latent vectors derived from
a deep-learning autoencoder designed to
capture the features and patterns of \sst\ imagery.
To formally calculate \ac{LL}, the latent space of
the autoencoder outputs is transformed to a 512-dimension
Gaussian manifold with a normalizing flow.  The 
\ac{LL} metric is then the sum of the log probability
from the 512 Gaussians and the log of the
determinant of the transformation.
These give the normalization of the
\ac{LL} distribution \cite{Morningstar2021DensityOS}.

%This \ac{LL} metric is constructed to 
%describe the relative probability of observing a 
%given image within the ensemble distribution.
%Unlike simple metrics (e.g., the standard
%deviation of \ssta), the distribution analyzed are
%the image ``patterns'' as encoded by the deep learning
%algorithm.  The \ac{LL} is then simply the evaluation
%of the probability product of a 512-dimension Gaussian
%space that describes the image latent vectors.  [challenging!]

\begin{figure}[ht]
\centering
\includegraphics[width=0.5\textwidth]{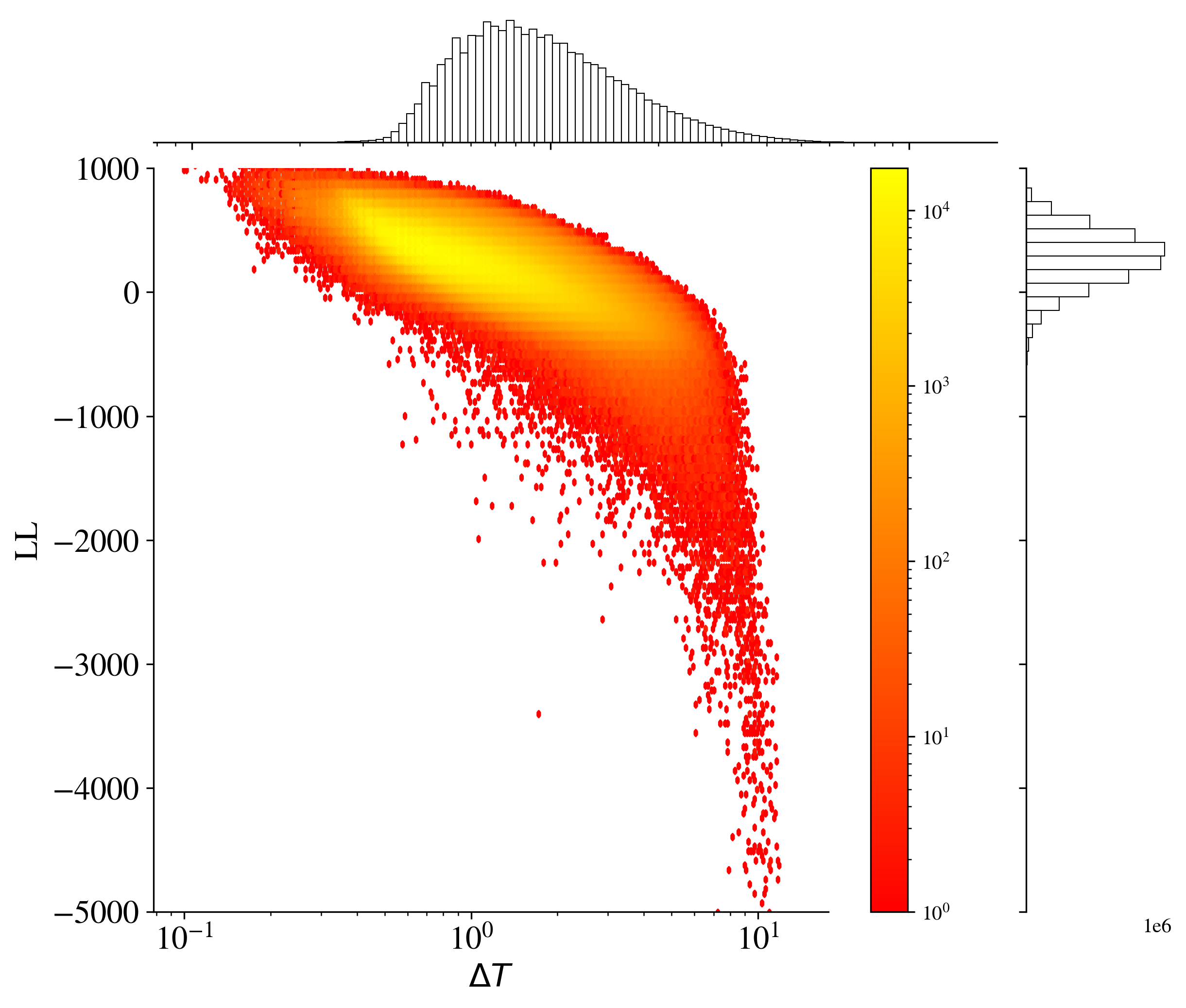}
\caption{2D histogram of the distribution of 
the \ac{LL} metric vs.\ temperature range \DT, with the 
color proportional to the log of the counts in each bin.
The two distributions are strongly correlated, with 
the highest \DT\ exhibiting
the lowest \LL\ values. 
The marginal distributions are shown as simple histograms.
While the \ac{LL} distribution is well-described by 
a Gaussian, it does exhibit an asymmetric tail to 
very low values which are the extreme outliers
of the dataset \cite{ulmo}. 
}
\label{fig:LLvsDT}
\end{figure}

During their analysis, \cite{ulmo} discovered that
the \ac{LL} values of the cutouts correlate strongly
with the temperature range.
Put simply, cutouts with high \DT\ are rare and yield
lower \LL\ values on average.
%$\DT \equiv T_{90} - T_{10}$ where $T_{X}$ is the
%X$^{th}$-percentile of the \ssta\ distribution.
We have calculated and tabulated \DT\ for all
of the cutouts (Table~\ref{tab:cutouts}),
recovering a LogNormal distribution after
allowing for a shift and scale, 

\begin{equation}
    p(\Delta T') = \frac{1}{\sigma_{\DT} \Delta T' \sqrt{2\pi}} \, 
     \exp \left ( - \frac{\ln^2 \Delta T'}{2 \sigma_{\DT}^2} \right )
\end{equation}
$\Delta T' = (\DT - l_{\DT})/S_{\DT}$
finding
$\sigma_{\DT} \approx 0.58$, 
$l_{\DT} \approx 0.1$, and
$S_{\DT} \approx 0.83$. 
In comparison, the \ac{LL} distribution is well-described
by a Gaussian with mean
$\bar x_{\rm LL} \approx 270$ and
standard deviation $\sigma_{\rm LL} \approx 227$, 
albeit with a long
tail to very low values that define the outlier
sample of \cite{ulmo}.

Figure~\ref{fig:LLvsDT} shows the \ac{LL} vs.\ \DT\ 
bivariate distribution; this reveals 
the strong correlation; 
e.g., cutouts with
high \DT\ exhibit the lowest \ac{LL}.
Despite the correlation, one should not 
over-interpret Figure~\ref{fig:LLvsDT}
to conclude that the \ac{LL} metric is solely 
tracking \DT\ and therefore could
be replaced by it.
First, the relationship is not 
strict; there
is a wide range of \ac{LL} values at a given \DT.
Furthermore, one expects a correlation with \ac{LL}
for any simple metric that describes 
the features in the imagery.  However, 
no single
metric can fully describe the complexity of 
\sst\ imagery. 
% and thereby replace \LL\ which 
%is itself a single valued metric.
Stated differently, the complexity of \sst\ imagery
belies any single, simple statistic -- including
the \ac{LL} measure of our previous work.
This motivates, in part, the application of 
an alternative, deep learning technique. 
As important, we found that the latent space
of \ulmo\ did not group together cutouts with
similar patterns.
This too motivates the contrastive learning
model adopted here.

% Figure slopes
\begin{figure}[ht]
\centering
\includegraphics[width=0.5\textwidth]{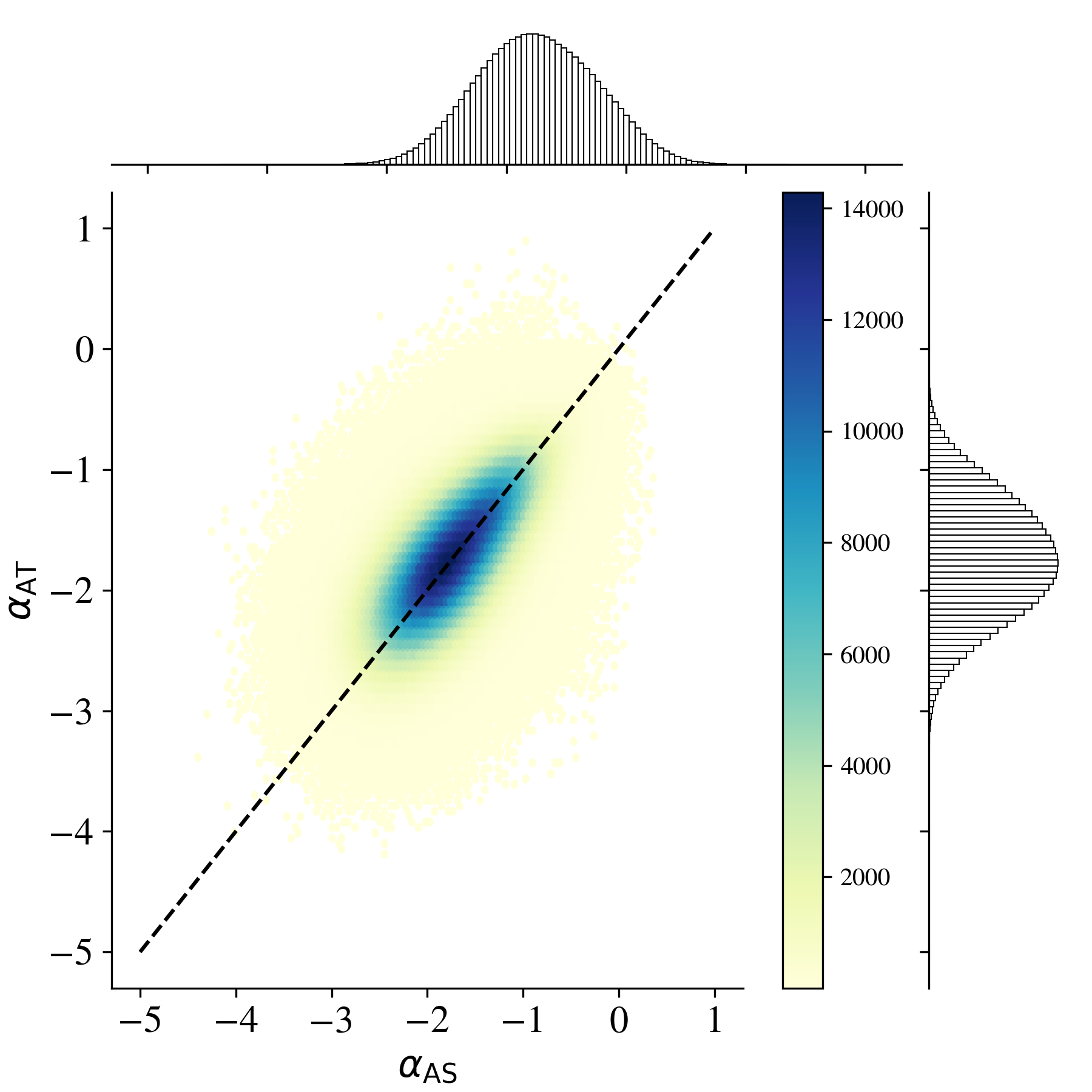}
\caption{
Bivariate distribution of the along-scan spectral slope
\zslope\ versus the along-track \mslope, each 
determined from the portion of the spectra for wavelengths between 12 and 50\,km.
Typical uncertainties in the measurements are one
to several tenths (fitting error only).
We find that $>99\%$ of the cutouts
have \zslope\ and \mslope\ within
one $\sigma$ of the estimated error
from the 1-to-1 line,  indicating 
their values are highly correlated.
The good correspondence between the \zslope\ and \mslope\ 
implies isotropic motions within the ocean on 
these physical scales.
}
\label{fig:slopes}
\end{figure}

% %%%%%%%%%%%%%%%%%%%%%%%%%%%%%%%%%%%%
% %%%%%%%%%%%%%%%%%%%%%%%%%%%%%%%%%%%%
\subsection{Slopes of the Power Spectrum}
\label{sec:power_spectrum}

Another metric often used to describe structure in \sst\ fields 
is the slope $\alpha$ of the 1D power spectrum;
i.e., the slope of the energy in wavenumber space. 
When estimated over fine horizontal scales, such as sub-mesoscales ($1-50$\,km),
the slope can be descriptive of underlying turbulent processes. For example, classical quasi-geostrophic (QG) turbulence predicts a spectral slope of $\alpha = -3$, surface quasi-geostrophic (SQG) turbulence predicts $\alpha = -5/3$, and an internal-wave continuum predicts $\alpha=-2$ (i.e.,~for scales less than
10\,km and flatter otherwise) \cite{charney71,gm72,callies13}.

For the cutouts analyzed here, the smallest scales are
expected to be dominated by instrument noise.
Therefore, we focus the analysis on larger scales
($\lambda \approx 12-50$\,km), which corresponds 
to the upper end of
wavelengths associated with sub-mesoscale processes
(depending on latitude). Focusing on this range also mitigates, to some extent, contributions to the spectral power at smaller scales due to the multi-detector nature of \ac{MODIS}. Note, however, that a wavelength of $12$\,km nominally resolves two eddy-like anomalies of opposite sign, each with radii of $3$\,km. Since the mixed layer deformation radius of sub-mesoscale anomalies typically exceeds $3$~km, this resolves most sub-mesoscale phenomena except perhaps those at polar latitudes.

%$\lambda_{ml} \sim (N/f)h_{ml}$, where $N/f \sim 50$~$100$ is a non-dimensional stratification and $h_{ml} \sim 50$-$100$~m is a mixed layer depth, 

For each cutout, we determine ensemble averaged power spectra independently for the along-scan (AS) and 
along-track (AT) directions.
Specifically, for each row (or column)
of the cutout, we detrend the data with a linear fit 
and subtract the mean.  We then calculate the 
fast fourier transform for each row/column and
ensemble average the results by row/column.  This signal is median
filtered with a window of 5 pixels and a power-law
is fit to the resultant power spectrum with related 
uncertainty.
We refer to these slopes as \zslope\ and \mslope\ for
the AS and AT directions, respectively.

% Figure slopes
\begin{figure}[ht]
\centering
\includegraphics[width=0.5\textwidth]{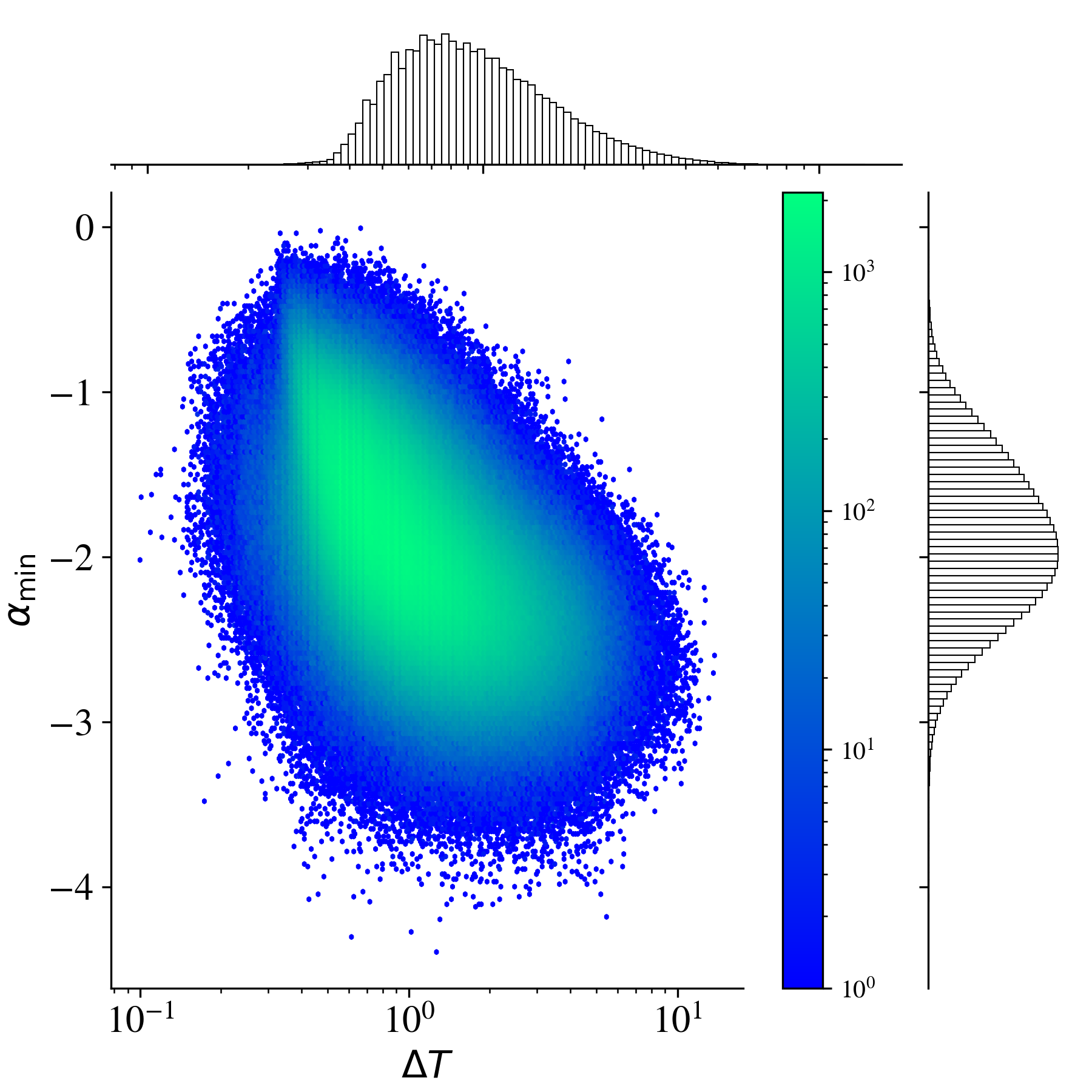}
\caption{
2D histogram of the characterstic slope \slope\ 
of each cutout vs.\ \DT. 
Regions of the ocean with larger temperature range
also tend to exhibit a steeper power spectrum 
on $\sim 10-50$km scales, 
albeit with substantial scatter.
}
\label{fig:slopevsDT}
\end{figure}

We applied this algorithm to all of the cutout images
to calculate \zslope\ and \mslope\ independently
(Table~\ref{tab:cutouts})
and also to estimate their uncertainties. 
Figure~\ref{fig:slopes} plots the two slopes against one 
another.  
The marginal distributions of \zslope\ and \mslope\ 
are well described by Gaussians with
nearly identical mean and standard deviations:
$\bar x_\alpha \approx -1.74$, $\sigma _\alpha \approx 0.5$.
Furthermore, we find the slopes for a given
cutout agree within uncertainty
for the overwhelming majority of cases 
(i.e.,~$99\%$ fall within one $\sigma$ of the mean).
This suggests a high degree of isotropy within 
the dataset.
%, unless there were an aniostropy oriented
%orthogonal to the nadir track which is 20\,degrees
%off the meridian.  One could perform a 2D fourier
%transform to investigate this further.

In the following, we characterize each cutout by a single
spectral slope $\slope \equiv \min(\zslope, \mslope)$.
Figure~\ref{fig:slopevsDT} shows this quantity versus
\DT, revealing a correlation but with large scatter.
Given Figure~\ref{fig:LLvsDT}, one notes the \slope\ is
also correlated with \LL\, with the most 
negative \slope\ (largest in magnitude)
cutouts having preferentially lower {\LL} (as one would expect).
%but we again emphasize
%that neither \DT\ nor \slope\ can capture the 
%complexity of structure in \sst\ imagery.

% %%%%%%%%%%%%%%%%%%%%%%%%%%%%%%%%%%%%%%%%%%%%%%%%%%%%%%%%%%%
% %%%%%%%%%%%%%%%%%%%%%%%%%%%%%%%%%%%%%%%%%%%%%%%%%%%%%%%%%%%
\section{Methodology}
\label{sec:methods}

%\subsection{Outline of Methodology section}
%\begin{itemize}
%\item big picture of the task

%\item brief introduction to the SSL: SimCLR
%
%\item Details of Analysis: 
%\begin{itemize}
%    \item (pre-processing), 
%    \item augmentations, 
%    \item model structure, 
%    \item training (validation) and evaluation, 
%    \item latents visualization,
%\end{itemize}
%\end{itemize}
%
\subsection{Overview}

Recently, several simple yet powerful {\ssl} 
frameworks have been proposed to learn useful representations 
(latents) from high dimensional data sets. This includes 
so-called contrastive losses which maximize the ``similarity" of the views from the same data samples in the latent space \cite{hadsell2006dimensionality, bachman2019learning, chen2020simple, chen2020big, chen2020improved, he2020momentum, tian2020makes}. 
In our work, we apply the SimCLR framework to analyze the preprocessed cutouts ($64\times64$\,pixel$^2$ arrays) from the MODIS granules by designing task-dependent augmentations based 
on our domain knowledge of \sst\ data. 
The latents obtained by our model are, as designed,
a transformation-invariant representation of the \sst\ data.
We then study structure in the latent space by projecting 
into a two-dimensional vector space 
using a \ac{UMAP} \cite{2018arXivUMAP}. 
The reader is also encouraged to study the methodology
developed by \cite{cheng21} using the scattering transform
\cite{mallat2012}.

\subsection{An Introduction to SimCLR}

SimCLR is an efficient but simple self-supervised learning model in which a base neural encoder is used to transform the augmented input data point to a latent vector \latentv. 
By minimizing the contrastive loss, 
an essential representation can be extracted \cite{chen2020simple}. The contrastive loss function defined for an augmented sample pair $\{ \mlatentv, \tilde{\mathbf{z}}^{(i^{\prime})}\}$ is given by:

\begin{align}
\label{eq: contrastive_loss}
   \text{loss}_{i, j} = -\text{sim}(\tilde{\mathbf{z}}^{(i)}, \tilde{\mathbf{z}}^{(j)})/\tau
   + \log{\sum_{k: k\neq i}^{2m}}\exp{(\text{sim}(\tilde{\mathbf{z}}^{(i)}, \tilde{\mathbf{z}}^{(k)})/\tau)},
\end{align}
where $\text{sim}(\tilde{\mathbf{z}}^{(i)}, \tilde{\mathbf{z}}^{(j)})$ is defined as the dot-product similarity of the latent pair $\{\tilde{\mathbf{z}}^{(i)},\tilde{\mathbf{z}}^{(j)}\}$, 
namely the  output of the neural encoder, and $\tau$ is a 
hyper-parameter used to fine-tune the amplitude of the similarity. In essence, the contrastive loss (\ref{eq: contrastive_loss}) is just the rescaled similarity of the sample pair with a batch normaliziation penalty term. 

For our model, we use ResNet-50 \cite{he2016deep} as the encoder 
because unsupervised learning favors models with larger capacity \cite{chen2020big}. 
We do not employ a projection head, and the latents generated by the encoder are passed directly to the contrastive loss function, without additional nonlinear transformation.
%We do not consider that the projection head and the latents produced by the encoder are fed into the contrastive loss function without further non-linear transformations.
The advantage of discarding the projection head is that a 
reasonable similarity can be defined for the latents obtained by the encoder during the evaluation 
step\footnote{If the project head is used, the similarity in the contrastive loss is defined by the 
dot-product of the transformed latents.}. 
After some experimentation, we have set the dimensionality of
the latent space to \ndim.

% %%%%%%%%%%%%%%%%%%%%%%%%%%%%%%%%%%%%%%%%%%%%%%%%%
% %%%%%%%%%%%%%%%%%%%%%%%%%%%%%%%%%%%%%%%%%%%%%%%%%
\begin{figure}[ht]
\centering
\includegraphics[width=0.5\textwidth]{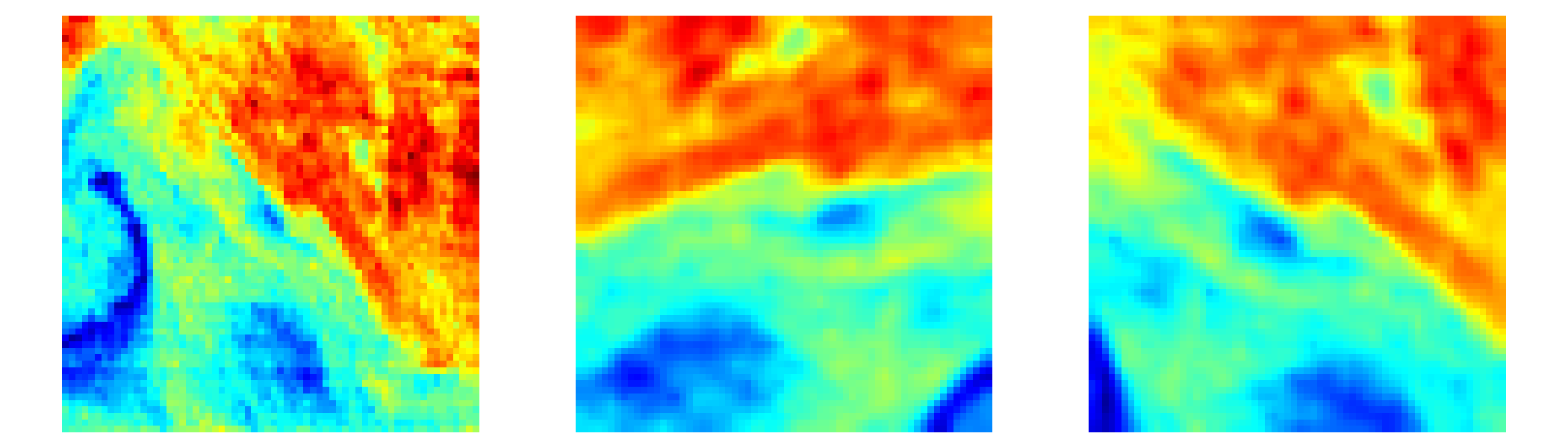}
\caption{The left panel is the `raw' cutout input to
the augmentation module and the other
two images are a pair of augmented samples.
The first (center image) was randomly flipped
about the vertical axis and then rotated by $\approx 320$\,deg
and jittered by 5~pixels vertically.
The second (right image) was not flipped, was
rotated by $\approx 18$\,deg and jittered by 2~pixels
horizontally and vertically.
Each augmented image was
then cropped by the inner $40 \times 40$~pixel$^2$, resized to $64 \times 64$\,pixel$^2$ and demeaned.
}
\label{fig:augmentation}
\end{figure}

\subsection{Augmentations}
\label{sec:augment}

The random augmentation module is crucial for model 
performance and affects the nature of the patterns
captured in the latent space.
After experimenting with standard augmentations intended
for natural images, we refined based 
on the following domain considerations: invariance to 
 (i) rotation, 
 (ii) translation, and
 (iii) reflection.
 The reason for the first consideration is that the MODIS L2 data product is not oriented
 parallel to either the zonal or meridional directions. 
Moreover, while there may be scientific motivations for preserving orientation when analyzing  \ssta\ fields more generally,   the north-south gradient in solar insolation and in beta has little impact on the gradient in SST at the scales of interest here 
($<80$\,km).
%sub-mesoscale process do not ``feel'' the effect of beta (i.e., the meridional gradient of Coriolis) and are therefore more or less ignorant of direction. 
This is supported by Figure~\ref{fig:slopes}. 
The reason for the second consideration is that we desired a description of the
 \sst\ patterns that is insensitive to small spatial shifts. That is,
 viewing the region offset by, for example, $10$\,km typically results in a similar pattern. Finally, the impetus for the third consideration is that we suspect a reflection or ``mirror image'' of 
 the original cutout should capture all of the same information. As an aside, we also considered inverting the \ssta\ field
 (i.e., turning warm to cold), but did not do so in the end since some
 dynamic features have a physical preference
 (e.g., mesoscale currents are generally warmer than
 the surrounding waters).

With these considerations in mind, we imposed the following
augmentations in this order:

\begin{itemize}
    \item Random Flip: We randomly flip the image in each 
    dimension with a 50\%\ probability (e.g.,\ a 25\%\ 
    probability of no alteration).
    \item Random Rotation: We rotate the cutouts with random 
    angles which are uniformly sampled 
    from $\theta = [0, 2\pi]$.
    The new image maintains $64 \times 64$\,pixel$^2$ with
    data rotated `off' now lost and otherwise empty regions
    imputed with 0 values.
    %a smaller physical size per pixel than the original.
    \item Random Jitter and Crop: 
    We randomly select a new image center uniformly offset from
    the original by up to 5 pixels in each dimension.
    %move the crop center around the center of the input image ($64\times64$) by $[0, 5]$ pixel with equal probability. Then 
    We crop the original image around the translated center 
    to $40 \times 40$\,pixel$^2$ to eliminate the inclusion of the zero-filled portions introduced by the rotation. 
    Last, we resize the cropped image back to the original grid size
    of $64 \times 64$\,pixel$^2$.
    \item Demean:
    The final image is demeaned.
    %\item Gaussian Noise: 
    %We sample noise from a normal distribution with zero mean
    %and standard deviation $\sigma_{\rm noise} = 0.05$ 
    %(using the $\approx 0.1$\,K random noise associated with the original pixels reduced by the four pixel average, neglecting the temperature dependence) and
    %add it to the sample.
\end{itemize}

Figure~\ref{fig:augmentation} shows a representative cutout
and two of its randomly augmented samples.
These were randomly flipped, rotated  and jittered, 
and then cropped to the inner $40 \times 40$\,pixel$^2$, resized to $64 \times 64$\,pixel$^2$ and demeaned.
Our \ssl\ model, which operates solely on  
augmented images, learned to recognize
these as similar and to distinguish their
underlying patterns from those of other cutouts.

\subsection{Training and Evaluation}

For training of the model, we randomly selected \ntrain\ cutouts from the full set.
For validation, we randomly selected another \nvalid\ cutouts.
%We split these data into two parts: a training set and a
%validation set. 
With $4$~GPUs, we train \mname\ with a batch 
size of \nbatch. 
To monitor the contrastive lost on the validation set, we find that after 
$\approx 5$~epochs of training, the model learns good 
representations of the training data set. 
SSL models benefit from bigger batch sizes; our experimentation 
showed good performance for sizes of \nbatch\ or more.

\begin{figure}[ht]
\includegraphics[width=0.5\textwidth]{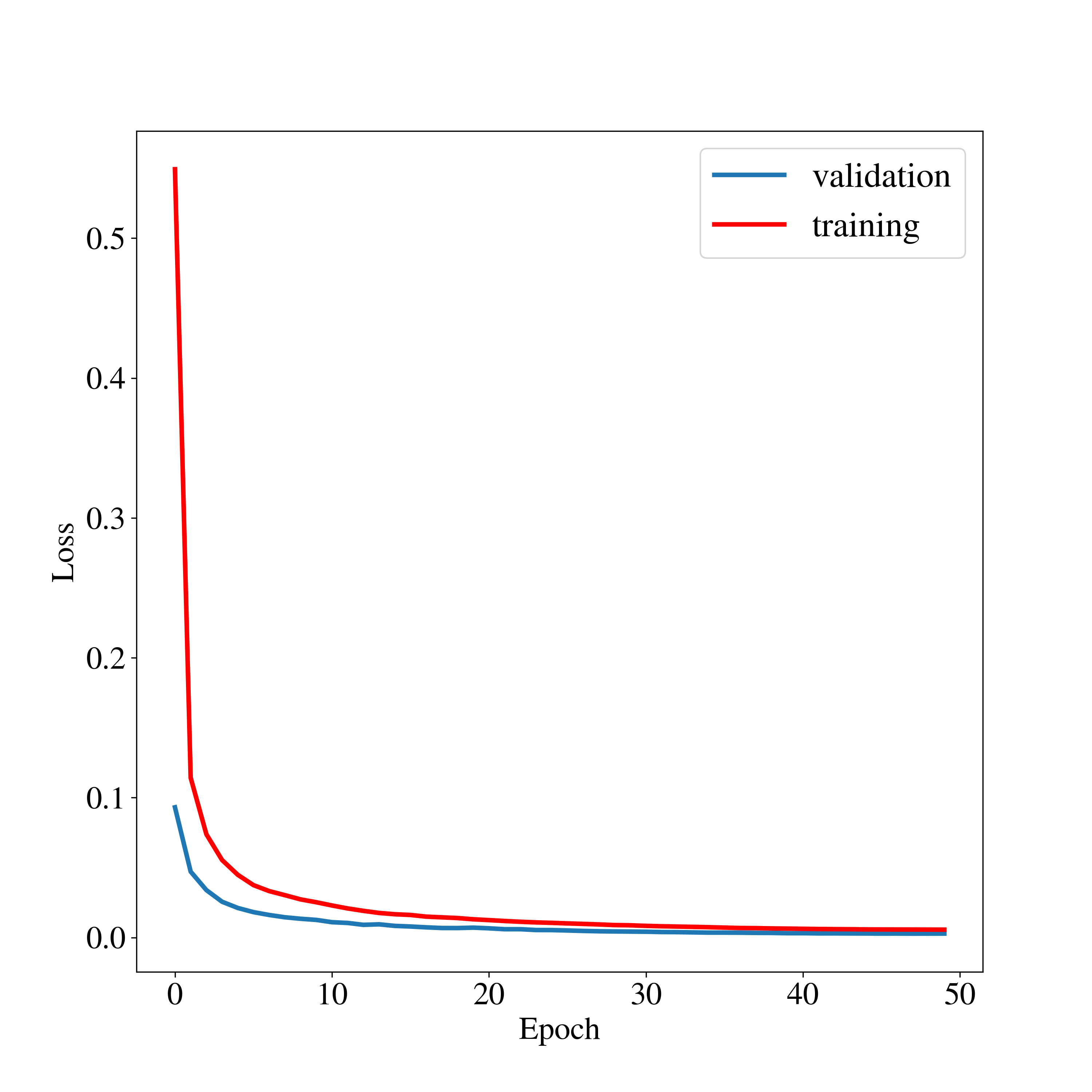}
\caption{Learning curves (batch losses) for the training set (red)
and validation set (blue) for our {\ssl} model.
The model achieves a low loss after $\approx 5$~epochs 
and has effectively converged after $\sim 20$.
}
\label{fig:learning_curve_ssl}
\end{figure}

In Fig.~\ref{fig:learning_curve_ssl}, we show the learning 
curves for the \mname\ model in the training 
and validation process. 
%Fig.~\ref{fig: learning_curve_train} shows 
%the learning process of the \ssl\ model on the training set. 
Learning loss decreases rapidly with training batch number suggesting that the model converges well. 
%However, in Fig.~\ref{fig: learning_curve_test} we find that the validation loss 
%of the \ssl\ model oscillates around a decreasing trend. We think the oscillation is due to the non-convexity of the contrastive objective in the \ssl\ model which corresponds 
%to the fruitful patterns in the 
%\ssta\ cutouts. 
%After the first several epochs, our \ssl\ model learns 
%most of the common and dominant patterns in the viewed 
%data so the loss decreases to a local optimum.
%By training more epochs, our \ssl\ model will get 
%to another local optimum which leads to the 
%oscillation of the validation learning curve. 

%The task of our model is to learn a representation of 
%\ssta\ cutouts, so unlike the supervised learning problems (classification, regression, etc) there is not an objective criterion to verify how good the representation is, so background knowledge is necessary to help us compare the models 
%at different local optimas, namely we should analyse the latents of the models with different 
%hyper-parameters according to their performance in our downstream tasks to determine the hyper-parameters (model comparing).  
%With about $6$ epochs, we find that good patterns are learned by the model after about $(6)$ epochs training by analysis of the patterns in the latents.

We then evaluated all 21~years of \ac{MODIS} 
cutouts, generating and recording a unique
\ndim-dimension latent vector for each.
These constitute the learned representations
of the fundamental {\sst} patterns of the
imagery on scales $\leq 80$\,km.

\subsection{UMAP Dimensionality Reduction}
\label{sec:umap}

While the latent vectors encode the features 
characterizing the \ac{SST} patterns, it is difficult
as humans
to visualize and explore this \ndim-dimension latent
space.  Therefore, we have implemented the \ac{UMAP}
algorithm to further reduce the latent space to two
dimensions ($U_0, U_1$).  
We adopt the default parameters of the 
\ac{UMAP} package \cite{McInnes2018} throughout.

We emphasize that neither of the parameters
$U_0, U_1$ need to have a direct, physical
meaning or analogue. These are only statistical in nature, although they
are tuned to separate the latent vector 
space and therefore track the fundamental 
patterns of \sst\ imagery. 
%The following section explores the outputs of this procedure and also their scientific implications.
%We first fitted a
%standard two-dimensional \ac{UMAP} model to 
%\numap~random latent vectors from \tyear\
%and then applied this transformation to the entire
%dataset.
Only the 256~dimension latent
vectors are the full model representation of the underlying
patterns--i.e., the two-dimensions of this 
\ac{UMAP} are (by design)
a limited representation of 
the full complexity.  Nevertheless, we find
that by applying the \ac{UMAP} to a series of 
\DT\ sub-samples that we describe a 
majority of the diversity within 
the dataset.
The following section explores the outputs of this 
procedure and also their scientific implications.

\begin{figure}[h!]
\centering
\includegraphics[width=0.5\textwidth]{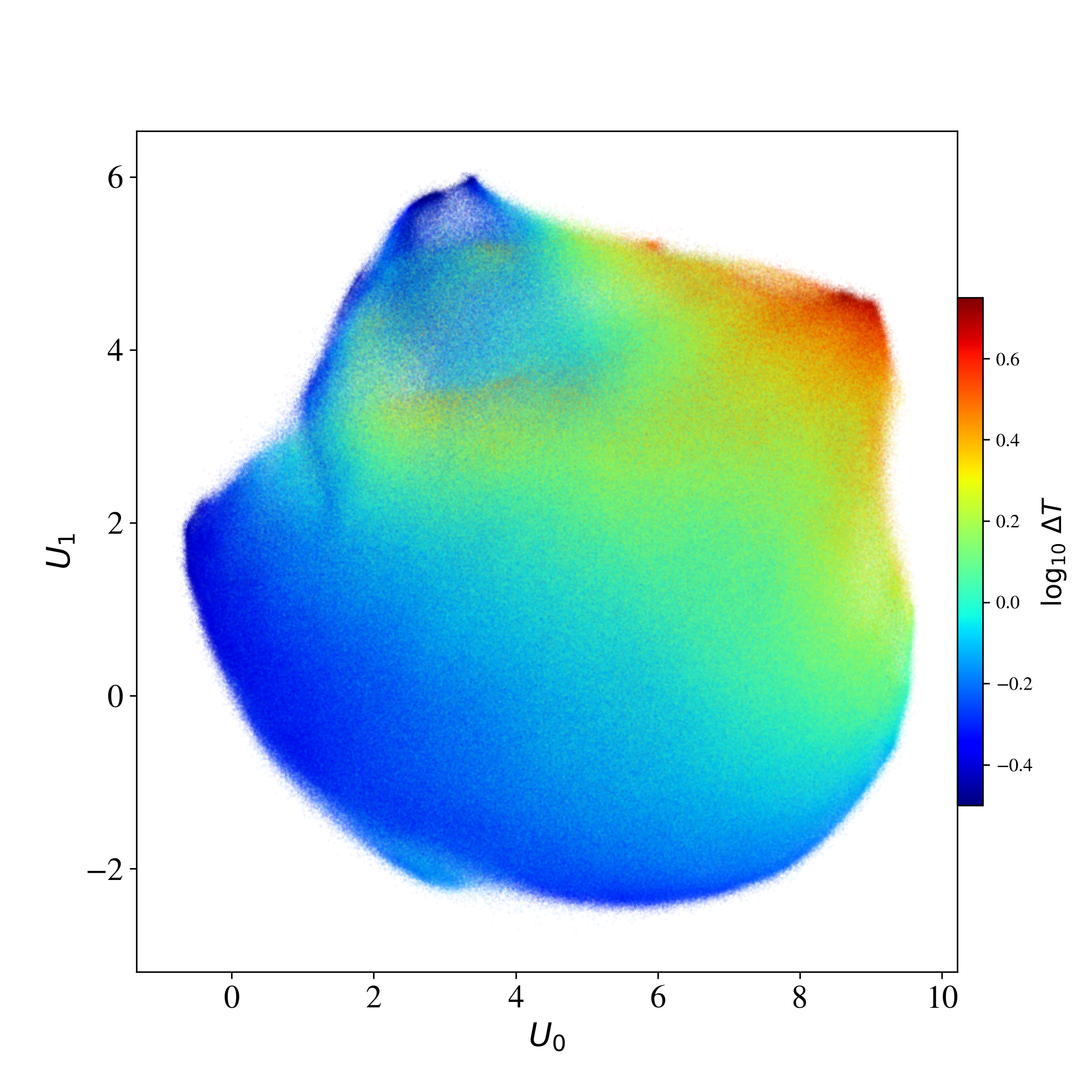}
\caption{\ac{UMAP} two-dimensional manifold description of 
the latent-vector space (\ndim-dimensions)
which represents the fundamental \sst\ patterns.  
Each cutout is shown as a dot in the space,
color-coded by $\log_{10} \DT$ of the cutout.
}
\label{fig:umap_full}
\end{figure}

% %%%%%%%%%%%%%%%%%%%%%%%
% %%%%%%%%%%%%%%%%%%%%%%%
% %%%%%%%%%%%%%%%%%%%%%%%
\begin{figure*}[ht]
\centering
\includegraphics[width=0.9\textwidth]{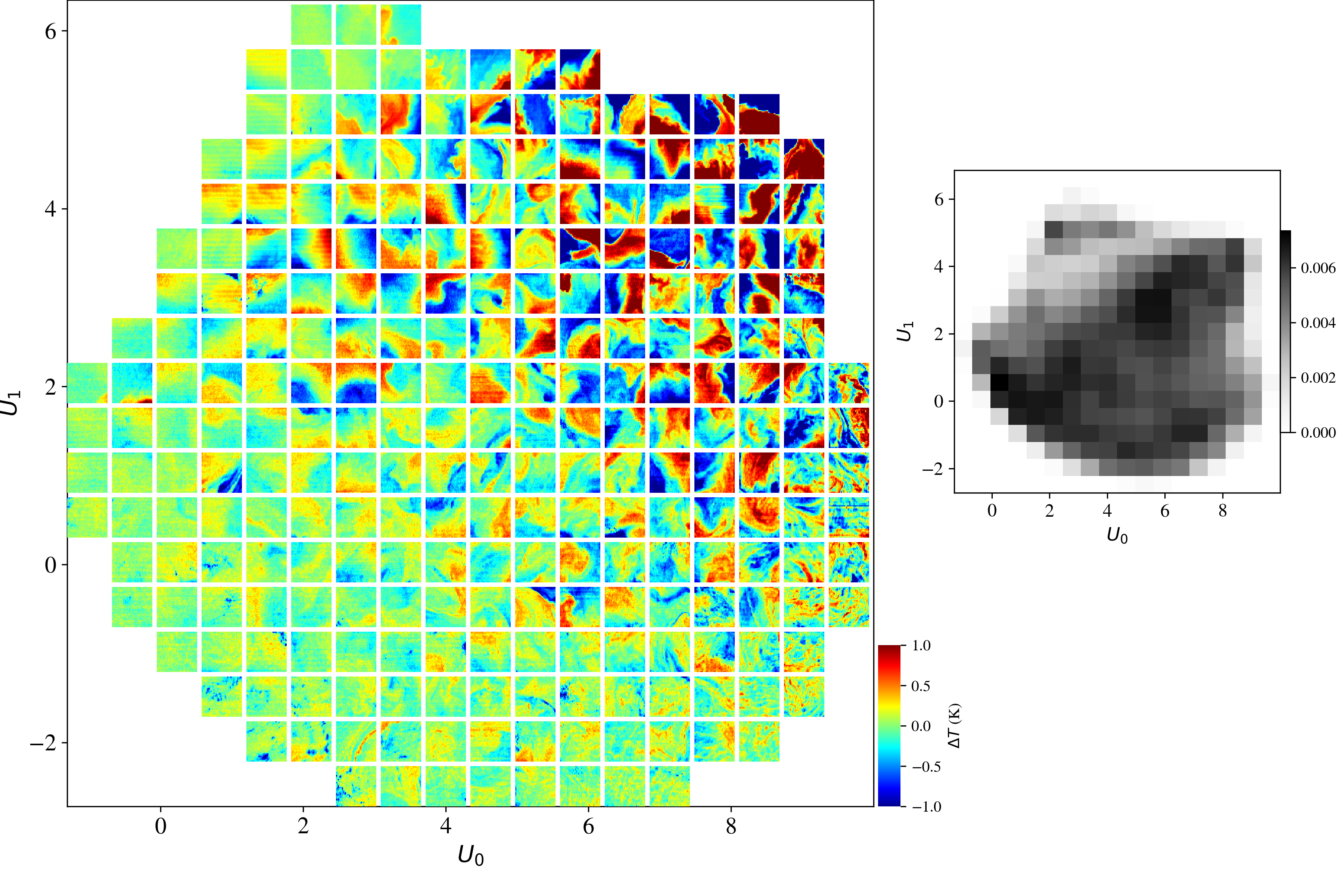}
\caption{
Gallery of cutouts, each one selected at random
from a gridded $U_0, U_1$ space for the full 
dataset.% (Figure~\ref{fig:umap_full}).
The data show increasing structure and complexity as one
transitions from the lower left
($U_0, U_1 \approx [0,-1]$) to the
upper right of the domain.  
Along this same direction the average \DT\ is increasing,
i.e., this embedding is especially sensitive to the cutout \DT.
The gray histogram on the right describes the relative
frequency of cutouts at each location in the \ac{UMAP} domain.
}
\label{fig:umap_gallery_all}
\end{figure*}
% %%%%%%%%%%%%%%%%%%%%%%%
% %%%%%%%%%%%%%%%%%%%%%%%
% %%%%%%%%%%%%%%%%%%%%%%%

% %%%%%%%%%%%%%%%%%%%%%%%%%%%%%%%%%%%%%%%%%%%%%%%%
\section{Results}
\label{sec:results}

The previous section described the deep learning 
methodology 
introduced to analyze the \ac{MODIS} \ac{L2} dataset.
Specifically, the \mname\ model was trained 
to associate images with similar \sst\ patterns on scales 
of $\approx 80 \times 80$\,km$^2$ and smaller, 
while separating these
from dissimilar images.  The resulting 256-dimension
latent space therefore represents a new method or \textit{vocabulary} for describing
\sst\ imagery on sub-mesoscales.
%The present section describes the principle results
%of applying the \mname\ model to the \ac{MODIS} \sst\ imagery.
This section examines a portion of the vocabulary
describing the rich complexity of \sst\ imagery.
To facilitate the exploration, we map the
256-dimension latent space to a reduced
basis (two dimensions) for the full dataset
and subsets in \DT.

% %%%%%%%%%%%%%%%%%%%%%%%%%%%%%%%%%
\subsection{Full Sample}

We begin by examining the full sample, applying
the \ac{UMAP} algorithm to all of
the \ncutouts~latent vectors.
Figure~\ref{fig:umap_full} shows 
the reduction of \ac{SST} patterns into
two dimensions for the full set of imagery,
color-coded by $\log_{10} \DT$. 
%which highlights a strong correlation between
%that metric and the $U_0$ dimension.  
%More precisely, the cutouts near the peak
%of the \ac{LL} distribution (i.e., most normal)
%lie at intermediate $U_0$ values with the extrema
%(high/low \ac{LL}) occurring at extreme $U_0$ 
%values, and a subset of very low \ac{LL} 
%at the lowest $U_1$. 
The relatively uniform distribution 
indicates the \sst\ imagery exhibits
a smooth continuum of patterns and features
as opposed to ``clusters'' of discrete classes.
Furthermore, by coloring the data by \DT, we emphasize
that this property is a dominant ``axis'' in separating
the \ac{SST} imagery
(here oriented at $\sim 45^\circ$ to the $U_0,U_1$ axes).  
The strong \DT\ dependence is relatively obvious in
hindsight; having removed the first moment in 
\ac{SST} (the mean), the second moment (variance)
remains a principle characteristic of each 
\ac{SSTa} cutout.
And while \DT\ has 
physical significance for the ocean (e.g., high
values preferentially arise in dynamically active regions),
we wish to discriminate the patterns
based on finer features.  These are captured
in the full 256-latent space and we access them
in the following section by sub-dividing the
full sample into \DT\ intervals and then reducing each
sub-sample with its own \ac{UMAP} algorithm.

%As detailed in the previous section, we applied a 
%\ac{UMAP} reduction primarily to provide
%a representation of the latent vector space that
%is tractable to human exploration.
%Figure~\ref{fig:umap_LL} demonstrates that 
%the imagery is well separated according to the
%\ac{LL} metric which tracks the full distribution 
%of patterns spanning the \ac{UMAP} space.

Before proceeding, we further
examine the \ac{UMAP} results for the full dataset
via a gallery of representative images.
We have gridded the $U_0,U_1$ space and 
selected a random cutout from within each grid cell;
these are plotted in Figure~\ref{fig:umap_gallery_all}.
%with a colormap that spans 
%\ssta=[-1,1]\,K. 
As designed, \mname\ has effectively separated the
\sst\ imagery by their patterns.
Qualitatively, the data is highly uniform
%($\DT \approx 0$\,K) 
in the lower left region of the \ac{UMAP} space 
$(U_0, U_1 \approx 0,-1)$ and one observes increasing 
complexity as one moves away from this ``origin''.
Greatest structure is evident 
at high $U_0, U_1$ values, with
patterns %at the upper right region of the 
%space are 
characterized by water masses of 
high positive and negative \ssta\ separated
by a sharp front.  These are likely associated with 
strong 
ocean currents, such as the Gulf Stream, Kuroshio, and Antarctic Circumpolar Current, regions of relatively large changes in shallow bathymetry often encountered near the edge of the continental shelf and in vertically stratified 
regions with strong winds such as off the Isthmus of Tehuantepec.

%Within that cohort, one identifies significantly
%higher temperature contrasts (\DT) at lower $U_1$.  
%Travelling to higher $U_1$ along the right edge,
%the patterns show progressively lower \DT\ but
%maintain regions with elevated temperatures
%throughout surrounded by a larger region with
%%\ssta~$\approx 0$\,K. 
%Nearly the opposite pattern -- regions of cooler
%water and otherwise \ssta~$\approx 0$\,K -- 
%are found along the bottom of the figure
%(i.e.\added[id=pc]{,} lowest $U_1$).
%These are rare (there are only XX\%\ of all
%cutouts in this region) and CLOUDS?.

%Figure~\ref{fig:umap_gallery_all} shows another gallery
%with a different random seed and with the
%color bar stretched to higher contrast
%($\delta T = [-1,1]\,\rm K)$.
%Starting at the origin $(U_0,U_1) = (-4,6)$,
%structure slowly emerges towards higher $U_0$ or $U_1$
%values and the diversity of patterns is striking.  
%On the left side (low $U_0$), the first
%structures are regions with isolated and small
%patches of cooler water.  At $U_0 \approx 0$,
%the patterns exhibit nearly equal portions of cool/warm
%water and these transition to larger patches with higher
%\ssta\ values by $U_0 \approx 2$.

% %%%%%%%%%%%%%%%%%%%%%%%
% %%%%%%%%%%%%%%%%%%%%%%%
% %%%%%%%%%%%%%%%%%%%%%%%
\begin{figure*}[h]
\centering
\includegraphics[width=\textwidth]{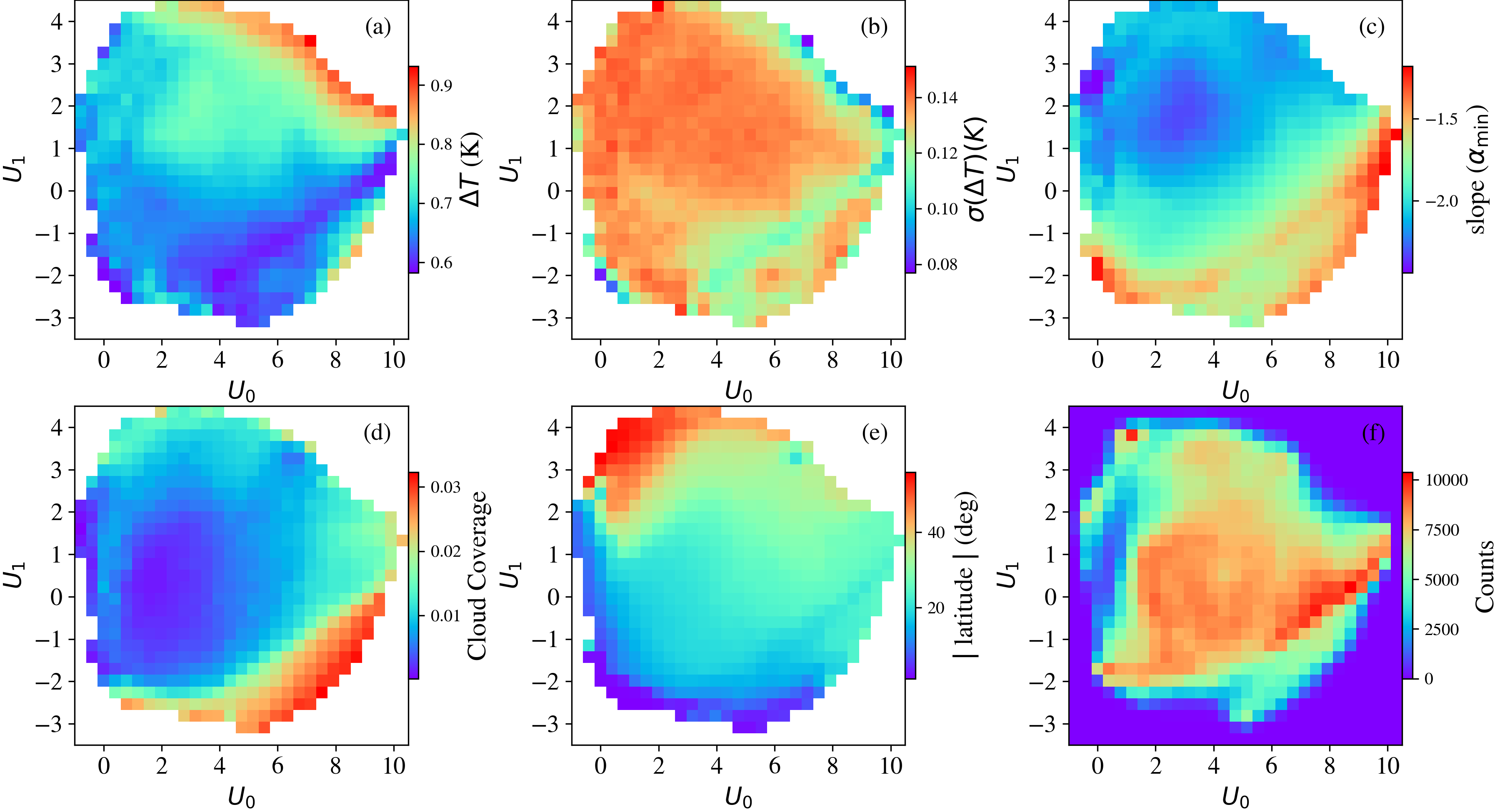}
\caption{
Binned evaluations in the \ac{UMAP}
embedding for properties of the cutout
subset with \DThalf.
Panels (a), (c), (d), and (e) show the median value
of {\DT} (with a linear color scale because of the relatively small temperature range compared with the log$_{10}$ scale used for Figure~\ref{fig:umap_full}), \slope, cloud coverage, and the absolute
latitude of the cutouts respectively.
Panel (b) presents the standard deviation in \DT\ 
and panel (f) records the number of cutouts in each bin.
%\ac{UMAP} embedding of the subset of cutouts with \DThalf,
%colored by \DT\ measured on the inner 
%$40 \times 40$\,pixel$^2$ of each cutout
%(i.e., the region analyzed by \mname).
%In contrast to the embedding of the full sample
%(Figure~\ref{fig:umap_full}), 
%we find the cutouts span the domain
%without a monotonic or
%uniform dependence on \DT.
%This is by design to differentiate the \sst\ patterns
%independently of \DT.
}
%\label{fig:umap_DT1}
\label{fig:umap_multi}
\end{figure*}
% %%%%%%%%%%%%%%%%%%%%%%%
% %%%%%%%%%%%%%%%%%%%%%%%
% %%%%%%%%%%%%%%%%%%%%%%%

% %%%%%%%%%%%%%%%%%%%%%%%%%%%%%%%%%%%%%%%%
\subsection{The Fundamental Patterns of \texorpdfstring{\ac{SST}}{}}

\begin{figure*}[ht]
\centering
\includegraphics[width=0.95\textwidth]{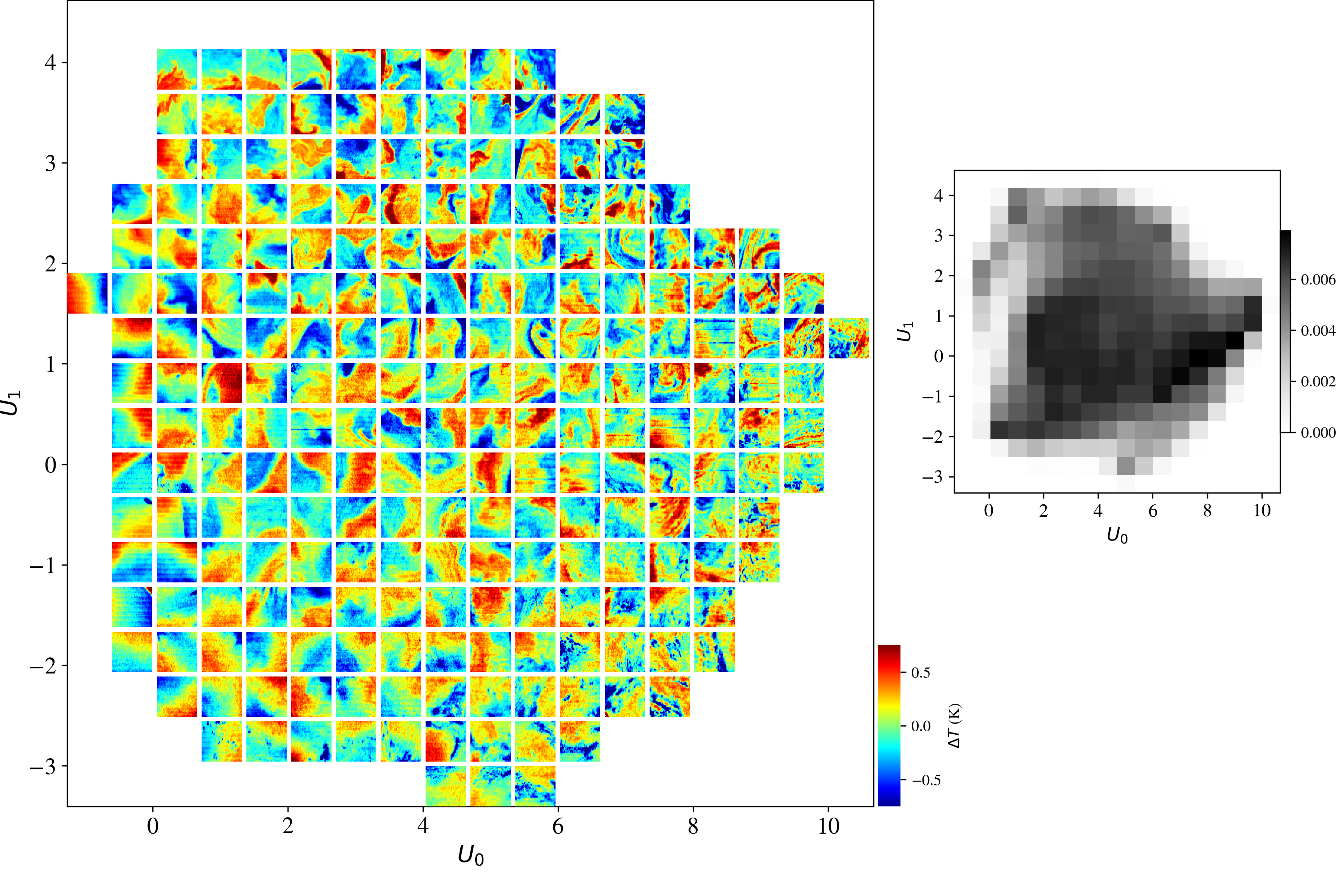}
\caption{
As in Figure~\ref{fig:umap_gallery_all} but for the \ac{UMAP}
embedding of the \DThalf\ sub-set of cutouts
(3,320,395 in total; a cutout is only
shown in $U_0,U_1$ bins
where there are at least 1,000).
In contrast to the full set where the \ac{UMAP} space was largely
defined
by the \DT\ of each cutout, this embedding separates the patterns
by finer features.
Spanning the domain, one identifies a diversity of
patterns with structure on the full range of scales.
As designed, \mname\ has effectively generated a
vocabulary to describe and distinguish 
between the fundamental patterns of \sst.
%By examining the gray histogram, one observes that several regions in $(U_0, U_1)$-space
%are sparsely populated 
%(e.g., the lower right and upper left regions of the domain).
%The black rectangles (dashed and dotted)
%in the left panel indicate 
%the portions of the \ac{UMAP} domain analyzed in 
%Figures~\ref{fig:global_shallow}
%and \ref{fig:global_turb}.
}
\label{fig:umap_gallery_DT1}
\end{figure*}

\begin{figure*}[ht]
\centering
\includegraphics[width=0.95\textwidth]{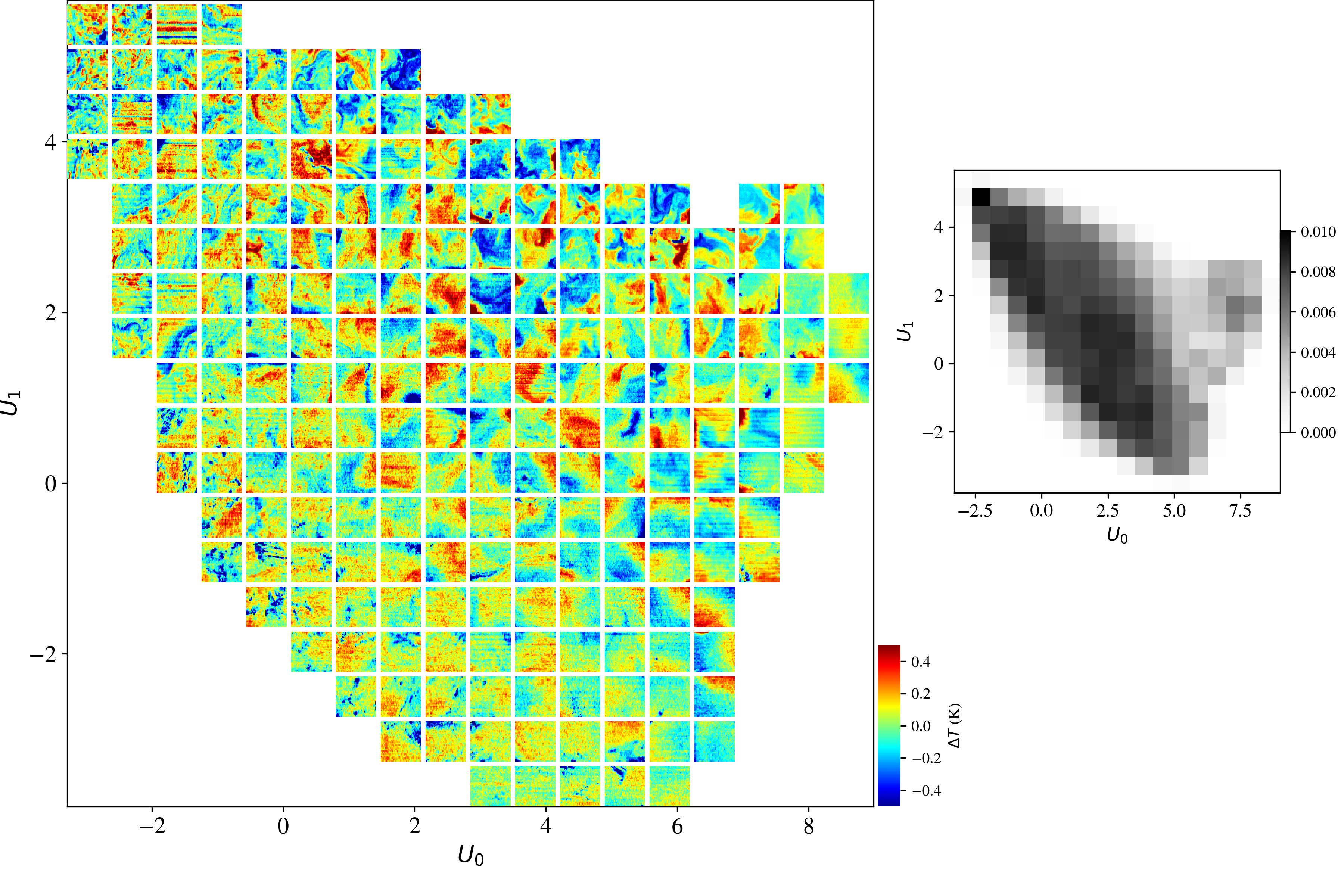}
\caption{
Same as Figure~\ref{fig:umap_gallery_all} but for the \ac{UMAP}
embedding of the \DTzero\ sub-set of cutouts.
}
\label{fig:umap_gallery_DT0}
\end{figure*}

\begin{figure*}[ht]
\centering
\includegraphics[width=0.95\textwidth]{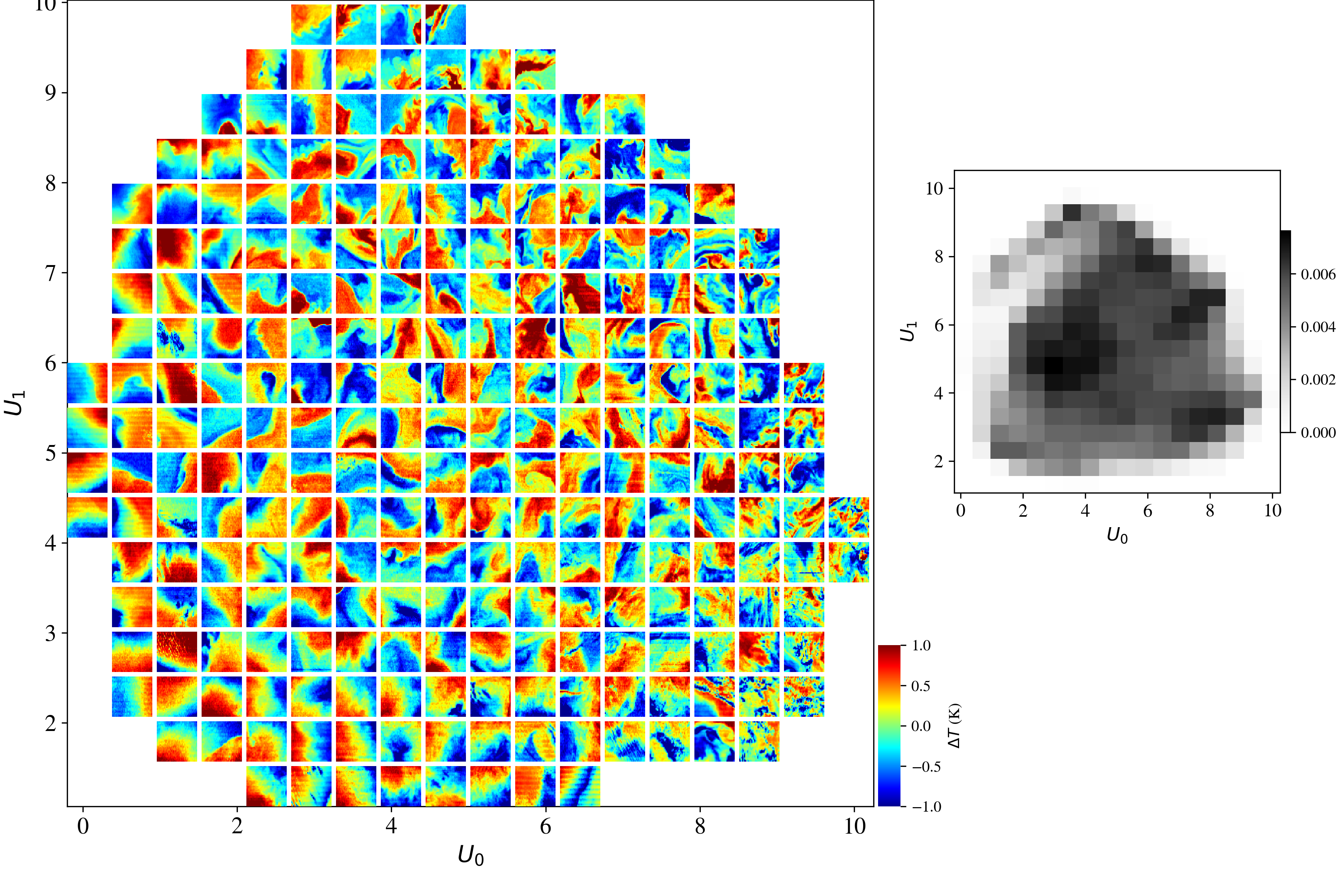}
\caption{
Same as Figure~\ref{fig:umap_gallery_all} but for the \ac{UMAP}
embedding of the \DTone\ sub-set of cutouts.
}
\label{fig:umap_gallery_DT15}
\end{figure*}

\begin{figure*}[ht]
\centering
\includegraphics[width=0.95\textwidth]{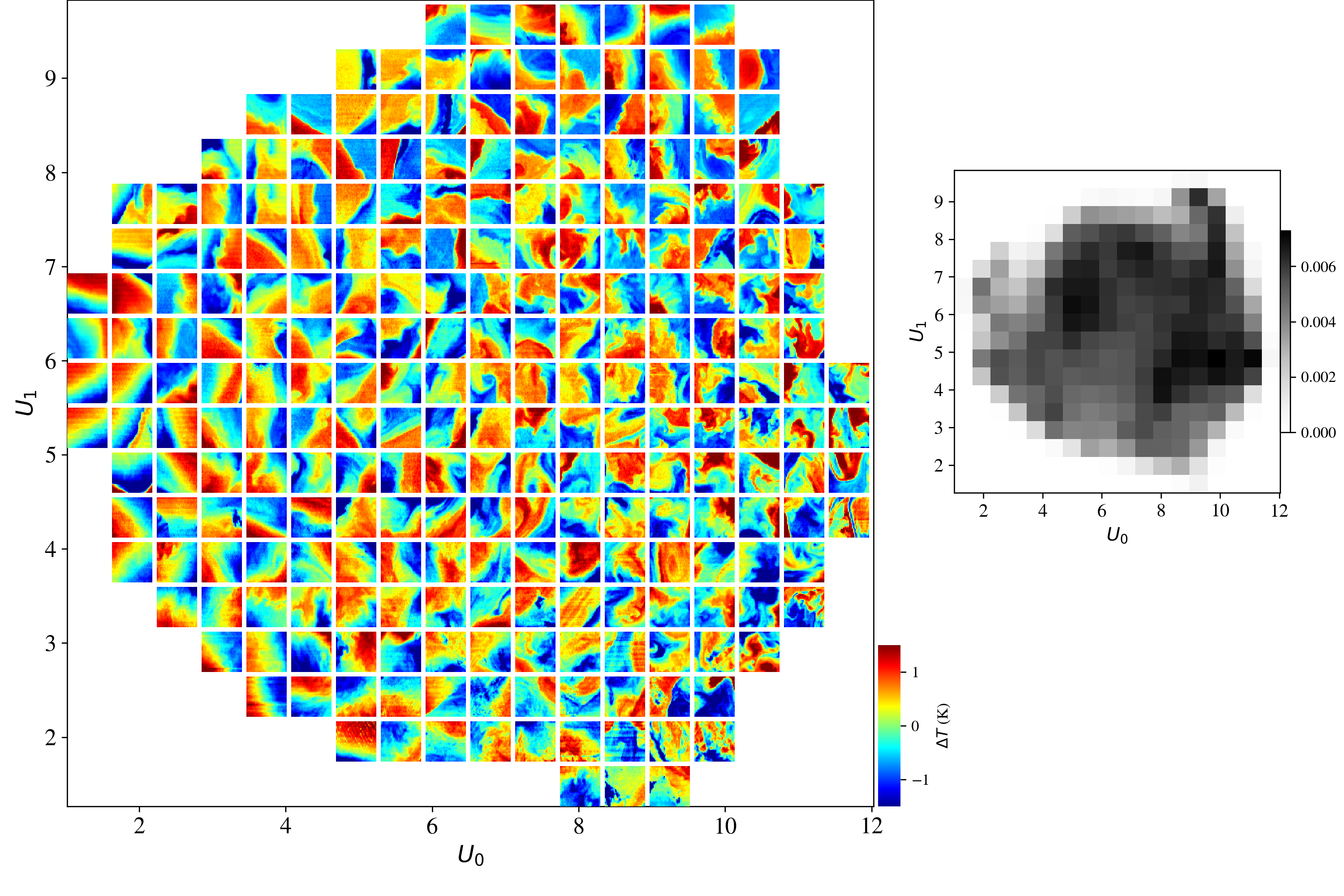}
\caption{
Same as Figure~\ref{fig:umap_gallery_all} but for the \ac{UMAP}
embedding of the \DTtwo\ sub-set of cutouts.
}
\label{fig:umap_gallery_DT2}
\end{figure*}

\begin{figure*}[ht]
\centering
\includegraphics[width=0.95\textwidth]{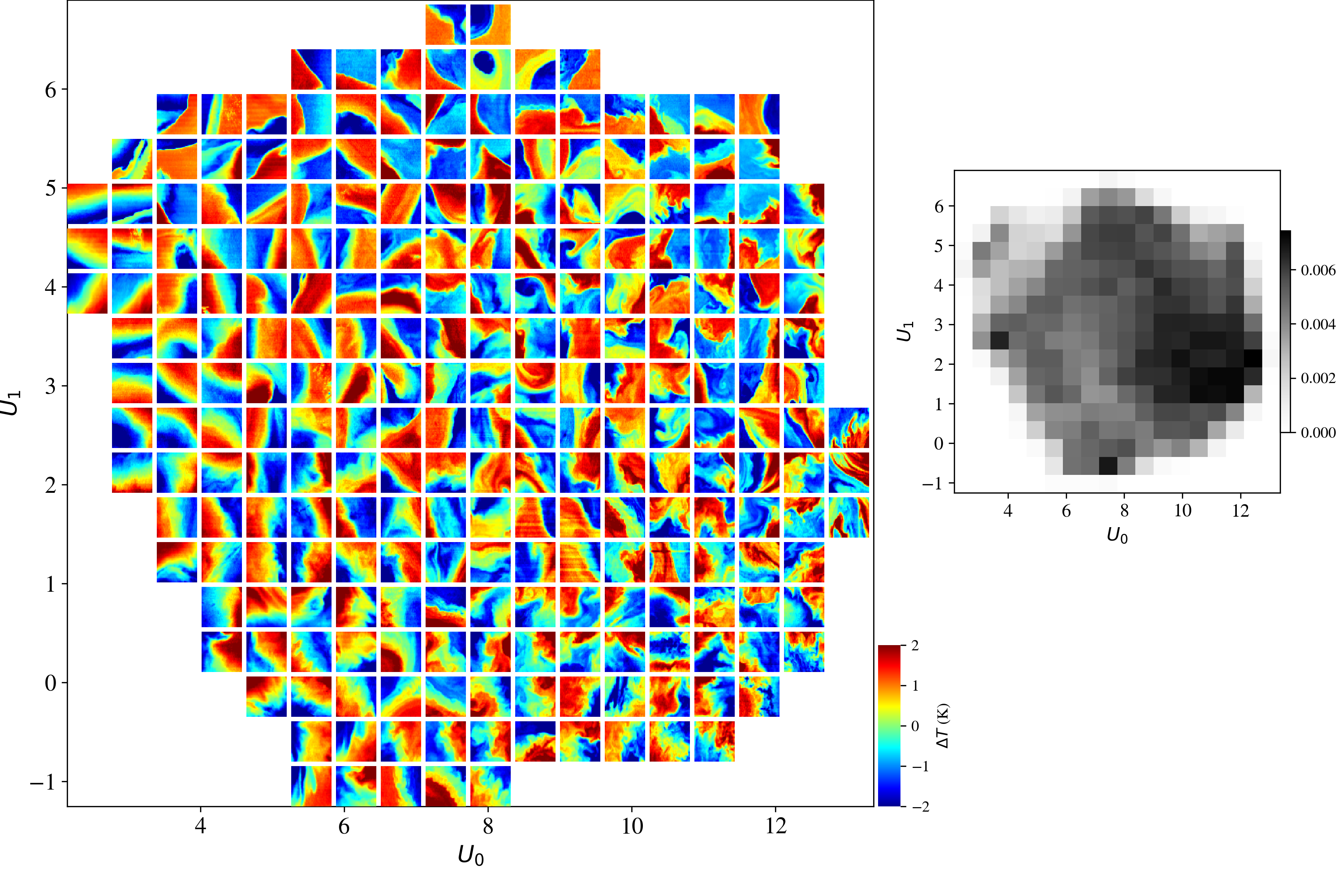}
\caption{
Same as Figure~\ref{fig:umap_gallery_all} but for the \ac{UMAP}
embedding of the \DTfour\ sub-set of cutouts.
}
\label{fig:umap_gallery_DT4}
\end{figure*}

\begin{figure*}[ht]
\centering
\includegraphics[width=0.95\textwidth]{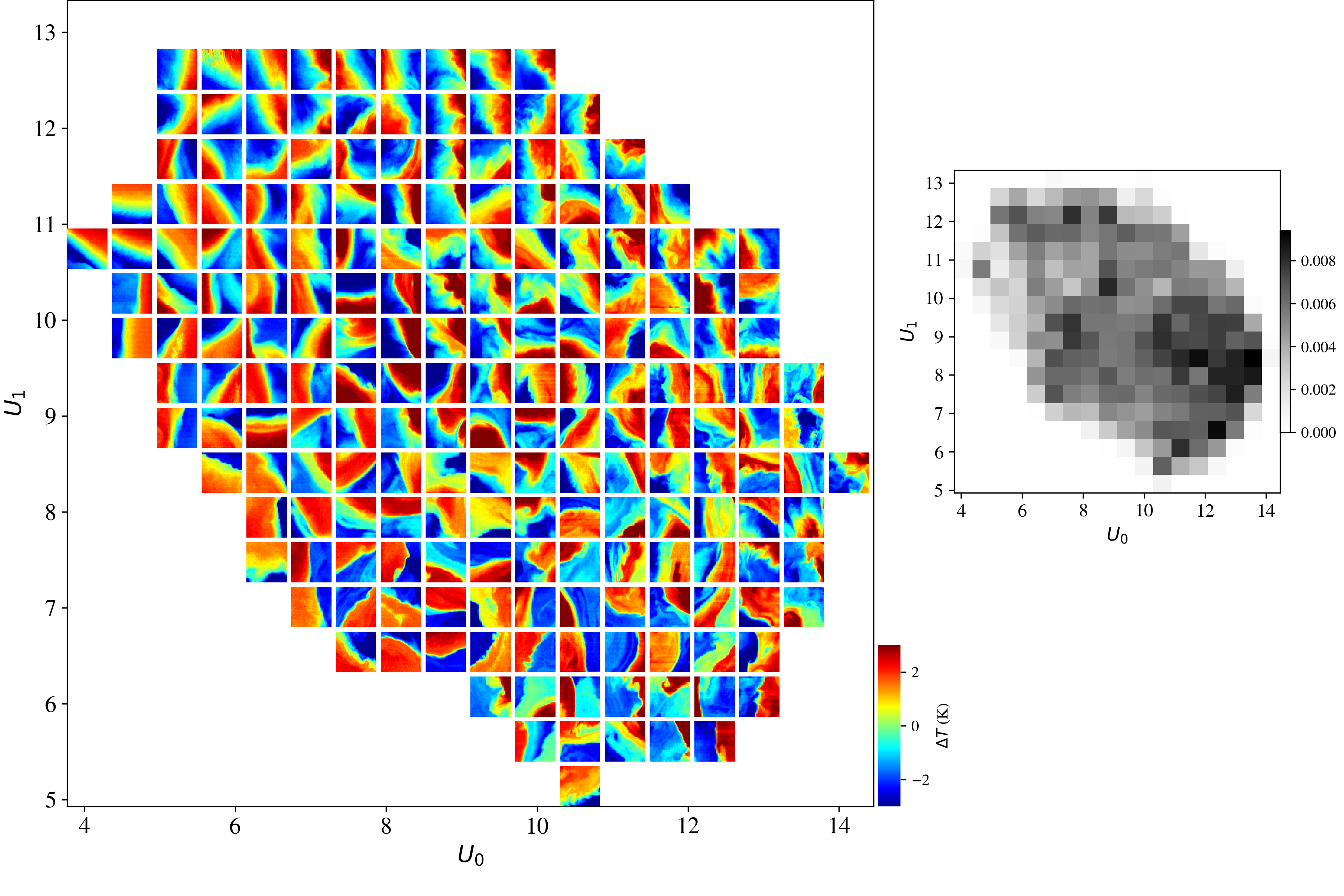}
\caption{
Same as Figure~\ref{fig:umap_gallery_all} but for the \ac{UMAP}
embedding of the \DTfive\ sub-set of cutouts.
}
\label{fig:umap_gallery_DT5}
\end{figure*}

In the previous sub-section, we showed that
reducing the full set of latent vectors to two dimensions
with a \ac{UMAP} embedding primarily separated the \sst\ 
patterns by \DT.  While informative, this does not
express well the great diversity within the dataset.
Therefore, we proceeded
to generate a unique \ac{UMAP} embedding
of the latent space for sub-samples of the
cutouts separated by \DT.
Specifically, we measure \DT\ from the inner 
$40 \times 40$\,pixel$^2$ regions of each cutout
which corresponds to the regions augmented by 
the \mname\ model (see $\S$~\ref{sec:augment}).
We then consider \DT\ bins as follows: %\textcolor{red}{CEB: feel free to eliminate the ordered list (original text in latex document)}
  % (i)~${\DT = 0-0.5}$\,K;
  % (ii)~${\DT = 0.5-1}$\,K;
  % (iii)~${\DT = 1-1.5}$\,K;
  % (iv)~${\DT = 1.5-2.5}$\,K;
  % (v)~${\DT = 2.5-4}$\,K;
  % and
  % (vi)~${\DT > 4}$\,K.
\begin{enumerate}
  \item[(a)]{${\DT = 0-0.5}$\,K; 2635727 cutouts; 
    Figure~\ref{fig:umap_gallery_DT0}}
  \item[(b)]{${\DT = 0.5-1}$\,K; 3288103 cutouts;
    Figure~\ref{fig:umap_gallery_DT1}}
  \item[(c)]{${\DT = 1-1.5}$\,K; 1126838 cutouts; Figure~\ref{fig:umap_gallery_DT15}}
  \item[(d)]{${\DT = 1.5-2.5}$\,K; 535038 cutouts;
    Figure~\ref{fig:umap_gallery_DT2}}
  \item[(e)]{${\DT = 2.5-4}$\,K; 126576 cutouts;
    Figure~\ref{fig:umap_gallery_DT4}}
  \item[(f)]{${\DT > 4}$\,K; 18317 cutouts;
    Figure~\ref{fig:umap_gallery_DT5}}
\end{enumerate}

This otherwise arbitrary
binning was chosen to minimize \DT\ as a 
defining factor in the analysis while maintaining
large samples of cutouts.
%(the first subset contains $\sim 3,000,000$
%cutouts and the last $\sim 20,000$).  
Figure~\ref{fig:umap_multi}a shows the \ac{UMAP}
representation of the  \DThalf\
sub-sample colored by \DT.  While there is structure
in the domain that correlates with \DT, it is no longer
monotonic and cutouts with the full range of \DT\ are
located throughout
(the standard deviation in \DT\ is 
$\sigma(\DT) \approx 0.13$\,K; 
see Figure~\ref{fig:umap_multi}b).

%\textcolor{red}{CEB: I wonder if it would be possible to form a three-dimensional plot of this all. For example, a cloud plot of observations in $(U_o,U_1,\Delta T)$-space where this would be similar to an $(x,y,z)$ plot? I understand that $U_o$ vs $U_1$ are not common amongst all \DT and this might cause a hurdle for this. But something along these lines to eliminate the subjective nature of our chosen \DT bins.}

% %%%%%%%%%%%%%%%%%%%%%%%
% %%%%%%%%%%%%%%%%%%%%%%%
% %%%%%%%%%%%%%%%%%%%%%%%
% %%%%%%%%%%%%%%%%%%%%%%%
% %%%%%%%%%%%%%%%%%%%%%%%
% %%%%%%%%%%%%%%%%%%%%%%%

Figure~\ref{fig:umap_gallery_DT1} shows
a gallery of randomly selected cutouts 
for the \DThalf\ subset at 
their binned $U_0, U_1$ values 
(restricted to areas
with at least 1000 cutouts per bin).
%By construction, these cutouts
%have features representative of all the other
%cutouts with similar $U_0, U_1$ values. 
Examining the figure, one observes a 
great diversity in the patterns as one
travels throughout the \ac{UMAP} space.
%The separation of the imagery by their patterns
%is visceral, as is the great diversity. 
One identifies
features for which SSTa contours are predominantly linear (e.g., at low $U_0, U_1$), 
curved (e.g., $U_0,U_1 \approx 8,2$) or 
amorphous (e.g., $U_0,U_1 \approx 4,4$), 
patterns with apparent symmetry about an axis 
(e.g., $U_0,U_1 \approx 0,2$)
and
those without any discernible structure 
(e.g., $U_0,U_1 \approx 9,1$).
%One also finds cutouts with temperature
%variance on only the largest scales
%(e.g., $U_0,U_1 \approx 1,-1.5$)
%versus those 
%with significant variance on small-scales
%($< \sim 10$\,km; $U_0,U_1 \approx 6,3.5$).

% %%%%%%%%%%%%%%%%%%%%%%%
% %%%%%%%%%%%%%%%%%%%%%%%
\begin{figure}[ht]
\centering
\includegraphics[width=0.5\textwidth]{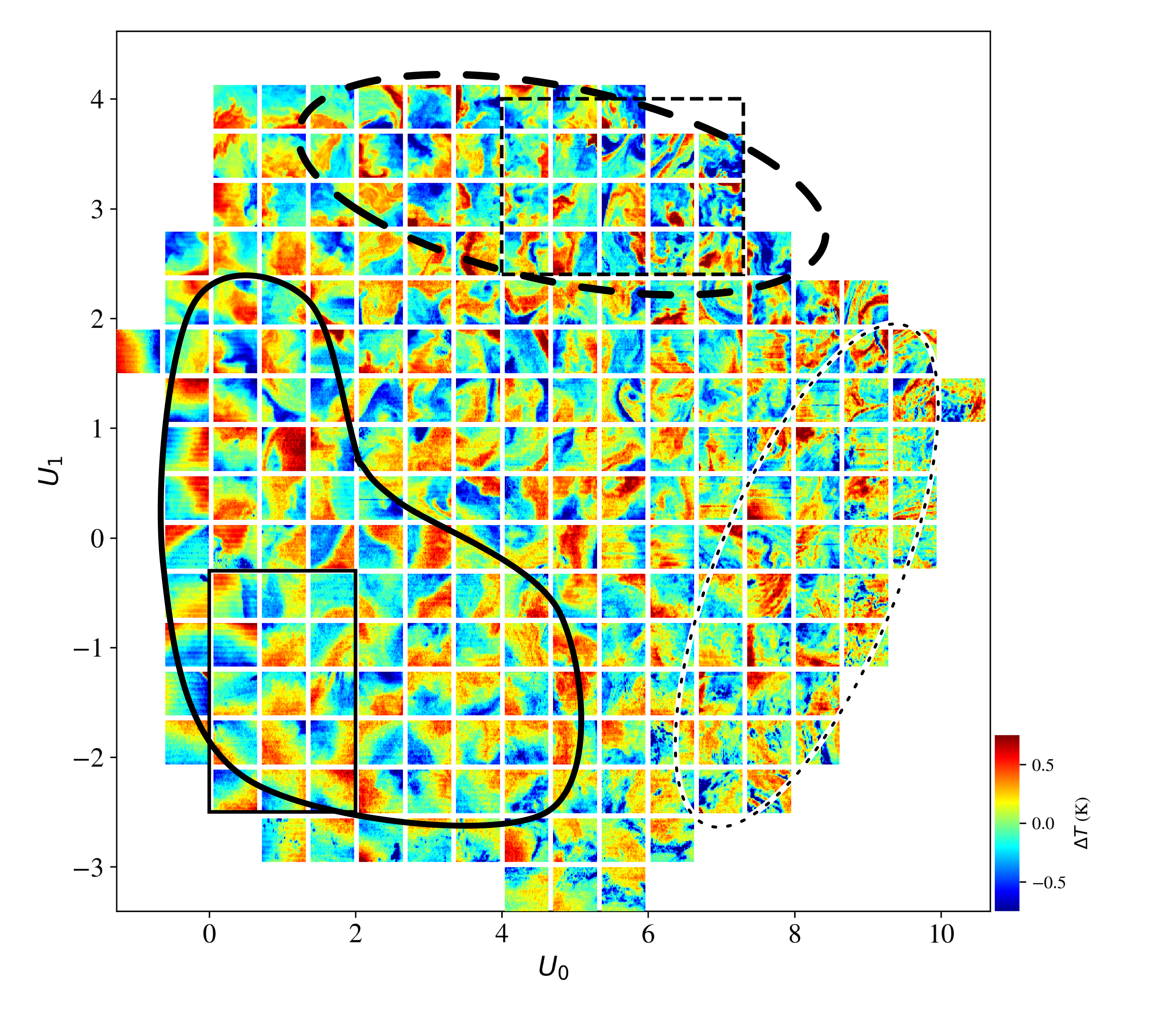}
\caption{
A reproduction of the \ac{UMAP} domain
in Figure~\ref{fig:umap_gallery_DT1}
but now annotated to highlight
regions of interest in the \ac{UMAP} space.
The annotations approximately designate
regions where the patterns show:
 (solid-line annotation) temperature variance
 predominantly at the largest wavelengths;
 (dashed-line ellipse) temperature variance
 reflective of an energetic spectrum, e.g.\ 
 non-linear, or turbulent motions;
 and 
 (dash-dot ellipse) patterns with ``noisy''
 features, predominantly on the smallest scales.
Meanwhile, the black rectangles (solid and dashed)
in the left panel indicate 
the portions of the \ac{UMAP} domain analyzed in 
Figures~\ref{fig:global_shallow}
and \ref{fig:global_turb}.
}
\label{fig:umap_annotated}
\end{figure}
To build additional intuition for the patterns resolved by \mname,
we show a series of properties of the cutouts 
in Figure~\ref{fig:umap_multi}, binned 
by $U_0$ and $U_1$.
The characteristic most highly correlated 
with the pattern
distribution in the \ac{UMAP} embedding is the 
absolute latitude \ablat\ (Figure~\ref{fig:umap_multi}e).
Given the \ablat\ dependence for the radius of deformation,
this suggests the patterns have been separated by
the dominant feature scale, especially along
the $U_1$ dimension.
Visual inspection supports this hypothesis to an extent,
e.g.\ the patterns with largest-scale variance have 
lower $U_1$.  On the other hand, cutouts with the smallest
scale features arise at high 
$U_0$ and low $U_1$
corresponding to mid-latitudes.
These also have the highest cloud coverage
fraction (Figure~\ref{fig:umap_multi}d)
and close inspection of individual cutouts
reveals small-scale blemishes 
characteristic of clouds.  
We believe these have been missed
by the standard screening algorithm
and future work may leverage \mname\
to flag corrupt data and/or improve the
screeing process.

Figure~\ref{fig:umap_multi}c shows the median
spectral slope \slope\ which varies across the domain. 
The values are all negative, 
indicative of reduced energy at high wavenumbers and, thus, consistent with all theories of geophysical turbulence \cite{charney71,gm72,callies13} (cf.~section~\ref{sec:power_spectrum}). 
% Moreover, the median \slope\ 
% within each portion of \ac{UMAP} space follows visual impressions gathered from the gallery of cutouts in Figure~\ref{fig:umap_gallery_DT1}--i.e., fields dominated by low-wavenumber energy (temperature variance at large-horizontal scales)
% have the the steepest spectral slopes (blue; largest in magnitude), whereas cutouts dominated by temperature variance at finer horizontal scales are characterized by shallow spectral slopes (green and yellow; smallest in magnitude). 
The pattern dependence on \slope\ is
complex with very differing imagery 
having similar median \slope.
Similarly the standard deviation of
\slope\ is large (not shown)
Together, these indicate \slope\ is 
an incomplete
descriptor of \sst\ patterns and features.

To better guide analysis presented in 
the following section, 
Figure~\ref{fig:umap_annotated}
shows the \ac{UMAP} with several sub-regions annotated
to approximately demarcate several
types of patterns: 
those with temperature variance predominantly
on the largest scales (solid annotation),
cutouts with an ``energetic'' spectrum
showing structure at a range of scales
(dashed ellipse), and noisy patterns with structure
predominantly on the smallest scales
(dash-dot ellipse).

Galleries for the other \DT\ sub-samples are
provided in the Appendix
(Figures~\ref{fig:umap_gallery_DT0}-\ref{fig:umap_gallery_DT5}).  Together with Figure~\ref{fig:umap_gallery_DT1}
these six \ac{UMAP} spaces span the full diversity of 
\ac{SST} patterns, as encoded in the full \ac{SSL}
latent space.
With \mname, we have constructed a vocabulary
that encodes the complex language
of \sst\ imagery on sub-mesoscales.
Any \ac{SST}
cutout drawn from the \ac{MODIS} \ac{L2} dataset
maps directly to one of the 6~galleries and is--by
design--very similar to its neighboring
cutouts in \ac{UMAP} space.  This enables novel exploration of 
the data--e.g., the
ability to define and investigate 
regions of the ocean with special or unique
\ac{SST} signatures.

Before proceeding, however, a cautionary note is warranted. 
Because a unique \ac{UMAP} embedding is generated for the 
latent vectors in each $\Delta T$ bin, 
the meaning of the $U_0$ and $U_1$ axes likely 
differ from one $\Delta T$ bin to the next. Having said this, we note that with the exception of the gallery for \DT\ $=0-0.5$\,K (Fig.~\ref{fig:umap_gallery_DT0}) all of the galleries 
%(Fig.~\ref{fig:umap_gallery_DT1}-\ref{fig:umap_gallery_DT5}) 
show the complexity (e.g., curvature of {\ssta} contours) increasing as $U_1$ increases. That is, setting aside the effect of \DT, this may be the primary distinguishing characteristic of the fields identified by \mname.

% %%%%%%%%%%%%%%%%%%%%%%%
% %%%%%%%%%%%%%%%%%%%%%%%
% %%%%%%%%%%%%%%%%%%%%%%%
\section{Applications and Discussion}
\label{sec:discuss}

In the previous section, we presented the fundamental patterns
of \sst\ imagery learned by the \mname\ model
on scales smaller than $\approx 80$\,km. 
We now demonstrate several applications enabled by 
constructing this vocabulary to describe the complex 
patterns of these data.
In particular, one may explore the nature and origins
of specific patterns across the ocean 
or the patterns that manifest in select regions,
and explore temporal variability.
In this section, we focus on the \DThalf\ subset 
and emphasize that the primary results are 
largely insensitive to this choice.

% %%%%%%%%%%%%%%%%%%%%%%%%%%%%%%%%%%%%%%%%%
\subsection{Geographical Exploration}
\label{sec:geo}

A defining characteristic of data collected by remote sensing satellites 
such as Aqua is their global coverage of the ocean.
Using the \mname\ model, we may select specific
patterns of \ac{SST} and reveal 
their global geographic locations.  
We may then assess commonality 
(or otherwise) between the processes 
that generate them.
Alternatively, we may select specific local regions
to highlight the \sst\ patterns that define them.
Here, we consider each of these approaches in turn.

\begin{figure*}[h]
\centering
\includegraphics[width=1.0\textwidth]{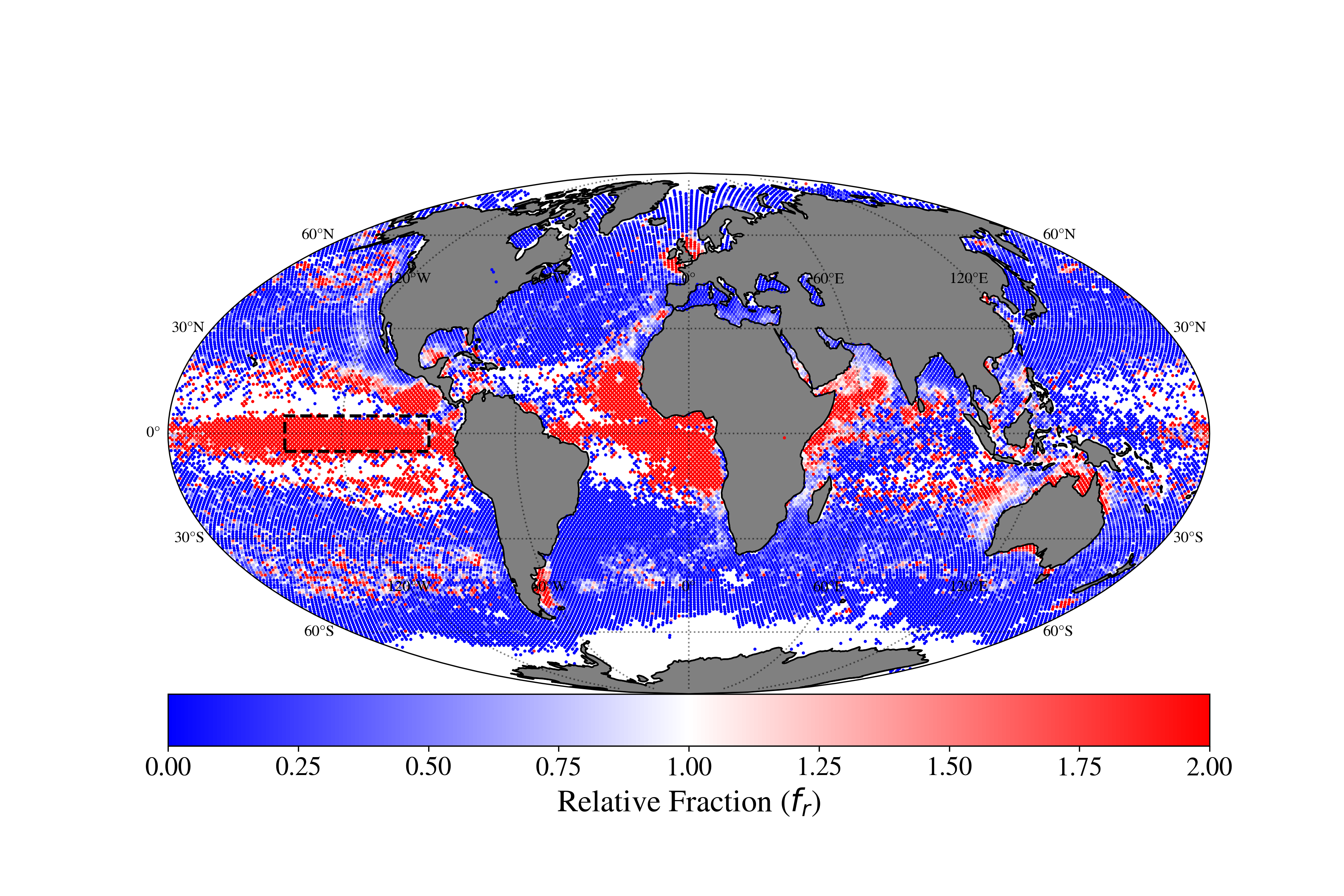}
\caption{
Relative fraction \relf\ of \sst\ cutouts with 
temperature variance predominantly on large scales 
($\approx 80$\,km)
from the \DThalf\ subset
(specified by the dashed rectangle 
in Figure~\ref{fig:umap_gallery_DT1}).
A low value (blue) implies that the geographic
location, with area of $\approx 100 \times 100$\,km$^2$, 
rarely exhibits such patterns while $\relf \ge 1$ (pink/red)
indicates an excess.
Gray regions indicate land. 
The figure reveals that large-scale patterns 
arise primarily at  lower latitudes (e.g., 
the equatorial Pacific and coastal Africa),
but also along the Patagonia shelf and Northern Sea.
Regions with fewer than 5~cutouts
in the \DT\ interval are indicated in white 
and have not been included in the analysis. 
The dashed black rectangle is the focus area 
for Fig.\,\ref{fig:local_eqpsa}.}
\label{fig:global_shallow}
\end{figure*}

\subsubsection{{\bf Global}}
\label{sec:geo_global}

To perform a global analysis, we have divided the surface
of the ocean using the \ac{HEALPix} procedure 
which tesselates a spherical surface into 
equal-area curvilinear quadrilaterals
\cite{healpix}.
Specifically, we adopted nside=64, which 
divides the globe into
$\approx 50,000$ \ac{HEALPix}
cells each with 
approximately 100$\times$100\,km$^2$ area.
We then perform statistics on the cutouts which occur
within each of these \ac{HEALPix} cells.

As a first example, we consider cutouts with 
patterns that exhibit temperature variance
predominantly on the largest scales of our
augmented images (i.e.,~$80$\,km).
Specifically, we select the portion of 
the \ac{UMAP} domain corresponding to \DThalf\ 
with \weak. This includes 
$\approx 225,000$ cutouts from the lower left 
corner of the domain (Figure~\ref{fig:umap_annotated}; solid
rectangle).

Let us now define a \textit{relative fraction} $f_r$ for each 
\ac{HEALPix} which is the ratio of two fractions:
 (i) $f_{H}$, the fraction of cutouts in the \ac{HEALPix}
 which lie within the \ac{UMAP} rectangle defined above
 and;
 (ii) $f_{T}$, the total fraction of all cutouts for the given \DT\ range that
 lie within the same rectangle.
We then calculate 

\begin{equation}
\relf \equiv \frac{f_H}{f_T}
\end{equation}
for each \ac{HEALPix} cell,
ignoring cells with fewer than 5 cutouts.
Small values of \relf\ indicate the pattern (characterized in \ac{UMAP}-space)
occurs rarely within the \ac{HEALPix} relative
to the average, and vice-versa.

Figure~\ref{fig:global_shallow} presents the geographic
distribution of \relf\ for \weak\ in the \DThalf\ bin (Figure~\ref{fig:umap_annotated}; solid
rectangle).
Focusing on geographical regions with an enhanced incidence of
these patterns ($\relf > 1$), we find the majority
occur in large, coherent regions ($\gg 100$\,km),
i.e., much larger than the cutouts analyzed by the \mname\ model.
In this case, the \sst\ patterns have tagged
processes on scales substantially larger than the sub-mesoscale.
One also notes that these areas are
primarily at low latitudes; this 
follows physical intuition 
that the high radius of deformation near the equator
leads to larger-scale features. 
%
%One identifies several distinct regions of the ocean 
%where these 
%\ac{SST} patterns preferentially manifest:
%  (1) in the central and eastern equatorial Pacific;
%  (2) at tropical and equatorial latitudes off the coasts of Africa,
%  especially within the Atlantic basin;
%  (3) in waters near the western coastline of Central America;
%  (4) on the Patagonia shelf; 
%  (5) off the northern coast of Australia; 
%  and
%  (6) in the North Sea.
We briefly describe processes that may
generate the underlying \sst\ patterns in several of 
these regions.
  
Two of the 
largest regions with high \relf\ 
in Figure~\ref{fig:global_shallow} 
are along the eastern equatorial
Pacific and Atlantic.  These areas are commonly referred
to as the Pacific and Atlantic \ac{ECT}, and are of interest to
climate studies owing to their impact on air-sea interactions
(e.g., \cite{mitchell1992,okumura2004}).
Briefly, \ac{ECT}s are believed to result from upwelling 
associated with
northward wind stresses in the eastern
portion of each basin.
This water is then forced by zonal winds to 
propagate westward \cite{mitchell1992,pedlosky1990,okumura2004}.
Figure~\ref{fig:global_shallow} indicates the \ac{ECT}s exhibit similar \sst\ patterns that span several thousands of kilometers along
the equator, stretching to $\approx 180^\circ$\,W
in the Pacific and nearly 
across the entire Atlantic.
The Pacific \ac{ECT} is confined to 
a meridional extent of approximately $\pm 5^\circ$. 
In the following section, we examine its
seasonal variation. 

To the north (at $\approx 10^\circ$\,N)
and distinct from each \ac{ECT} is
an additional region of high relative frequency,
one off the coast of Central America and another
off the coast of western Africa.
Each region is characterized by (seasonal) coastal upwelling
and poleward currents \cite{hagen2005,kounta2018}.
For example, in the Eastern Tropical Pacific, off the coast of Central America,
seasonally-varying and intense wind ``jets''  
generate westward-propagating Rossby waves and/or mesoscale eddies in this region \cite{willett06}.
The \sst\ patterns accentuated in this region, however,
are not characteristic of such dynamics (see below);
we expect the large-scale temperature variance
arises from different processes hereto not explored
by previous studies.
The Atlantic region, meanwhile, corresponds approximately
to the ``shadow zone'' of the eastern tropical
North Atlantic and the so-called Guinea dome below it
\cite{stramma2005,kounta2018}.
These have weak circulation and our results suggest,
unexpectedly, that
the prevailing \sst\ signatures within them
are large-scale temperature variations.

Distinct from the enhanced regions at lower latitudes
are the waters off the Patagonia shelf.
This region is characterized by a shallow continental shelf
and the nearby confluence of two strong currents:
the equatorward-flowing Malvanis current
which meets the poleward-traveling Brazilian current 
at $\approx 40^\circ$S.
The circulation on the shelf is predicted (and observed)
to be relatively weak \cite{combes2018}.
We hypothesize that bathymetry plays a 
leading role in the generation of the observed
\sst\ patterns, which is supported by the 
orientation of the large-scale temperature variance
in the cutouts. 
We also note that the waters off eastern Africa
at $\approx 5^\circ$N share features common to 
the Patagonia shelf \cite{painter2020}; 
perhaps bathymetry is a driving factor for its
high \relf, as well.

\begin{figure*}[h]
\centering
\includegraphics[width=1.0\textwidth]{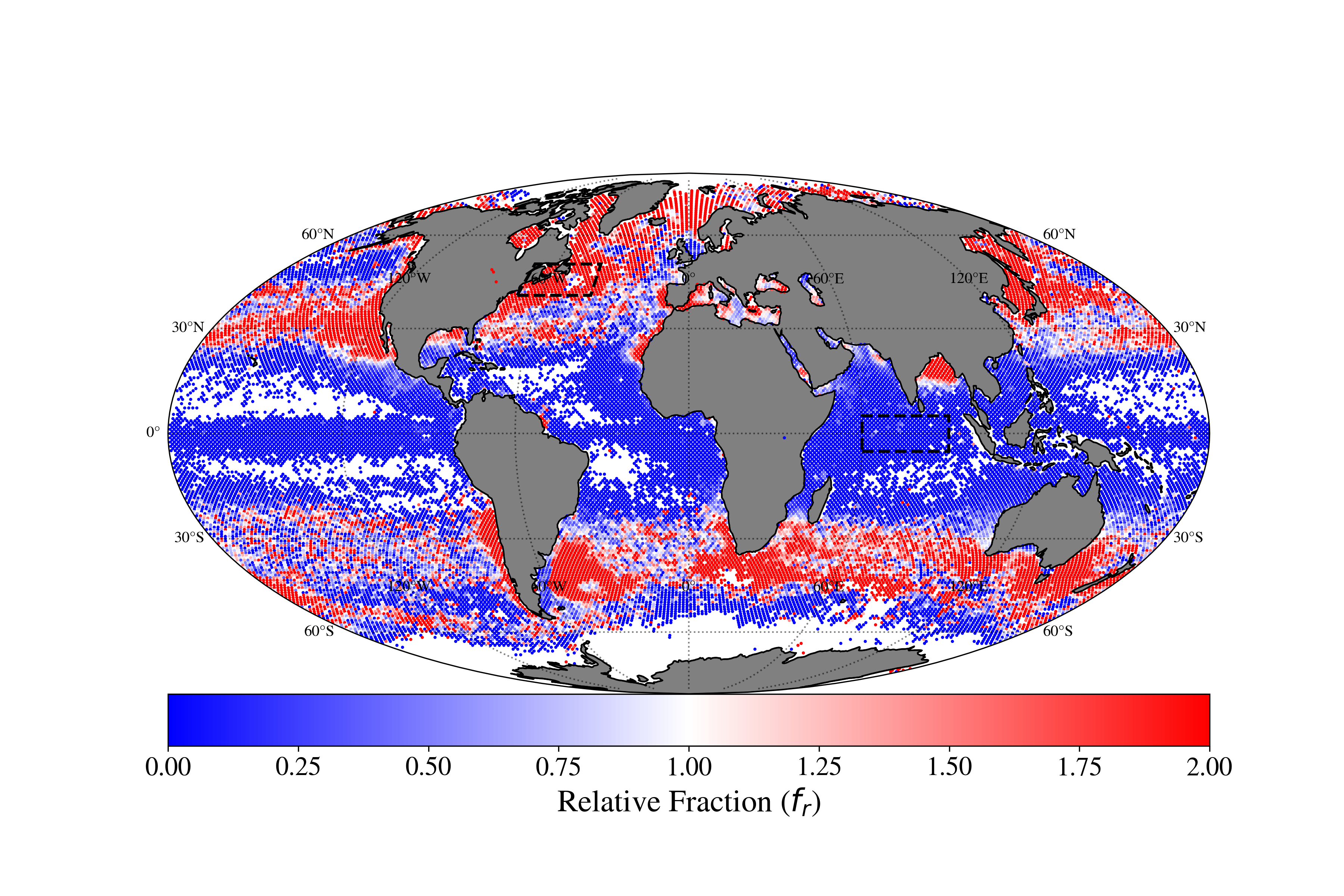}
\caption{
Same as Figure~\ref{fig:global_shallow} but for \sst\ patterns
with structure across a range of scales and
from the \DThalf\ subset (specifically \strong). 
The regions that show a high fraction of these turbulent
patterns are predominantly those influenced by western
boundary currents and areas generally known to show
frequent eddy formation and propagation \cite{chelton2011}.
Strong upwelling
may also generate these patterns in select
regions: coastal California, western Africa, Chile.
We also emphasize the 
the Bay of Bengal whose dynamics are driven by 
qualitatively different processes 
\cite{chatterjee2017dynamics}.
Dashed black rectangles on the Gulf Stream near New England
and the equatorial Indian Ocean
are the focus areas for Fig.\,\ref{fig:local_strong}.
}
\label{fig:global_turb}
\end{figure*}

%\begin{figure*}[h]
%\centering
%\includegraphics[width=1.0\textwidth]{Figures/fig_umap_geo_global_DT1_center_96clear_v4_S1.png}
%\caption{
%Same as Figure~\ref{fig:global_shallow} but for
%intermediate $U_0, U_1$ values.
%[Should we show this figure??
%It is currently not discussed]
%}
%\label{fig:global_inter}
%\end{figure*}

In an effort to contrast the geographical distribution of 
the large-scale \ac{SST} patterns of Figure~\ref{fig:global_shallow} 
with that for cutouts suggesting a significantly shorter range of scales, we now focus on \ac{SST} patterns 
for which \DThalf\ with \strong\
(Figure~\ref{fig:umap_annotated}; dashed rectangle).
Their complex patterns are 
suggestive of regions in which non-linear,
sub-mesoscale dynamics (e.g., eddies and filaments)
and/or ageostrophic turbulence 
are to be found.

Figure~\ref{fig:global_turb} presents the relative
frequency \relf\ of these patterns. 
All of the geographical regions 
in Figure~\ref{fig:global_shallow} 
with high relative fractions
now have low \relf, 
and an entirely distinct set of locations with high \relf\ appear.
These are primarily poleward of $20^\circ$\,deg and may be divided into four main groups. 
  (1) waters associated with strong open-ocean currents (e.g., 
        the Brazil-Malvanis confluence,
        the Gulf Stream,
        the Kuroshio, 
        the Agulhas Retroflection); 
        %and
        %\item large portions of the Antarctic Circumpolar Current.
   (2) enclosed or semi-enclosed seas
   (e.g., 
        parts of the Mediterranean, 
        %\item Baffin Bay, 
        %\item Hudson Bay, 
        the Sea of Japan, 
        the Okhotsk Sea,
        the southern end of the Red Sea,
        the northern half of the Gulf of Mexico,
        the Baltic Sea);
        %\item the Black Sea, and
        %\item the southern half of the Caspian Sea.
   (3) eastern boundaries of most ocean basins in the subtropics 
   (e.g., off 
        California and Mexico,
        Chile,
        Northwest Africa); %, and
        %Portugal.
and (4) anomalous regions equatorward of $22.5^\circ$:
        the Bay of Bengal and 
        the retroflection of a portion of the South Equatorial Current into the Equatorial Countercurrent off of northern Brazil and French Guiana. 
        %the southern end of the Red Sea (also listed above as part of a semi-enclosed sea). 

As noted above the first group is associated with 
strong currents and comprise regions
where eddies frequently form  
(e.g., \cite{chelton2011}).
Several are also noted regions of strong upwelling,
e.g., coastal California \cite{huyer1983coastal}. Neither of these are a surprise. By contrast, a significant fraction of virtually all major enclosed or semi-enclosed seas comprise cutouts for which \sst\ contours show signs of relatively small spatial scales while these regions show little to no signs of long wavelength contours (Fig.~\ref{fig:global_shallow}). Is this because topography is complex in these regions exerting significant control over circulation? Or, maybe land constrains the flow both of currents and winds resulting in smaller scale motion? Regardless, the implications here are intriguing.

We may speculate--given the randomly-selected cutouts presented in \ac{UMAP}-space--that the \ac{SST} patterns with $\relf >> 1$
Figure~\ref{fig:global_turb} trace regions of active
frontogensis, frontoloysis, and frontal instabilities at scales defined as sub-mesoscale
($\lambda < 50 \, \rm km$)
and therefore enhanced ageostrophic turbulence.
It is an active area of interest for the authors
to connect such processes to the
specific \sst\ patterns defined by \mname.
%If true,
%one may track such activity 
%across the globe and in time
%to gauge aspects of the global
%energy budget and frequency of mixing.

Last, a visual inspection of the cutouts with high $U_0$ and low $U_1$
in the dotted ellipse of Figure~\ref{fig:umap_annotated} 
shows signatures of 
clouds, e.g., patterns of speckled, cooler patches.
These are primarily cases that passed through the 
cloud-screening algorithms  \cite{FreundMason1999} of the R2019.0 \ac{MODIS} processing.
In this respect, our \mname\ model has efficiently identified 
corrupt fields.  We have examined the geographic distribution
of these cutouts and found they generally track the overall
distribution of data, although there 
is a preference for mid-latitudes.

\begin{figure}[ht]
\centering
\includegraphics[width=0.5\textwidth]{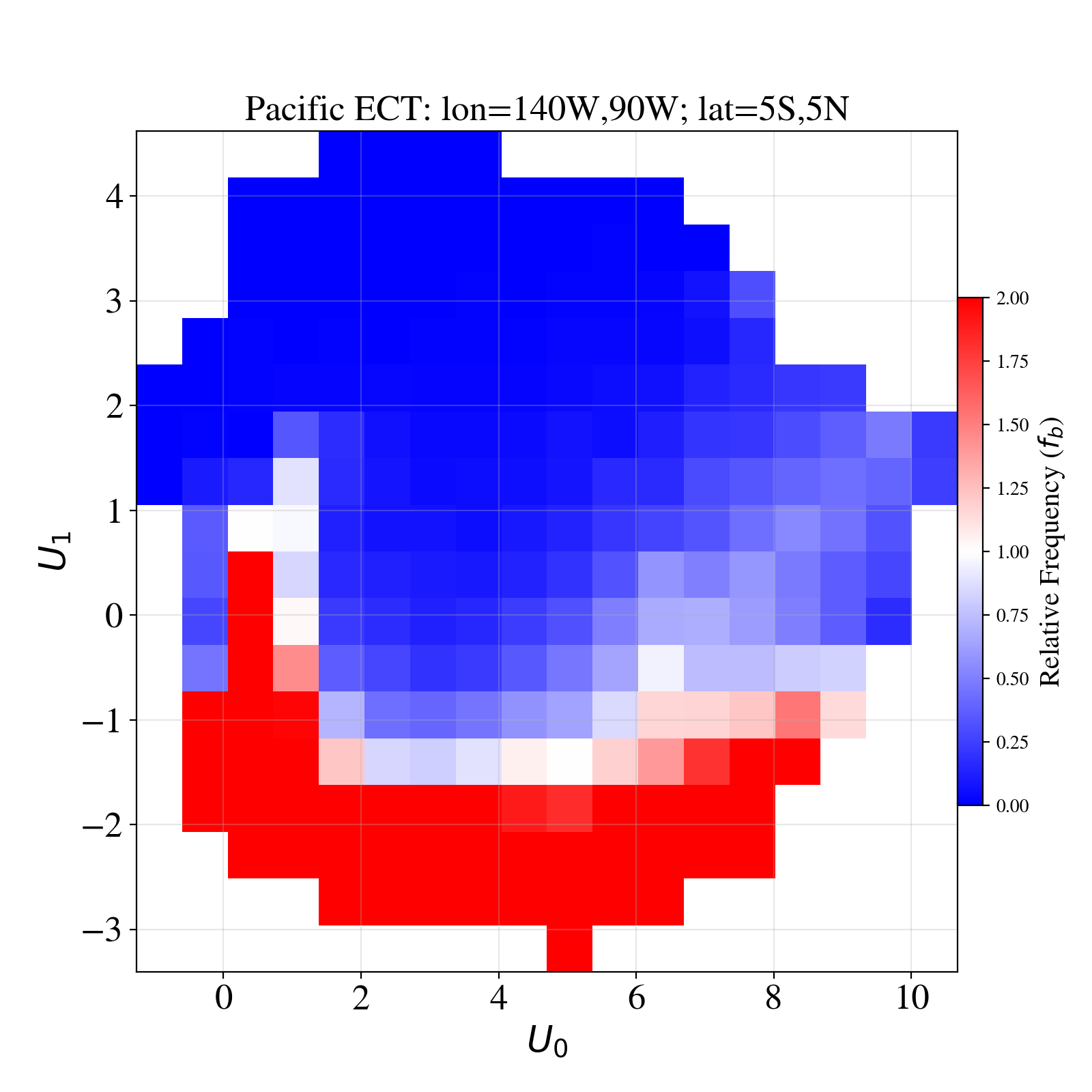}
\caption{
Two-dimensional histogram of the
relative frequency of specific \sst\ patterns for 
\DThalf\ in the Pacific \ac{ECT}
%(top\sout{; lon=[140W,90W]; lat=[5S,5N]})
%and
%South Atlantic %Mediterranean Sea.  
%(bottom\sout{; lon=[35W,10E]; lat=[20S,30S]}). 
This region is identified by the dashed black rectangle in Fig.~\ref{fig:global_shallow}
and contains 81,675 cutouts.
The Pacific region is dominated by patterns 
with temperature variance at large scales 
(low $U_1$ and low $U_0$) and patterns 
with blemishes suggestive of unmasked clouds
(low $U_1$ and high $U_0$).
%In contrast, the South Atlantic region exhibits patterns
%dominated by variance on the smallest scales
%and those that most resemble white noise
%in the domain.
Bins with fewer than 200 cutouts have been excluded from the panels. % 200 is correct
}
\label{fig:local_eqpsa}
\end{figure}

% %%%%%%%%%%%%%%%%%%%%%%%%%%%%%%%%%%%%%%%%%%%
% %%%%%%%%%%%%%%%%%%%%%%%%%%%%%%%%%%%%%%%%%%%
\subsubsection{{\bf Local}}
\label{sec:geo_local}

We may effectively invert the above analysis to 
examine the \sst\ patterns that are dominant in
select geographical regions.
As an example,
Figure~\ref{fig:local_eqpsa} shows the relative 
frequency, corresponding to the \DThalf\ subset, of \ac{SST} patterns in the Pacific \ac{ECT}
%and a region of the Southern Atlantic
(black dashed rectangle in 
Figure~\ref{fig:global_shallow}).
For this analysis,
we have divided the \ac{UMAP}
domain into $18 \times 18$ equal-spaced bins that 
span 99.9\%\ of each dimension
($\Delta U_0 \approx 0.65$ and $\Delta U_1 \approx 0.45$).
We then calculate a \textit{relative frequency} of occurrence \relb,
defined as the frequency of cutouts
for the region in each bin $f_R$
divided by the frequency measured 
for the entire ocean $f_O$:

\begin{equation}
\relb \equiv \frac{f_R}{f_O} \;\; .
\end{equation}
High $f_b$ values therefore indicate \sst\ patterns
that are more characteristic of the region 
than that of the global ocean.
These \relb\ maps are the \sst\ ``fingerprints''
of the regions specifying the \sst\ patterns
they preferentially manifest.

Examining Figure~\ref{fig:local_eqpsa},
we find the Pacific \ac{ECT} shows an excess
of \sst\ patterns 
%\sout{within the solid-line annotation
%in Figure~\ref{fig:umap_annotated},
%%$U_0, U_1 \approx 0.5,-1.5$, 
%i.e., 
%those with large-scale temperature variance as anticipated
%from Figure~\ref{fig:global_shallow}.
%In addition, this region shows a propensity 
%for all of the patterns} 
with low $U_1$;  
these are a combination of cutouts
with large-scale structure 
and cutouts
with significant ``blemishes'', which we 
identify as unmasked clouds.
%by the shallow, large-scale gradient patterns
%that span the left and upper edges of the \ac{UMAP}
%domain.  
The patterns with $\relb < 1$ for the Pacific \ac{ECT},
meanwhile, are those with temperature variance across
a range of scales including small scales.
We conclude
it is very rare for such fluctuations
to occur in this region, i.e., 
the westward propagation of the \ac{ECT}
is stable against such instabilities.
Additional exploration reveals that features
likely associated with tropical instability
waves that occur along the meridional periphery of 
the ECT \cite{tanaka2019} 
have larger scale gradients and
higher \DT\ (e.g., at $U_0,U_1 \approx 6,8$
in Figure~\ref{fig:umap_gallery_DT2}).

\begin{figure}[ht]
\centering
\includegraphics[width=0.5\textwidth]{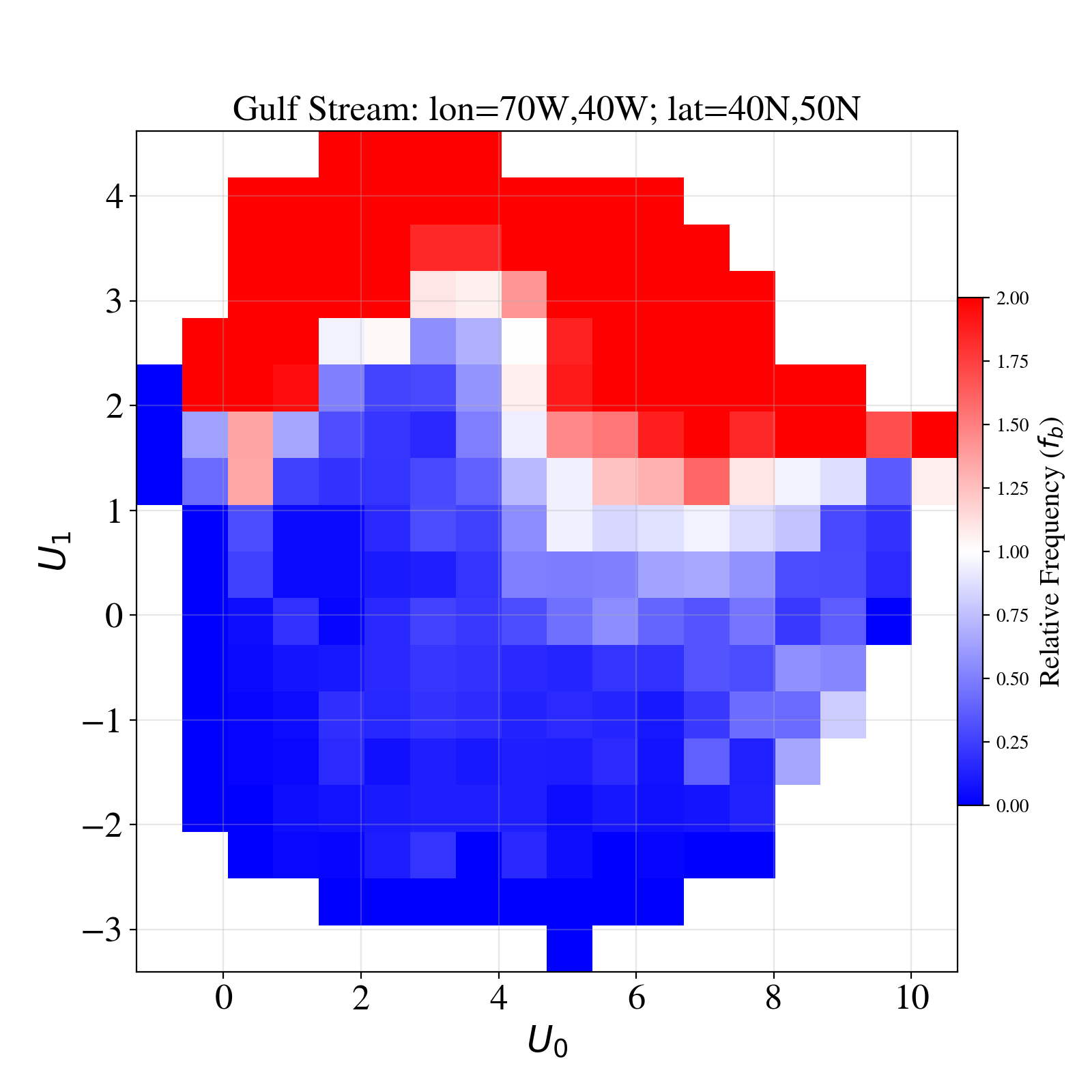}
\includegraphics[width=0.5\textwidth]{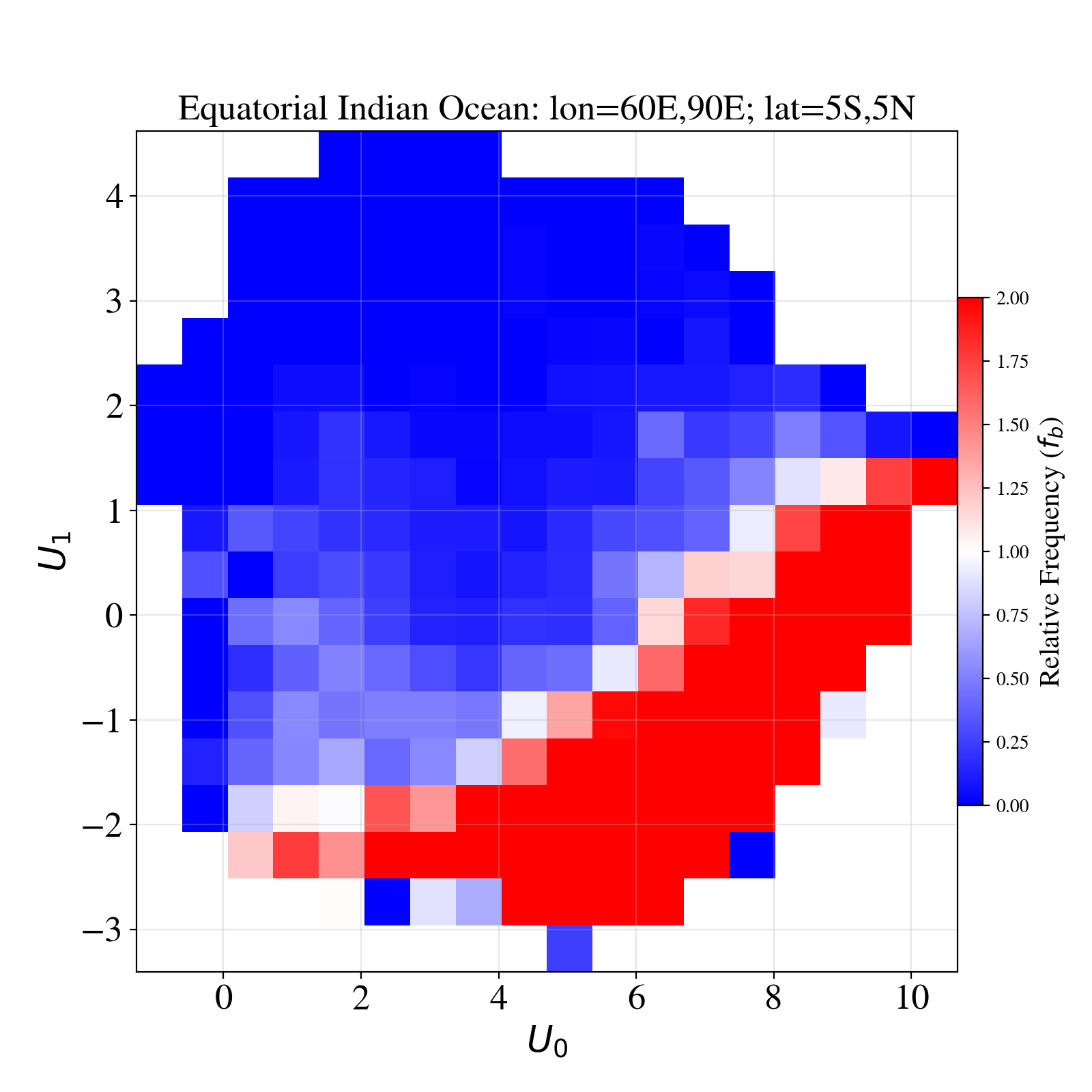}
\caption{
Similar to Figure~\ref{fig:local_eqpsa} but for \sst\ patterns
of the \DThalf\ subset and for
the Gulf Stream near New England
(top)
%; lon=[$70^{\circ}$W,$40^{\circ}$W]; lat=[$40^{\circ}$N,$50^{\circ}$N]).
%(lat=[10,20]\,deg S)
and the Equatorial Indian Ocean 
(bottom).
The precise regions evaluated are labeled 
on the figures and shown 
in Figure~\ref{fig:global_turb}
and each contains approximately 10,000 cutouts.
%; lon=[60E,90E]; lat=[5S,5N]).
The Gulf Stream shows an excess
of patterns within the energetic
region of the \ac{UMAP} space
(dashed ellipse in Figure~\ref{fig:umap_annotated})
which  have temperature variance at a wide
range of scales.
In contrast, the Equatorial Indian Ocean
primarily exhibits patterns with 
small-scale variations, including 
those most similar to white noise
(described by the dotted ellipse in Figure~\ref{fig:umap_annotated}).
}
\label{fig:local_strong}
\end{figure}

%In Figure~\ref{fig:local_south} we consider two large 
%geographical regions (dashed black rectangles in Fig.\,\ref{fig:global_turb}) not discussed above:
%  portions of 
%  the tropical oceans of the South Atlantic and Pacific
%  (specifically lat=[10, 30]\,deg S and [10, 20]\,deg S, respectively).
%The dominant SST patterns in these regions exhibit 
%very little
%structure indicating quiescent upper ocean circulation
%and dynamics.
%This suggests that these large portions of the
%major basins play a limited role in energy dissipation
%and/or ocean mixing.

Two more local regions are presented
in Figure~\ref{fig:local_strong}. These highlight
two areas from Figure~\ref{fig:global_turb}: 
one with a high relative fraction (the Gulf Stream
near New England)
indicative of a highly dynamic region, and 
one with low \relf\ (the equatorial Indian Ocean).
The dominant \sst\ patterns of the Gulf Stream
are drawn from the high $U_1$ portion of the 
\ac{UMAP} domain (Figure~\ref{fig:local_strong}),
especially the energetic region annotated
in Figure~\ref{fig:umap_annotated} (dashed
ellipse).
%Examining Figure~\ref{fig:umap_gallery_DT1},
These patterns show the highest degree of complexity
with temperature variance at a range of scales,
and features with strong curvature, and strong gradients.
By the same token, the Gulf Stream avoids
the regions of the \ac{UMAP} domain
where cutouts show predominantly 
large-scale or small-scale structures
(solid and dotted annotations 
respectively in 
Figure~\ref{fig:umap_annotated}).
%at least in this \DT\ subset.
Together, the \relb\ distribution in the upper
panel of Figure~\ref{fig:local_strong} identifies
the \sst\ patterns that manifest in highly dynamic
regions (e.g., near and within western boundary currents).

The lower panel of Figure~\ref{fig:local_strong}
describes the \sst\ patterns prevalent in the
equatorial Indian Ocean.  These occur predominantly
within the lower-right portion of 
the \ac{UMAP} domain (dotted annotation
in Figure~\ref{fig:umap_annotated}).
This region contains primarily cutouts
%are a mix of cutouts with large-scale features
with isolated, cooler patches (e.g., unmasked clouds)
and those dominated by small-scale variance.
%Throughout this dataset
%blemishes characteristic
%of unmasked clouds are present.
Likewise,
it is very rare for this portion of the Indian
Ocean to exhibit temperature variations on 
medium to larger scales.

\begin{figure}[h]
\centering
\includegraphics[width=0.5\textwidth]{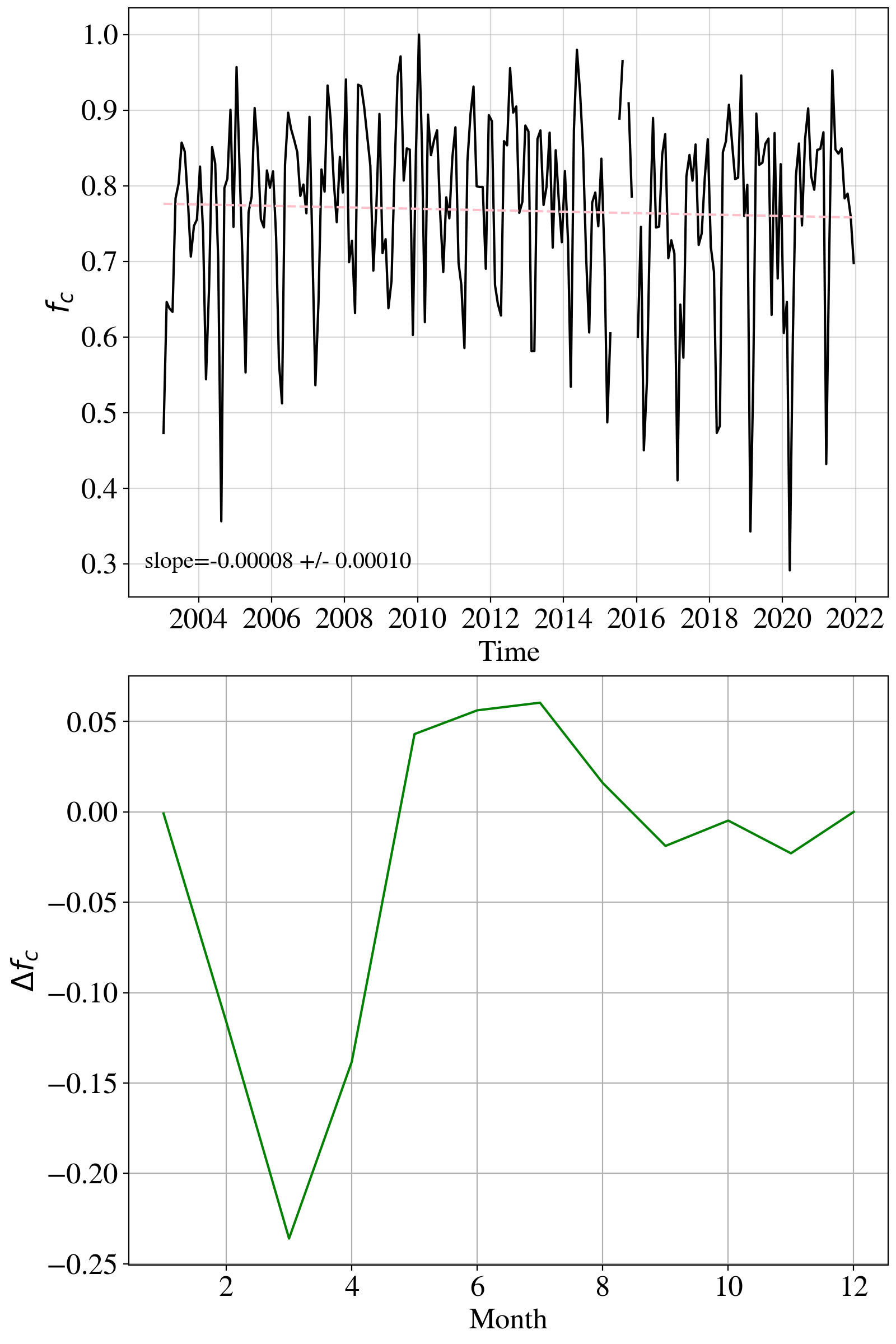}
\caption{
Time series analysis of the cutout fraction
for the prominent \sst\ 
patterns of the Pacific \ac{ECT}
(limited to \DThalf). These are defined by 
the portion of the \ac{UMAP} domain
where $\relb > 1.5$ in Figure~\ref{fig:local_eqpsa}.
A cutout fraction \fracc=1 indicates all of the
cutouts in that month and year have these
patterns.
%(limited to $\DT = [1,1.5]$\,K).
The upper panel shows the full time-series
of \fracc\ and
the dashed line is a fit to the inter-annual
trend accounting for seasonal variations.
There is a strong seasonal signature (lower panel)
associated with 
the intensification of the \ac{ECT} by 
an increase in northward wind stress
\cite{mitchell1992}.
}
\label{fig:time_series_eq}
\end{figure}

% %%%%%%%%%%%%%%%%%%%%%%%%%%%%%%%%%%%%%%%%%
% %%%%%%%%%%%%%%%%%%%%%%%%%%%%%%%%%%%%%%%%%
% %%%%%%%%%%%%%%%%%%%%%%%%%%%%%%%%%%%%%%%%%
\subsection{Temporal Change}
\label{sec:temporal}

In addition to the extenstive (global) 
geographic coverage afforded by remote
sensing observations, the \ac{MODIS} dataset
now spans nearly two decades of daily observations.
This permits a time-series analysis of the incidence
of \ac{SST} patterns characteristic of particular
dynamics. 
We consider both long-term (i.e., inter-annual) and 
short-term (seasonal) trends in
select regions. 

Returning to the Pacific \ac{ECT}, 
we perform a time-series analysis for its
prominent \ac{SST} patterns. 
We define these patterns as the portion of the \ac{UMAP}
domain for the \DThalf\ sample that have 
a relative frequency $f_b > 1.5$,
i.e., the red regions in % the upper panel of  
Figure~\ref{fig:local_eqpsa}.
For each month of each year in the dataset, 
we calculate the
fraction of cutouts \fracc\
in that portion of the \ac{UMAP} domain.
If there are fewer than 10~cutouts in a given month
and year, that time-stamp is ignored.

The time-series of \fracc\
and its seasonal and inter-annual analysis 
is plotted in Figure~\ref{fig:time_series_eq}.
The time-series analysis models the data
assuming a linear trend for any inter-annual 
variation and 11 free parameters for the
months January-November 
to assess seasonal variations
(the results are
relative to December). 
A strong seasonal component is apparent 
with the boreal summer months (May-September) 
dominant and with the \ac{ECT} suppressed
from February-March inclusive.
The enhancement has been associated with 
the intensification of the Pacific \ac{ECT} by 
air-sea interactions
\cite{chelton2001,ray2018}.
There is no significant
linear inter-annual trend, but we do
observe that the near-disappearance 
of the \ac{ECT} (low \fracc)
in February/March has
been intensifying over the 
past $\sim 10$\,years.

\begin{figure}[ht]
\centering
\includegraphics[width=0.5\textwidth]{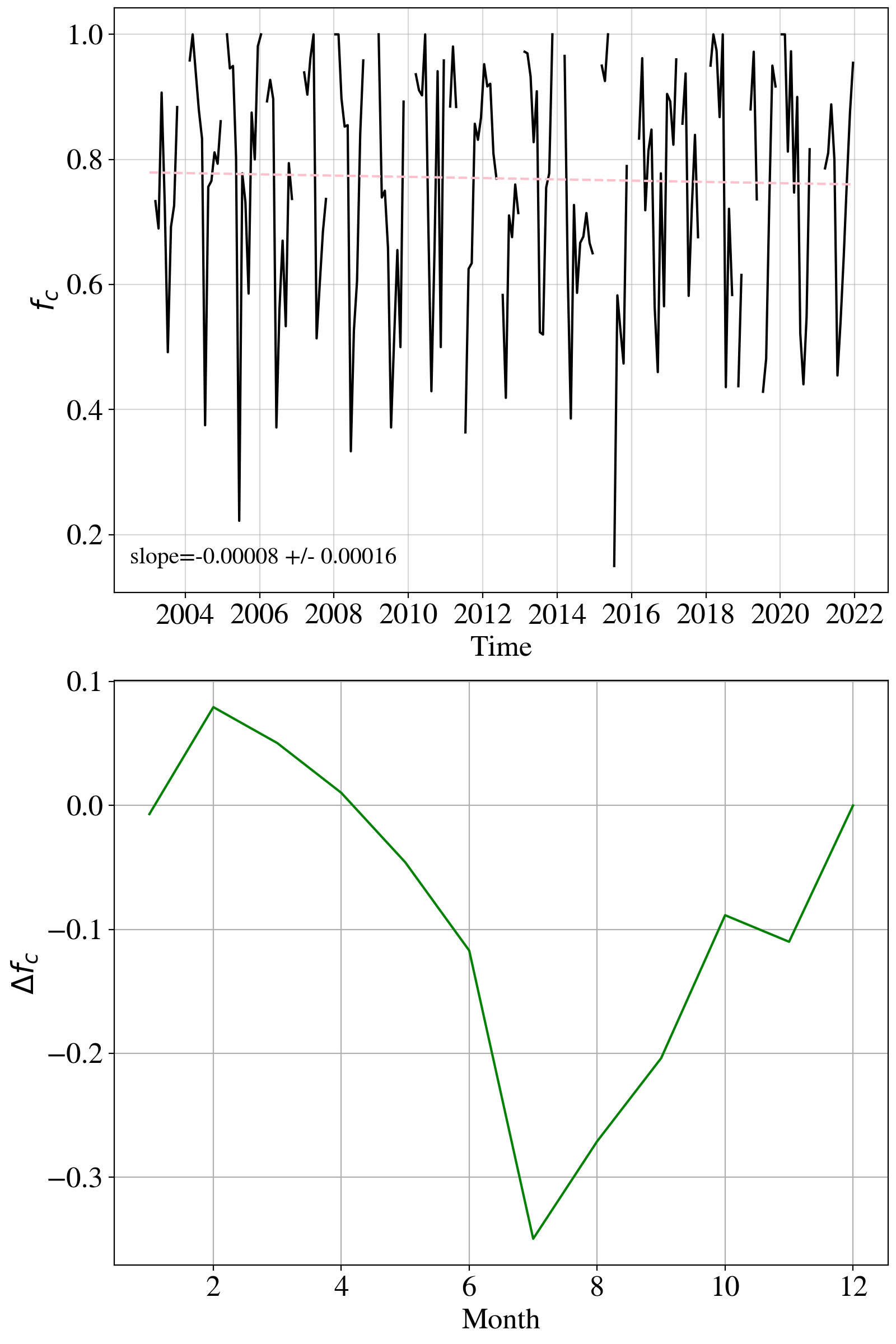}
\caption{
Same as for Figure~\ref{fig:time_series_eq} but for
the enhanced patterns of the
Gulf Stream near New England,
those with $f_b > 1.25$ in the upper panel
of Figure~\ref{fig:local_strong}.
The incidence of these complex patterns exhibits
a strong seasonal dependence, with a strong suppression
during summer months possibly due to masking 
by the seasonal thermocline.
%The strong seasonal dependence tracks the warming
%of the sea and the associated shallowing of the
%mixed layer.
%In addition, there is evidence for
%an inter-annual trend
%($>90\%$ c.l.) of increasing
%dynamics, perhaps associated with global warming.
}
\label{fig:time_series_gulf}
\end{figure}

Turning to a portion of Gulf Stream near New England,
we isolate the cutouts with 
$\relb > 1.25$ in the upper panel of
Figure~\ref{fig:local_strong}
which are those with complex patterns.
Similar to the Pacific \ac{ECT},
the primary \sst\ patterns of the Gulf Stream
exhibit a strong seasonal dependence (Figure~\ref{fig:time_series_gulf}).
Its seasonal trend, however, is (coincidentally)
out-of-phase with the \ac{ECT} showing the
highest $f_c$ in winter months and a great
deficit in mid-summer.
The latter may occur as a significant seasonal thermocline is formed in the area to the north and west of the Gulf Stream effectively obscuring the more turbulent waters beneath.

%\begin{figure}[ht]
%\centering
%\includegraphics[width=0.5\textwidth]{Figures/fig_yearly_geo_DT15_med_96clear_v4_S1.png}
%\caption{
%Same as for Figure~\ref{fig:time_series_eq} but for
%strong gradients in the Mediterranean Sea.
%The strong seasonal dependence tracks the warming
%of the sea and the associated shallowing of the
%mixed layer.
%In addition, there is evidence for
%an inter-annual trend
%($>90\%$ c.l.) of increasing
%dynamics, perhaps associated with global warming.
%}
%\label{fig:time_series_med}
%\end{figure}

% %%%%%%%%%%%%%%%%%%%%%%%%%%%%%%%%%%%%%%%%%%%%%%%%
\section{Summary and Concluding Remarks}
\label{sec:conclusion}

We have designed a deep, contrastive learning model
to reveal and resolve the fundamental patterns
of \sst\ for the global ocean on scales of
$\approx 80 \times 80 \, \rm km^2$ (i.e., sub-mesoscale).
This \mname\ model is intentionally invariant 
to vertical and horizontal flips, 
rotations, and small translations.
We trained \mname\ on \ntrain\ random cutouts
from the \ac{MODIS} \ac{L2} night-time dataset
in the years 2003-2021 inclusive.
To readily explore the 256-dimension latent
space that defines the complexity
of \sst\ patterns,
we bin the cutouts by \DT--the size
of its 90\%\ \sst\ interval--and generated a \ac{UMAP} transformation 
of the latent vectors to 2-dimensions.

We confirmed that \mname\ has successfully separated
the \sst\ patterns in this 3-dimensional space
(one for \DT, and 2 for each \ac{UMAP}) 
to describe the full complexity of \sst\ imagery.
We identify \sst\ patterns with structures on scales 
ranging from the $\approx 2$\,km sampling of the
data to the full extent of the cutouts.
Furthermore, \mname\ separates patterns
with predominantly linear structures 
from those with significant
curvature.
In essence, \mname\ defines a vocabulary
for the language of \sst\ imagery at the 
sub-mesoscale.

Leveraging this vocabulary, we examined the preferred
geographical distribution of several sets of
\sst\ patterns.  
\mname\ reveals that \sst\ patches with 
temperature variance on scales larger 
than the sub-mesoscale
occur preferentially in several
distinct and large regions at low 
latitudes. These include
the Pacific and Atlantic \ac{ECT}s
where the westward propagation of cool waters
upwelled from eastern areas in the basin
yield large temperature gradients along
the equator.
Additionally, we find that complex \sst\ patterns with
significant temperature variance on scales
ranging from a few to tens of kilometers
arise predominantly in dynamically active
regions (e.g., western boundary currents)
at high latitudes.
These results confirm the direct mapping
from \sst\ patterns to distinct ocean processes.

Another application of \mname\ is to resolve
the principal \sst\ patterns 
in select geographical regions.  
We recover highly distinct distributions of
\sst\ patterns when comparing the 
Pacific \ac{ECT} (temperature variations 
primarly on large scales),
%the Southern Atlantic (patterns with 
%small-scale variance in temperature),
a portion of the Gulf Stream near New England 
(complex \sst\ patterns with temperature
variance at a range of scales),
and the Equatorial Indian Ocean 
(primarily small-scale temperature variations).
The distributions of these \sst\ patterns 
are a fingerprint that
define the regions and the dominant near-surface  processes
in these regions.

Last, we performed a time-series analysis
of the defining patterns of the Pacific \ac{ECT}
and the Gulf Stream near New England to search for
inter-annual and seasonal trends.
We identify seasonal trends for each:
 the strengthening of the \ac{ECT} in mid-summer months
 and the depression of dynamic features
 in/near the Gulf Stream during the summer.

We conclude, at least statistically, 
that a vocabulary that defines \sst\ patterns
provides a novel approach to tracking physical
processes across the global ocean and dissecting
such processes within distinct regions.
Emboldened by this success,
we will regress these patterns,
specifically the full latent space representation
of \mname, against specific dynamic processes
(e.g., frontogenesis) to leverage \sst\ 
as an indirect descriptor of such dynamics.
This will require performing analysis and training 
on ocean model output with known dynamics,
e.g., the LLC4320 ocean general circulation model \cite{Rocha2016a,Rocha2016b,Arbic2018}.
Ultimately, we may
track the roles of sub-mesoscale
processes in mixing and energy dissipation
across the globe over the past two~decades.

With this publication, we provide all of the data,
outputs, and software related to \mname\ and this
manuscript.  We also provide on-line
visualization tools that enable the community
to explore \sst\ patterns from the dataset
and those provided by the user.
See XXX for full details.

\begin{table*}
\centering
\caption{MODIS Cutouts\label{tab:cutouts}}
\begin{tabular}{cccccccccc}
\hline 
lon & lat & date & $\Delta T$ & \slope & LL & $U_{0,\rm all}$ & $U_{1,\rm all}$& $U_0$ & $U_1$ \\ 
(deg) & (deg) & & (K) 
\\ 
\hline 
8.688 & -32.425& 2003-01-01 00:35:00& 0.495& -1.86& 172.7& 3.9 & 3.5& -1.6 & 5.3\\ 
9.434 & -33.705& 2003-01-01 00:35:00& 0.655& -1.77& 204.6& 0.1 & 2.7& 6.8 & 2.2\\ 
10.311 & -33.231& 2003-01-01 00:35:00& 0.689& -1.83& 90.3& 1.8 & 2.0& 9.1 & 1.6\\ 
8.381 & -33.559& 2003-01-01 00:35:00& 0.712& -2.18& 188.9& -0.3 & 3.6& 7.4 & 0.9\\ 
9.303 & -32.807& 2003-01-01 00:35:00& 0.754& -1.85& 137.0& 2.0 & 1.2& 9.0 & 1.2\\ 
9.151 & -33.372& 2003-01-01 00:35:00& 0.536& -1.68& 212.0& -0.7 & 2.6& 8.0 & 0.9\\ 
11.091 & -33.327& 2003-01-01 00:35:00& 0.470& -1.87& 274.6& 2.7 & 2.2& -0.0 & 0.8\\ 
10.385 & -32.951& 2003-01-01 00:35:00& 0.605& -1.77& 149.0& 2.4 & 2.0& 8.9 & 0.2\\ 
11.023 & -33.612& 2003-01-01 00:35:00& 0.374& -1.93& 312.6& 3.3 & 3.6& -0.4 & 1.1\\ 
9.030 & -32.474& 2003-01-01 00:35:00& 0.645& -1.78& 183.6& 1.0 & 2.1& 8.1 & 1.6\\ 
9.940 & -33.183& 2003-01-01 00:35:00& 0.679& -1.95& 110.6& 1.5 & 3.3& 9.2 & 1.5\\ 
9.176 & -31.906& 2003-01-01 00:35:00& 0.415& -2.05& 242.9& 2.3 & 3.1& 0.3 & 4.3\\ 
9.506 & -33.420& 2003-01-01 00:35:00& 0.707& -1.63& 187.1& 0.3 & 2.8& 8.3 & 1.0\\ 
9.869 & -33.468& 2003-01-01 00:35:00& 0.715& -2.07& 108.1& 0.8 & 1.9& 8.9 & 1.9\\ 
8.613 & -32.710& 2003-01-01 00:35:00& 0.518& -1.83& 161.0& 3.7 & 3.7& 9.4 & 0.8\\ 
8.348 & -32.376& 2003-01-01 00:35:00& 0.418& -1.86& 162.4& 4.4 & 3.9& -2.0 & 5.3\\ 
10.016 & -32.903& 2003-01-01 00:35:00& 0.565& -1.82& 152.3& 2.6 & 3.7& 8.9 & 0.6\\ 
10.241 & -33.515& 2003-01-01 00:35:00& 0.748& -2.05& 111.9& 1.2 & 1.7& 9.0 & 1.9\\ 
10.454 & -32.666& 2003-01-01 00:35:00& 0.412& -1.67& 249.4& 3.8 & 3.1& -0.6 & 4.6\\ 
11.876 & -33.711& 2003-01-01 00:35:00& 0.386& -1.95& 384.9& 3.9 & 4.6& 1.6 & 4.0\\ 
10.085 & -32.618& 2003-01-01 00:35:00& 0.420& -1.69& 250.1& 3.8 & 3.5& -0.4 & 4.4\\ 
\hline 
\end{tabular} 
\\ 
Notes: The \DT\ value listed here is measured from the inner $40 \times 40$\,pixel$^2$ region of the cutout. \\ 
LL is the log-likelihood metric calculated from the \ulmo\ algorithm. \\ 
$U_{0,\rm all}, U_{1,\rm all}$ are the UMAP values for the UMAP analysis on the full dataset. \\ 
$U_0, U_1$ are the UMAP values for the UMAP analysis in the \DT\ bin for this cutout. \\ 
\end{table*}

\newpage
\noindent{\bf Acronyms}
{\smaller \begin{acronym}[12345678901234]
\acro{(A)ATSR}{one or all of \textsmaller{ATSR}, \textsmaller{ATSR-2} and \textsmaller{AATSR}}
\acro{AATSR}{Advanced Along Track Scanning Radiometer}
\acro{ACC}{Antarctic Circumpolar Current}
\acro{ACCESS}{Advancing Collaborative Connections for Earth System Science}
\acro{ACL}{Access Control List}
\acro{ACSPO}{Advanced Clear Sky Processor for Oceans}
\acro{ADA}{Automatic Detection Algorithm}
\acro{ADCP}{Acoustic Doppler Current Profiler}
\acro{ADT}{absolute dynamic topography}
\acro{AESOP}{Assessing the Effects of Submesoscale Ocean Parameterizations}
\acro{AGU}{American Geophysical Union}
\acro{AI}{Artificial Intelligence}
\acro{AIRS}{Atmospheric Infrared Sounder}
\acro{AIS}{Ancillary Information Service}
\acro{AIST}{Advanced Information Systems Techonology}
\acro{AISR}{Applied Information Systems Research}
\acro{ADL}{Alexandria Digital Library}
\acro{API}{Application Program Interface}
\acro{APL}{Applied Physics Laboratory}
\acro{API}{Application Program Interface}
\acro{AMSR}{Advanced Microwave Scanning Radiometer}
\acro{AMSR2}{Advanced Microwave Scanning Radiometer 2}
\acro{AMSR-E}{Advanced Microwave Scanning Radiometer - \textsmaller{EOS}}
\acro{ANN}{ Artificial Neural Network}
\acro{AOOS}{Alaska Ocean Observing System}
\acro{APAC}{Australian Partnership for Advanced Computing}
\acro{APDRC}{Asia-Pacific Data-Research Center}
%\acro{ARC}{\acs{ATSR} Reprocessing for Climate}
\acro{ARC}{{\smaller ATSR} Reprocessing for Climate}
\acro{ASCII}{American Standard Code for Information Interchange}
\acro{AS}{Aggregation Server}
\acro{ASFA}{Aquatic Sciences and Fisheries Abstracts}
\acro{ASTER}{Advanced Spaceborne Thermal Emission and Reflection Radiometer}
%{ -- \it http://asterweb.jpl.nasa.gov}
\acro{ATBD}{Algorithm Theoretical Basis Document}
\acro{ATSR}{Along Track Scanning Radiometer}
\acro{ATSR-2}{Second \textsmaller{ATSR}}
\acro{AVISO}{Archiving, Validation and Interpretation of Satellite Oceanographic Data}
\acro{ANU}{Australian National University}
\acro{AVHRR}{Advanced Very High Resolution Radiometer}
\acro{AzC}{Azores Current}

\acro{BAA}{Broad Agency Announcement}
\acro{BAO}{bi-annual oscillation}
\acro{BES}{Back-End Server}
\acro{BMRC}{Bureau of Meteorology Research Centre}
\acro{BOM}{Bureau of Meteorology}
\acro{BT}{brightness temperature}
\acro{BUFR}{Binary Universal Format Representation}
%{ -- \it http://www.wmo.ch/web/www/WDM/Guides/Guide-on-DataMgt-1.htm}

\acro{CAN}{Cooperative Agreement Notice}
\acro{CAS}{Community Authorization Service}
\acro{CC}{cloud cover}
\acro{CCA}{Cayula-Cornillon Algoritm}
\acro{CCI}{Climate Change Initiative}
\acro{CCLRC}{Council for the Central Laboratory of the Research Councils}
%{ --- \it http://www.cclrc.ac.uk/}
\acro{CCMA}{Center for Coastal Monitoring and Assessment}
\acro{CCR}{cold core ring}
\acro{CCS}{California Current System }
\acro{CCSM}{Community Climate System Model}
\acro{CCSR}{Center for Climate System Research}
\acro{CCV}{Center for Computation and Visualization}
\acro{CDAT}{Climate Data Analysis Tools}
\acro{CDC}{Climate Diagnostics Center}
\acro{CDF}{Common Data Format}
\acro{CDR}{Common Data Representation}
\acro{CEDAR}{Coupled Energetic and Dynamics and Atmospheric Regions}
%{ -- \it http://cedarweb.hao.ucar.edu/}
\acro{CEOS}{Committee on Earth Observation Satellites}
\acro{CERT}{Computer Emergency Response Team}
\acro{CenCOOS}{Central \& Northern California Ocean Observing System}
% \acro{CF}{NetCDF Climate and Forecast Metadata Conventions}
\acro{CF}{clear fraction}
\acro{CGI}{Common Gateway Interface}
%\acro{CHAP}{\textsmaller{CISL} \ac{HPC} Advisory Panel}
\acro{CHAP}{\textsmaller{CISL} High Performance Computing Advisory Panel}
\acro{CIFS}{Common Internet File System}
\acro{CIMSS}{Cooperative Institute for Meteorological Satellite Studies}
\acro{CIRES}{Cooperative Institute for Research (in) Environmental Sciences}
\acro{CISL}{Computational \& Information Systems Laboratory}
\acro{CLASS}{Comprehensive Large Array-data Stewardship System}
\acro{CLIVAR}{Climate Variability and Predictability}
\acro{CLS}{Collecte Localisation Satellites}
\acro{CME}{Community Modeling Effort}
\acro{CMS}{Centre de M\'et\'eorologie Spatiale}
\acro{CNN}{Convolutional Neural Network}
\acro{COA}{Climate Observations and Analysis}
\acro{COARDS}{Cooperative Ocean-Atmosphere Research Data Standard}
\acro{COAPS}{Center for Ocean-Atmospheric Prediction Studies}
\acro{COBIT}{Control Objectives for Information and related Technology}
%\acro{COCO}{\acs{CCSR} Ocean Component model}
\acro{COCO}{{\smaller CCSR} Ocean Component model}
\acro{CODAR}{Coastal Ocean Dynamics Applications Radar}
\acro{CODMAC}{Committee on Data Management, Archiving, and Computing}
\acro{Co-I}{Co-Investigator}
\acro{CORBA}{Common Object Request Broker Architecture}
\acro{COLA}{Center for Ocean-Land-Atmosphere Studies}
%{ -- \it http://grads.iges.org/cola.html}
\acro{CPU}{Central Processor Unit}
\acro{CRS}{Coordinate Reference System}
\acro{CSA}{Cambridge Scientific Abstracts}
\acro{CSC}{Coastal Services Center}
\acro{CSIS}{Center for Strategic and International Studies}
\acro{CSL}{Constraint Specification Language}
\acro{CSP}{Chermayeff, Sollogub and Poole, Inc.}
%{ -- \it http://csp-architects.com/contact.htm}
\acro{CSDGM}{Content Standard for Digital Geospatial Metadata}
%\acro{CTD}{Conductivity, Temperature and Salinity probes}
\acro{CSV}{Comma Separated Values}
\acro{CTD}{Conductivity, Temperature and Salinity}
\acro{CVSS}{Common Vulnerability Scoring System}
\acro{CZCS}{Coastal Zone Color Scanner}

\acro{DAAC}{Distribute Active Archive Center}
\acro{DAARWG}{Data Archiving and Access Requirements Working Group}
\acro{DAP}{Data Access Protocol}
\acro{DAS}{Data set Attribute Structure}
\acro{DBMS}{Data Base Management System}
\acro{DBDB2}{Digital Bathymetric Data Base} 
\acro{DChart}{Dapper Data Viewer}
\acro{DDS}{Data Descriptor Structure}
\acro{DDX}{\textsmaller{XML} version of the combined \textsmaller{DAS} and \textsmaller{DDS}}
\acro{DFT}{Discrete Fourier Transform}
\acro{DIF}{Directory Interchange Format}
\acro{DISC}{Data and Information Services Center}
\acro{DIMES}{Diapycnal and Isopycnal Mixing Experiment:  Southern Ocean}
\acro{DMAC}{Data Management and Communications committee}
\acro{DMR}{Department of Marine Resources}
\acro{DMSP}{Defense Meteorological Satellite Program}
\acro{DoD}{Department of Defense}
\acro{DODS}{Distributed Oceanographic Data System}
%{ -- \it http://www.unidata.ucar.edu/packages/dods}
\acro{DOE}{Department of Energy}
%{ -- \it http://www.energy.gov}
\acro{DSP}{U. Miami satellite data processing software}
\acro{DSS}{direct statistical simulation}

\acro{EASy}{Environmental Analysis System}
%{ -- \it http://members.cox.net/fjobrien/global/layout/people.htm}
\acro{ECCO}{Estimating the Circulation and Climate of the Ocean}
\acro{ECCO2}{{\smaller  Estimating the Circulation and Climate of the Ocean} Phase II}
\acro{ECS}{\textsmaller{EOSDIS} Core System}
\acro{ECT}{equatorial cold tongue}
\acro{ECHO}{Earth Observing System Clearinghouse}
\acro{ECMWF}{European Centre for Medium-range Weather Forecasting}
\acro{ECV}{Essential Climate Variable}
\acro{EDC}{Environmental Data Connector}
\acro{EDJ}{Equatorial Deep Jet}
\acro{EDFT}{Extended Discrete Fourier Transform}
\acro{EDMI}{Earth Data Multi-media Instrument}
%{ -- \newline \it http://www.newmediastudio.org/Homepage/TNMSHomeFramset.htm}
\acro{EEJ}{Extra-Equatorial Jet}
\acro{EIC}{Equatorial Intermediate Current}
\acro{EICS}{Equatorial Intermediate Current System}
\acro{EJ}{Equatorial Jets}
\acro{EKE}{eddy kinetic energy}
\acro{EMD}{Empirical Mode Decomposition}
\acro{EOF}{Empirical Orthogonal Function}
\acro{EOS}{Earth Observing System}
\acro{EOSDIS}{Earth Observing System Data Information System}
\acro{EPA}{Environmental Protection Agency}
\acro{EPSCoR}{Experimental Program to Stimulate Competitive Research}
\acro{EPR}{East Pacific Rise}
\acro{ERD}{Environmental Research Division}
\acro{ERS}{European Remote-sensing Satellite}
\acro{ESA}{European Space Agency}
\acro{ESDS}{Earth Science Data Systems}
%{ -- \it http://www.esdswg.org/}
\acro{ESDSWG}{Earth Science Data Systems Workign Group}
\acro{ESE}{Earth Science Enterprise}
\acro{ESG}{Earth System Grid}
%{ -- \it http://www.earthsystemgrid.org/}
\acro{ESG II}{Earth System Grid -- II}
%{ -- \it  http://www.earthsystemgrid.org}
\acro{ESIP}{Earth Science Information Partner}
%{ -- \it http://www.esipfed.org}
\acro{ESMF}{Earth System Modeling Framework}
%{ -- \texttt{http://www.esmf.ucar.edu}}
\acro{ESML}{Earth System Markup Language}
%{ -- \it http://esml.itsc.uah.edu/index.jsp}
\acro{ESP}{eastern South Pacfic}
\acro{ESRI}{Environmental Systems Research Institute}
%{-- \it http://www.esri.com}
\acro{ESR}{Earth and Space Research}
\acro{ETOPO}{Earth Topography}
\acro{EUC}{Equatorial Undercurrent}
\acro{EUMETSAT}{European Organisation for the Exploitation of Meteorological Satellites}
\acro{Ferret}{}
%{ -- \it http://ferret.pmel.noaa.gov/Ferret}

\acro{FASINEX}{Frontal Air-Sea Interaction Experiment}
\acro{FDS}{Ferret Data Server}
\acro{FFT}{Fast Fourier Transform}
\acro{FGDC}{Federal Geographic Data Committee}
%{ -- \it http://www.fgdc.gov}
\acro{FITS}{Flexible Image (or Interchange) Transport System}
\acro{FLOPS}{FLoating point Operations Per Second} 
\acro{FRTG}{Flow Rate Task Group}
\acro{FreeForm}{}
%{ -- \it http://www.ngdc.noaa.gov/seg/freeform/freeform.shtml}
\acro{FNMOC}{Fleet Numerical Meteorology and Oceanography Center}
\acro{FSU}{Florida State University}
\acro{FTE}{Full Time Equivalent}
\acro{ftp}[\normalsize  ftp]{File Transport Protocol}
\acro{FTP}[\normalsize  ftp]{File Transport Protocol}

\acro{GAC}{Global Area Coverage}
\acro{GAN}{Generative Adversarial Network}
\acro{GB}{GigaByte - $10^{9}$ bytes}
\acro{GCMD}{Global Change Master Directory}
%{ -- \it http://gcmd.nasa.gov}
\acro{GCM}{general circulation model}
\acro{GCOM-W1}{Global Change Observing Mission - Water}
\acro{GCOS}{Global Climate Observing System}
\acro{GDAC}{Global Data Assembly Center}
\acro{GDS}{\textsmaller{GrADS} Data Server}
\acro{GDS2}{GHRSST Data Processing Specification v2.0}
%{ -- \it http://grads.iges.org/grads/gds}
\acro{GEBCO}{General Bathymetric Charts of the Oceans}
\acro{GeoTIFF}{Georeferenced Tag Image File Format}
\acro{GEO-IDE}{Global Earth Observation Integrated Data Environment}
\acro{GES DIS}{Goddard Earth Sciences Data and Information Services Center}
\acro{GEMPACK}{General Equilibrium Modelling PACKage}
\acro{GEOSS}{Global Earth Observing System of Systems}
\acro{GFDL}{Geophysical Fluid Dynamics Laboratory}
\acro{GFD}{Geophysical Fluid Dynamics}
\acro{GHRSST}{Group for High Resolution Sea Surface Temperature}
\acro{GHRSST-PP}{\textsmaller{GODAE} High Resolution Sea Surface Temperature Pilot Project}
%{ -- \it  http://www.ghrsst-pp.org}
%\acro{GHRSST-PP}{\acs{GODAE}High Resolution Sea Surface Temperature Pilot Project}
%{ -- \it  http://www.ghrsst-pp.org}
\acro{GINI}{\textsmaller{GOES} Ingest and \textsmaller{NOAA/PORT} Interface}
\acro{GIS}{Geographic Information Systems} %{ -- \it http://www.gis.com}
\acro{Globus}{} %{ -- {http://www.globus.org}}
\acro{GMAO}{Global Modeling and Assimilation Office}
\acro{GML}{Geography Markup Language}
\acro{GMT}{Generic Mapping Tool}
\acro{GODAE}{Global Ocean Data Assimilation Experiment} %{ -- \it http://www.bom.gov.au/bmrc/ocean/GODAE}
\acro{GOES}{Geostationary Operational Environmental Satellites} %{ -- \it http://www.oso.noaa.gov/goes}
\acro{GOFS}{Global Ocean Forecasting System}
\acro{GoMOOS}{Gulf of Maine Ocean Observing System}
\acro{GOOS}{Global Ocean Observing System}
\acro{GOSUD}{Global Ocean Surface Underway Data}
\acro{GPFS}{ General Parallel File System}
\acro{GPU}{Graphics Processing Unit}
\acro{GRACE}{Gravity Recovery and Climate Experiment} %{ -- \it http://www.csr.utexas.edu/grace/spacecraft/config.html}
\acro{GRIB}{GRid In Binary} %{ -- \it http://www.wmo.ch/web/www/WDM/Guides/Guide-on-DataMgt-1.htm}
\acro{GrADS}{Grid Analysis and Display System} %{ -- \it http://grads.iges.org/grads/index.html}
\acro{GridFTP}{\textsmaller{FTP} with GRID enhancements}
\acro{GRIB}{GRid in Binary} %{ -- \it http://www.wmo.ch/web/www/DPS/grib-2.html}
\acro{GPS}{Global Positioning System}
\acro{GSFC}{Goddard Space Flight Center} %{ -- \it http://www.gsfc.nasa.gov}
\acro{GSI}{Grid Security Infrastructure}
\acro{GSO}{Graduate School of Oceanography}
\acro{GTSPP}{Global Temperature and Salinity Profile Program}
\acro{GUI}{Graphical User Interface}
\acro{GS}{Gulf Stream}

\acro{HAO}{High Altitude Observatory} %{ -- \it http://www.hao.ucar.edu/public/inside/data.html}
\acro{HLCC}{Hawaiian Lee Countercurrent}
\acro{HCMM}{Heat Capacity Mapping Mission}
\acro{HDF}{Hierarchical Data Format}
\acro{HDF-EOS}{Hierarchical Data Format - \textsmaller{EOS}} %{ -- \it http://hdfeos.gsfc.nasa.gov}
\acro{HEALPix}{Hierarchical Equal Area isoLatitude Pixelation}
\acro{HEC}{High-End Computing}
\acro{HF}{High Frequency}
\acro{HGE}{High Gradient Event}
\acro{HPC}{High Performance Computing}
\acro{HPCMP}{High Performance Computing Modernization Program}
\acro{HPSS}{High Performance Storage System} %{ -- \it http://www.sdsc.edu/hpss/hpss1.html}
\acro{HR DDS}{High Resolution Diagnostic Data Set}
\acro{HRPT}{High Resolution Picture Transmission}
\acro{HTML}{Hyper Text Markup Language}
\acro{html}{Hyper Text Markup Language}
\acro{http}{the hypertext transport protocol}
\acro{HTTP}{Hyper Text Transfer Protocol}
\acro{HTTPS}{Secure Hyper Text Transfer Protocol} %\acro{http}{Hypertext Transport Protocol} It seems that the acronym is all caps.
\acro{HYCOM}{HYbrid Coordinate Ocean Model} %{ -- \it http://oceanmodeling.rsmas.miami.edu/hycom/}

\acro{I-band}{imagery resolution band}
\acro{IDD}{Internet Data Distribution}
\acro{IB}{Image Band}
\acro{IBL}{internal boundary layer}
\acro{IBM}{Internation Business Machines}
%{ -- \it http://www.ibm.com/us}
\acro{ICCs}{Intermediate Countercurrents}
\acro{IDE}{Integrated Development Environment}
\acro{IDL}{Interactive Display Language}
%{ -- \it  http://www.rsinc.com/idl/index.asp}
\acro{IDLastro}{\textsmaller{IDL} Astronomy User's Library}
\acro{IDV}{Integrated Data Viewer}
%{ -- \it http://my.unidata.ucar.edu/content/software/metapps/index.html}
\acro{IEA}{Integrated Ecosystem Assessment}
\acro{IEEE}{Institute (of) Electrical (and) Electronic Engineers}
%{ -- \it http://www.ieee.org/portal/index.jsp}
\acro{IETF}{Internet Engineering Task Force}
\acro{IFREMER}{Institut Fran\c{c}ais de Recherche pour l'Exploitation de la MER}
%\acro{IMF}{Interplanetary Magnetic Field}
\acro{IMAPRE}{El Instituto del Mar del Per\'u}
\acro{IMF}{Intrinsic Mode Function}
\acro{IOOS}{Integrated Ocean Observing System}
\acro{ISAR}{Infrared Sea surface temperature Autonomous Radiometer}
\acro{ISO}{International Organization for Standardization}
\acro{ISSTST}{Interim Sea Surface Temperature Science Team}
\acro{IT}{Information Technology}
\acro{ITCZ}{Intertropical Convergence Zone} 
\acro{IP}{Internet Provider}
\acro{IPCC}{Intergovernmental Panel on Climate Change}
\acro{IPRC}{International Pacific Research Center}
\acro{IR}{infrared}
\acro{IRI}{International Research Institute for Climate and Society}
\acro{ISO}{International Standards Organization}

\acro{JASON}{JASON Foundation for Education}
%\acro{JASON}{NASA Altimeter}
%{ -- \it http://www.jason.org}
\acro{JDBC}{Java Database Connectivity}
\acro{JFR}{Juan Fern\'andez Ridge}
\acro{JGOFS}{Joint Global Ocean Flux Experiment}
%{ -- \it http://puddle.mit.edu/datasys/jgsys.html}
\acro{JHU}{Johns Hopkins University}
\acro{JPL}{Jet Propulsion Laboratory}
\acro{JPSS}{Joint Polar Satellite System}
%{ -- \it http://www.jpl.nasa.gov}

\acro{KDE}{Kernel Density Estimation}
\acro{KVL}{Keyword-Value List}
\acro{KML}{Keyhole Markup Language}
\acro{KPP}{K-Profile Parameterization}

\acro{LAC}{Local Area Coverage}
\acro{LAN}{Local Area Network}
\acro{LAS}{Live Access Server}
%{ -- \it http://www.ferret.noaa.gov/nopp/main.pl? }
\acro{LASCO}{Large Angle and Spectrometric Coronagraph Experiment}
\acro{LatMIX}{Scalable Lateral Mixing and Coherent Turbulence}
\acro{LDAP}{Lightweight Directory Access Protocol}
\acro{LDEO}{Lamont Doherty Earth Observatory}
\acro{LEAD}{Linked Environments for Atmospheric Discovery}
\acro{LEIC}{Lower Equatorial Intermediate Current}
\acro{LES}{Large Eddy Simulation}
\acro{L1}{Level-1}
\acro{L2}{Level-2}
\acro{L3}{Level-3}
\acro{L4}{Level-4}
\acro{LL}{log-likelihood}
\acro{LLC}{Latitude/Longitude/polar-Cap}
\acro{LLC4320}[LLC-4320]{\ac{LLC}-4320}
\acro{LLC2160}[LLC-2160]{\ac{LLC}-2160}
\acro{LLC1080}[LLC-1080]{\ac{LLC}-1080}
\acro{LHF}{Latent Heat Flux}
\acro{LST}{local sun time}
\acro{LTER}{Long Term Ecological Research Network}
\acro{LTSRF}{Long Term Stewardship and Reanalysis Facility}
\acro{LUT}{Look Up Table}

\acro{M-band}{moderate resolution band}
\acro{MABL}{marine atmospheric boundary layer}
\acro{MADT}{Maps of Absolute Dynamic Topography}
\acro{MapServer}{MapServer}
%{ -- \it http://mapserver.gis.umn.edu/}
\acro{MAT}{Metadata Acquisition Toolkit}
\acro{MATLAB}{}
%{ -- \it http://www.mathworks.com/products/}
\acro{MARCOOS}{Mid-Atlantic Coastal Ocean Observing System}
\acro{MARCOORA}{Mid-Atlantic Coastal Ocean Observing Regional Association}
\acro{MB}{MegaByte - $10^{6}$ bytes}
\acro{MCC}{Maximum Cross-Correlation}
\acro{MCR}{\textsmaller{MATLAB} Component Runtime}
\acro{MCSST}{Multi-Channel Sea Surface Temperature}
\acro{MDT}{mean dynamic topography}
\acro{MDB}{Match-up Data Base}
\acro{MDOT}{mean dynamic ocean topography}
\acro{MEaSUREs}{Making Earth System data records for Use in Research Environments}
\acro{MERRA}{Modern Era Retrospective-Analysis for Research and Applications}
\acro{MERSEA}{Marine Environment and Security for the European Area}
\acro{MTF}{Modulation Transfer Function}
\acro{MICOM}{Miami Isopycnal Coordinate Ocean Model}
\acro{MIRAS}{Microwave Imaging Radiometer with Aperture Synthesis}
\acro{MITgcm}{{\smaller MIT} General Circulation Model}
\acro{MIT}{Massachusetts Institute of Technology}
\acro{mks}{meters, kilograms, seconds}
\acro{MLP}{Multilayer Perceptron}
\acro{ML}{machine learning}
\acro{MLSO}{Mauna Loa Solar Observatory}
%{ -- \it http://mslo.\ac{HAO}ucar.edu/}
\acro{MM5}{Mesoscale Model}
\acro{MMI}{Marine Metadata Initiative}
\acro{MMS}{Minerals Management Service}
\acro{MODAS}{Modular Ocean Data Assimilation System}
\acro{MODIS}{MODerate-resolution Imaging Spectroradiometer}
%{ -- \it http://modis.gsfc.nasa.gov}
\acro{MOU}{Memorandum of Understanding}
\acro{MPARWG}{Metrics Planning and Reporting Working Group}
\acro{MSE}{mean square error}
\acro{MSG}{Meteosat Second Generation}
\acro{MTPE}{Mission To Planet Earth}
\acro{MUR}{Multi-sensor Ultra-high Resolution}
\acro{MV}{Motor Vessel}

\acro{NAML}{National Association of Marine Laboratories}
\acro{NAHDO}{National Association of Health Data Organizations}
%{ -- \it http://www.nahdo.org}
\acro{NAS}{Network Attached Storage}
\acro{NASA}{National Aeronautics and Space Administration}
%{ -- \it http://www.nasa.gov}
\acro{NCAR}{National Center for Atmospheric Research}
%{ -- \it http://www.ncar.ucar.edu/ncar/index.html}
\acro{NCEI}{National Centers for Environmental Information}
\acro{NCEP}{National Centers for Environmental Prediction}
\acro{NCDC}{National Climatic Data Center}
\acro{NCDDC}{National Coastal Data Development Center}
\acro{NCL}{NCAR Command Language}
\acro{ncBrowse}{}
%{ -- \it http://www.epic.noaa.gov/java/ncBrowse}
\acro{NcML}{\textsmaller{netCDF} Markup Language}
\acro{NCO}{\textsmaller{netCDF} Operator}
\acro{NCODA}{Navy Coupled Ocean Data Assimilation}
\acro{NCSA}{National Center for Supercomputing Applications}
\acro{NDBC}{National Data Buoy Center}
\acro{NDVI}{Normalized Difference Vegetation Index}
\acro{NEC}{North Equatorial Current}
\acro{NECC}{North Equatorial Countercurrent}
\acro{NEFSC}{Northeast Fisheries Science Center}
\acro{NEIC}{North Equatorial Intermediate Current}
\acro{netCDF}{NETwork Common Data Format}
\acro{NEUC}{North Equatorial Undercurrent}
\acro{NGDC}{National Geophysical Data Center}
%{ -- \it http://www.ngdc.noaa.gov}
\acro{NICC}{North Intermediate Countercurrent}
\acro{NIST}{National Institute of Standards and Technology}
\acro{NLSST}{Non-Linear Sea Surface Temperature}
\acro{NMFS}{National Marine Fisheries Service}
\acro{NMS}{New Media Studio}
%{ -- \it http://newmediastudio.org/Homepage/TNMSHomeFramset.htm}
\acro{NN}{Neural Network}
\acro{NOAA}{National Oceanic and Atmospheric Administration}
%{ -- \it http://www.noaa.gov}
\acro{NODC}{National Oceanographic Data Center}
\acro{NOGAPS}{Navy Operational Global Atmospheric Prediction System}
\acro{NOMADS}{\textsmaller{NOAA} Operational Model Archive Distribution System}
%{ -- \it http://www.ncdc.noaa.gov/oa/climate/nomads/nomads.html}
\acro{NOPP}{National Oceanographic Parternership Program}
%{ -- \it http://www.nopp.org}
\acro{NOS}{National Ocean Service}
\acro{NPP}{National Polar-orbiting Partnership}
\acro{NPOESS}{National Polar-orbiting Operational Environmental Satellite System}
\acro{NSCAT}{\textsmaller{NASA} SCATterometer}
%{ -- \it http://winds.jpl.nasa.gov}
%\acro{NSEN}{\acs{NASA}Science and Engineering Network}
\acro{NSEN}{\textsmaller{NASA} Science and Engineering Network}
\acro{NSF}{National Science Foundation}
%\acro{NSIPP}{\acs{NASA}Seasonal-to-Interannual Prediction Project}
\acro{NSIPP}{NASA Seasonal-to-Interannual Prediction Project}
\acro{NRA}{NASA Research Announcement}
\acro{NRC}{National Research Council}
\acro{NRL}{Naval Research Laboratory}
\acro{NSCC}{North Subsurface Countercurrent}
\acro{NSF}{National Science Foundation}
\acro{NSIDC}{National Snow and Ice Data Center}
\acro{NSPIRES}{\textsmaller{NASA} Solicitation and Proposal Integrated Review and Evaluation System}
\acro{NSSDC}{National Space Science Data Center}
\acro{NVODS}{National Virtual Ocean Data System}
%{ -- \it http://nvods.org}
\acro{NWP}{Numerical Weather Prediction}
\acro{NWS}{National Weather Service}

\acro{OBPG}{Ocean Biology Processing Group}
\acro{OB.DAAC}{Ocean Biology \textsmaller{DAAC}}
\acro{ODC}{\textsmaller{OPeNDAP} Data Connector}
\acro{OC}{ocean color}
\acro{OCAPI}{\textsmaller{OPeNDAP C API}}
\acro{ODSIP}{Open Data Services Invocation Protocol}
\acro{OFES}{Ocean Model for the Earth Simulator}
\acro{OCCA}{OCean Comprehensive Atlas}
\acro{OGC}{Open Geospatial Consortium}
%{ -- \it http://www.opengis.org}
\acro{OGCM}{ocean general circulation model}
\acro{OISSTv1}{Optimally Interpolated SST Version 1}
\acro{ONR}{Office of Naval Research}
\acro{OLCI}{Ocean Land Colour Instrument}
\acro{OLFS}{\textsmaller{OPeNDAP} Lightweight Front-end Server}
\acro{OOD}{out-of-distribution}
\acro{OOPC}{Ocean Observation Panel for Climate}
\acro{OPeNDAP}{Open source Project for a Network Data Access Protocol}
%{ -- \it http://opendap.org}
\acro{OPeNDAPg}{\textsmaller{GRID}-enabled \textsmaller{OPeNDAP} tools}
\acro{OpenGIS}{OpenGIS}
\acro{OSI SAF}{Ocean and Sea Ice Satellite Application Facility}
\acro{OSS}{Office of Space Science}
\acro{OSTM}{Ocean Surface Topography Mission }
\acro{OSU}{Oregon State University}
\acro{OS X}{}
\acro{OWL}{Web Ontology Language}
\acro{OWASP}{Open Web Application Security Project}

\acro{PAE}{probabilistic autoencoder}
\acro{PBL}{planetary boundary layer}
\acro{PCA}{Principal Components Analysis}
\acro{PDistF}[PDF]{probability distribution function}
\acro{PDF}{probability density function}
\acro{PF}{Polar Front}
\acro{PFEL}{Pacific Fisheries Environmental Laboratory}
\acro{PI}{Principal Investigator}
\acro{PIV}{Particle Image Velocimetry}
\acro{PL}{Project Leader}
\acro{PM}{Project Member}
\acro{PMEL}{Pacific Marine Environmental Laboratory}
%{ -- \it http://pmel.noaa.gov}
\acro{POC}{particulate organic carbon}
\acro{PO-DAAC}{Physical Oceanography -- Distributed Active Archive Center}
%{ -- \it http://podaac.jpl.nasa.gov}
\acro{POP}{Parallel Ocean Program }
\acro{PSD}{Power Spectral Density}
\acro{PSPT}{Precision Solar Photometric Telescope}
\acro{PSU}{Pennsylvania State University}
\acro{PyDAP}{Python Data Access Protocol}
\acro{PV}{potential vorticity}

\acro{QC}{quality control}
\acro{QG}{quasi-geostrophic}
\acro{QuikSCAT}{Quick Scatterometer}
%{ -- \it http://winds.jpl.nasa.gov/missions/quikscat/quikindex.html}
\acro{QZJ}{quasi-zonal jet}

\acro{R2HA2}{Rescue \& Reprocessing of Historical AVHRR Archives }
\acro{RAFOS}{{\smaller SOFAR}, SOund Fixing And Ranging, spelled backward}
\acro{RAID}{Redundant Array of Independent Disks}
\acro{RAL}{Rutherford Appleton Laboratory}
\acro{RDF}{Resource Description Language}
\acro{REASoN}{Research, Education and Applications Solutions Network}
\acro{REAP}{Realtime Environment for Analytical Processing}
\acro{ReLU}{Rectified Linear Unit}
\acro{REU}{Research Experiences for Undergraduates}
\acro{RFA}{Research Focus Area}
\acro{RFI}{Radio Frequency Interference}
\acro{RFC}{Request For Comments}
\acro{R/GTS}{Regional/Global Task Sharing}
\acro{RSI}{Research Systems Inc.}
\acro{RISE}{Radiative Inputs from Sun to Earth}
\acro{rms}{root mean square}
\acro{RMI}{Remote Method Invocation}
\acro{ROMS}{Regional Ocean Modeling System}
\acro{ROSES}{Research Opportunities in Space and Earth Sciences}
\acro{RSMAS}{Rosenstiel School of Marine and Atmospheric Science}
\acro{RSS}{Remote Sensing Sytems}
\acro{RTM}{radiative transfer model}

%\acro{SACCF}{Southern \acs{ACC} Front}
\acro{SACCF}{Southern {\smaller ACC} Front}
\acro{SAC-D}[(SAC)-D]{Sat\'elite de Aplicaciones Cient\'ificas-D}
%\acro{SAF}{Satellite Application Facility}
\acro{SAF}{Subantarctic Front}
\acro{SAIC}{Science Applications International Corporation}
\acro{SANS}{SysAdmin, Audit, Networking, and Security}
\acro{SAR}{synthetic aperature radar}
\acro{SATMOS}{Service d'Archivage et de Traitement M\'et\'eorologique des Observations Spatiales}
\acro{SBE}{Sea-Bird Electronics}
\acro{SciDAC}{Scientific Discovery through Advanced Computing}
\acro{SCC}{Subsurface Countercurrent}
\acro{SCCWRP}{Southern California Coastal Water Research Project}
\acro{SDAC}{Solar Physics Data Analysis Center}
%\acro{SeaWinds}{NASA Scatterometer}
\acro{SDS}{Scientific Data Set}
\acro{SDSC}{San Diego Supercomputer Center}
\acro{SeaDAS}{\textsmaller{SeaWiFS} Data Analysis System}
\acro{SeaWiFS}{Sea-viewing Wide Field-of-view Sensor}
%\acro{SEC}{Sun Earth Connection}
\acro{SEC}{South Equatorial Current}
\acro{SECC}{South Equatorial Countercurrent}
\acro{SECDDS}{Sun Earth Connection Distributed Data Services}
\acro{SEEDS}{Strategic Evolution of \textsmaller{ESE} Data Systems}
%{ -- \it http://lennier.gsfc.nasa.gov/seeds}
\acro{SEIC}{South Equatorial Intermediate Current}
\acro{SEUC}{Southern Equatorial Undercurrent}
\acro{SEVIRI}{Spinning Enhanced Visible and Infra-Red Imager}
\acro{SeRQL}{SeRQL}
\acro{SGI}{Silican Graphics Incorporated}
%{ -- \it http://www.sgi.com}
\acro{SHF}{Sensible Heat Flux}
\acro{SICC}{South Intermediate Countercurrent}
\acro{SIED}{single image edge detection}
\acro{SIPS}{Science Investigator--led Processing System}
\acro{SIR}{Scatterometer Image Reconstruction}
\acro{SISTeR}{Scanning Infrared Sea Surface Temperature Radiometer}
\acro{SLA}{sea level anomaly}
\acro{SMAP}{Soil Moisture Active Passive}
\acro{SMMR}{Scanning Multichannel Microwave Radiometer}
\acro{SMOS}{Soil Moisture and Ocean Salinity}
\acro{SMTP}{Simple Mail Transfer Protocol}
\acro{SOAP}{Simple Object Access Protocol}
\acro{SOEST}{School of Ocean and Earth Science and Technology}
\acro{SOFAR}{SOund Fixing And Ranging}
\acro{SOFINE}{Southern Ocean Finescale Mixing Experiment}
\acro{SOHO}{Solar and Heliospheric Observatory}
\acro{SPARC}{Space Physics and Aeronomy Research Collaboratory}
%{ -- \it http://intel.si.umich.edu/SPARC/}
\acro{SPARQL}{Simple Protocol and \textsmaller{RDF} Query Language}
\acro{SPASE}{Space Physics Archive Search Engine}
\acro{SPCZ}{South Pacific Convergence Zone}  
\acro{SPDF}{Space Physics Data Facility}
\acro{SPDML}{Space Physics Data Markup Language}
%{ -- \it http://sd-www.jhuapl.edu/SPDML/}
\acro{SPG}{Standards Process Group}
%{ -- \it http://www.esdswg.org/spg/}
\acro{SQL}{Structured Query Language}
\acro{SSCC}{South Subsurface Countercurrent}
%\acro{SSL}{Secure Sockets Layer}
\acro{SSL}{self-supervised learning}
\acro{SSO}{Single sign-on}
\acro{SSES}{Single Sensor Error Statistics}
\acro{SSH}{sea surface height}
%\acro{SSHA}{\ac{SSH} anomaly}
\acro{SSHA}{sea surface height anomaly}
\acro{SSMI}{Special Sensor Microwave/Imager}
\acro{SST}{sea surface temperature}
\acro{SSTa}{\ac{SST} anomaly}
\acro{SSS}{sea surface salinity}
\acro{SSTST}{Sea Surface Temperature Science Team}
\acro{STEM}{science, technology, engineering and mathematics}
\acro{STL}{Standard Template Library}
\acro{STCZ}{Subtropical Convergence Zone}
\acro{Suomi-NPP}[{\larger Suomi}-NPP]{Suomi-\acl{NPP}}
\acro{SWEET}{Semantic Web for Earth and Environmental Terminology}
\acro{SWFSC}{Southwest Fisheries Science Center}
\acro{SWOT}{Surface Water and Ocean Topography}
\acro{SWRL}{Semantic Web Rule Language}
\acro{SubEx}{Submesoscale Experiment}
\acro{SURA}{Southeastern Universities Research Association}
\acro{SURFO}{Summer Undergraduate Research Fellowship Program in Oceanography}
\acro{SuperDARN}{Super Dual Auroral Radar Network}

\acro{TAMU}{Texas A\&M University}
\acro{TB}{TeraByte - $10^{12}$ bytes}
\acro{TCASCV}{Technology Center for Advanced Scientific Computing and Visualization}
%{ -- \newline \it http://www.cascv.brown.edu/aboutus.html}
\acro{TCP}{Transmission Control Protocol}
\acro{TCP/IP}{Transmission Control Protocol/Internet Protocol}
\acro{TDS}{\textsmaller{THREDDS} Data Server}
\acro{TEX}{external temperature or T-External}
\acro{THREDDS}{Thematic Realtime Environmental Data Distributed Services}
%{ -- \newline \it http://my.unidata.ucar.edu/content/projects/THREDDS/index.html}
\acro{TIDI}{\textsmaller{TIMED} Doppler Interferometer}
\acro{TIFF}{Tag Image File Format}
\acro{TIMED}{Thermosphere, Ionosphere, Mesosphere, Energetics and Dynamics}
\acro{TLS}{Transport Layer Security}
\acro{TRL}{Technology Readiness Level}
\acro{TMI}{\textsmaller{TRMM} Microwave Imager}
\acro{TOPEX/Poseidon}{TOPography EXperiment for Ocean Circulation/Poseidon}
\acro{TRMM}{Tropical Rainfall Measuring Mission}
%{ -- \it http://trmm.gsfc.nasa.gov/}
\acro{TSG}{thermosalinograph}

\acro{UCAR}{University Corporation for Atmospheric Research}
%{ -- \it http://www.ucar.edu}
\acro{UCSB}{University of California, Santa Barbara}
\acro{UCSD}{University of California, San Diego}
\acro{uCTD}{Underway Conductivity, Temperature and Salinity or Underway \acs{CTD}}
\acro{UDDI}{Universal Description, Discovery and Integration}
\acro{UMAP}{Uniform Manifold Approximation and Projection}
\acro{UMiami}{University of Miami}
\acro{Unidata}{}
%{ -- \it http://unidata.ucar.edu}
\acro{URI}{University of Rhode Island}
%{ -- \it http://www.uri.edu}
\acro{UPC}{Unidata Program Committee}
\acro{URL}{Uniform Resource Locator}
\acro{USGS}{United States Geological Survey}
\acro{UTC}{Coordinated Universal Time}
\acro{UW}{University of Washington}

\acro{VCDAT}{Visual Climate Data Analysis Tools}
\acro{VIIRS}{Visible-Infrared Imager-Radiometer Suite}
\acro{VR}{Virtual Reality}
\acro{VSTO}{Virtual Solar-Terrestrial Observatory}

\acro{WCR}{Warm Core Ring}
\acro{WCS}{Web Coverage Service}
%{ -- \it http://www.ogcnetwork.org}
\acro{WCRP}{World Climate Research Program}
\acro{WFS}{Web Feature Service}
%{ -- \it http://www.ogcnetwork.org}
\acro{WMS}{Web Map Service}
%{ -- \it http://www.ogcnetwork.org}
\acro{W3C}{World Wide Web Consortium}
\acro{WJ}{Wyrtki Jets}
\acro{WHOI}{Woods Hole Oceanographic Institution}
\acro{WKB}{Well Known Binaries}
\acro{WIMP}{Windows, Icons, Menus, and Pointers}
\acro{WIS}{World Meteorological Organisation Information System}
\acro{WOA05}{World Ocean Atlas 2005}
\acro{WOCE}{World Ocean Circulation Experiment}
\acro{WP-ESIP}{Working Prototype Earth Science Information Partner}
\acro{WRF}{Weather \& Research Forecasting Model}
\acro{WSDL}{Web Services Description Language}
\acro{WSP}{western South Pacfic}
\acro{WWW}{World Wide Web}

\acro{XBT}{Expendable BathyThermograph}
\acro{XML}{Extensible Markup Language}
%{ -- \it http://www.w3.org/XML}
\acro{XRAC}{eXtreme Digital Request Allocation Committee}
\acro{XSEDE}{Extreme Science and Engineering Discovery Environment}

\acro{YAG}{yttrium aluminium garnet}

\end{acronym}
}

%\clearpage

% or


\appendices
\section{Additional Galleries}

% %%%%%%%%%%%%%%%%%%%%%%%

% %%%%%%%%%%%%%%%%%%%%%%%

% %%%%%%%%%%%%%%%%%%%%%%%

% %%%%%%%%%%%%%%%%%%%%%%%

% %%%%%%%%%%%%%%%%%%%%%%%

\section{}
Appendix two text goes here.

\section*{Acknowledgment}

JXP acknowledges future support from the Simons Foundation
and help from Claudie Beaulieu for the time-series analysis. EG acknowledges J. Xavier Prochaska and P. C. Cornillon's invaluable guidance on this project, as well as David Draper's insightful discussions. 

The authors acknowledge use of the Nautilus
cloud computing system which is supported
by the following
US National Science Foundation (NSF) awards:
CNS-1456638, CNS-1730158, CNS-2100237, CNS-2120019, ACI-1540112, ACI-1541349, OAC-1826967, OAC-2112167.

\ifCLASSOPTIONcaptionsoff
  \newpage
\fi



\bibliographystyle{IEEEtran}
\bibliography{bibtex/bib/our_bibli.bib}

\begin{thebibliography}{10}
\providecommand{\url}[1]{#1}
\csname url@samestyle\endcsname
\providecommand{\newblock}{\relax}
\providecommand{\bibinfo}[2]{#2}
\providecommand{\BIBentrySTDinterwordspacing}{\spaceskip=0pt\relax}
\providecommand{\BIBentryALTinterwordstretchfactor}{4}
\providecommand{\BIBentryALTinterwordspacing}{\spaceskip=\fontdimen2\font plus
\BIBentryALTinterwordstretchfactor\fontdimen3\font minus
  \fontdimen4\font\relax}
\providecommand{\BIBforeignlanguage}[2]{{%
\expandafter\ifx\csname l@#1\endcsname\relax
\typeout{** WARNING: IEEEtran.bst: No hyphenation pattern has been}%
\typeout{** loaded for the language `#1'. Using the pattern for}%
\typeout{** the default language instead.}%
\else
\language=\csname l@#1\endcsname
\fi
#2}}
\providecommand{\BIBdecl}{\relax}
\BIBdecl

\bibitem{thomas2008submesoscale}
L.~N. Thomas, A.~Tandon, and A.~Mahadevan, ``Submesoscale processes and
  dynamics,'' \emph{Ocean Modeling in an Eddying Regime, Geophys. Monogr. Ser},
  vol. 177, pp. 17--38, 2008.

\bibitem{mcwilliams2016review}
J.~C. McWilliams, ``Submesoscale currents in the ocean,'' \emph{Proceedings of
  the Royal Society of London A: Mathematical, Physical and Engineering
  Sciences}, vol. 472, no. 2189, pp. 1--32, 2016.

\bibitem{ulmo}
\BIBentryALTinterwordspacing
J.~X. Prochaska, P.~C. Cornillon, and D.~M. Reiman, ``Deep learning of sea
  surface temperature patterns to identify ocean extremes,'' \emph{Remote
  Sensing}, vol.~13, no.~4, 2021. [Online]. Available:
  \url{https://www.mdpi.com/2072-4292/13/4/744}
\BIBentrySTDinterwordspacing

\bibitem{pae}
V.~Böhm and U.~Seljak, ``Probabilistic auto-encoder,'' 2020.

\bibitem{hadsell2006dimensionality}
R.~Hadsell, S.~Chopra, and Y.~LeCun, ``Dimensionality reduction by learning an
  invariant mapping,'' in \emph{2006 IEEE Computer Society Conference on
  Computer Vision and Pattern Recognition (CVPR'06)}, vol.~2.\hskip 1em plus
  0.5em minus 0.4em\relax IEEE, 2006, pp. 1735--1742.

\bibitem{bachman2019learning}
P.~Bachman, R.~D. Hjelm, and W.~Buchwalter, ``Learning representations by
  maximizing mutual information across views,'' \emph{arXiv preprint
  arXiv:1906.00910}, 2019.

\bibitem{chen2020simple}
T.~Chen, S.~Kornblith, M.~Norouzi, and G.~Hinton, ``A simple framework for
  contrastive learning of visual representations,'' in \emph{International
  conference on machine learning}.\hskip 1em plus 0.5em minus 0.4em\relax PMLR,
  2020, pp. 1597--1607.

\bibitem{chen2020big}
T.~Chen, S.~Kornblith, K.~Swersky, M.~Norouzi, and G.~Hinton, ``Big
  self-supervised models are strong semi-supervised learners,'' \emph{arXiv
  preprint arXiv:2006.10029}, 2020.

\bibitem{chen2020improved}
X.~Chen, H.~Fan, R.~Girshick, and K.~He, ``Improved baselines with momentum
  contrastive learning,'' \emph{arXiv preprint arXiv:2003.04297}, 2020.

\bibitem{he2020momentum}
K.~He, H.~Fan, Y.~Wu, S.~Xie, and R.~Girshick, ``Momentum contrast for
  unsupervised visual representation learning,'' in \emph{Proceedings of the
  IEEE/CVF Conference on Computer Vision and Pattern Recognition}, 2020, pp.
  9729--9738.

\bibitem{tian2020makes}
Y.~Tian, C.~Sun, B.~Poole, D.~Krishnan, C.~Schmid, and P.~Isola, ``What makes
  for good views for contrastive learning?'' \emph{arXiv preprint
  arXiv:2005.10243}, 2020.

\bibitem{hayat2021}
M.~A. {Hayat}, G.~{Stein}, P.~{Harrington}, Z.~{Luki{\'c}}, and M.~{Mustafa},
  ``{Self-supervised Representation Learning for Astronomical Images},''
  \emph{The Astrophysical Journal Letters}, vol. 911, no.~2, p. L33, Apr. 2021.

\bibitem{Morningstar2021DensityOS}
W.~Morningstar, C.~Ham, A.~G. Gallagher, B.~Lakshminarayanan, A.~A. Alemi, and
  J.~V. Dillon, ``Density of states estimation for out-of-distribution
  detection,'' in \emph{AISTATS}, 2021.

\bibitem{charney71}
J.~Charney, ``Geostrophic turbulence,'' \emph{J. Atmos. Sci.}, vol.~28, pp.
  1087--1095, 1971.

\bibitem{gm72}
C.~Garrett and W.~Munk, ``Space–time scales of internal waves,''
  \emph{Geophys. Fluid Dyn.}, vol.~3, pp. 225--264, 1 1972.

\bibitem{callies13}
\BIBentryALTinterwordspacing
J.~Callies and R.~Ferrari, ``Interpreting energy and tracer spectra of
  upper-ocean turbulence in the submesoscale range ($1$-$200$~km),''
  \emph{Journal of Physical Oceanography}, vol.~43, pp. 2456--2474, 2013.
  [Online]. Available:
  \url{https://journals.ametsoc.org/view/journals/phoc/43/11/jpo-d-13-063.1.xml}
\BIBentrySTDinterwordspacing

\bibitem{2018arXivUMAP}
L.~{McInnes}, J.~{Healy}, and J.~{Melville}, ``{UMAP: Uniform Manifold
  Approximation and Projection for Dimension Reduction},'' \emph{ArXiv
  e-prints}, Feb. 2018.

\bibitem{cheng21}
S.~{Cheng} and B.~{M{\'e}nard}, ``{How to quantify fields or textures? A guide
  to the scattering transform},'' \emph{arXiv e-prints}, p. arXiv:2112.01288,
  Nov. 2021.

\bibitem{mallat2012}
\BIBentryALTinterwordspacing
S.~Mallat, ``Group invariant scattering,'' \emph{Communications on Pure and
  Applied Mathematics}, vol.~65, no.~10, pp. 1331--1398, 2012. [Online].
  Available: \url{https://onlinelibrary.wiley.com/doi/abs/10.1002/cpa.21413}
\BIBentrySTDinterwordspacing

\bibitem{he2016deep}
K.~He, X.~Zhang, S.~Ren, and J.~Sun, ``Deep residual learning for image
  recognition,'' in \emph{Proceedings of the IEEE conference on computer vision
  and pattern recognition}, 2016, pp. 770--778.

\bibitem{McInnes2018}
\BIBentryALTinterwordspacing
L.~McInnes, J.~Healy, N.~Saul, and L.~Großberger, ``Umap: Uniform manifold
  approximation and projection,'' \emph{Journal of Open Source Software},
  vol.~3, no.~29, p. 861, 2018. [Online]. Available:
  \url{https://doi.org/10.21105/joss.00861}
\BIBentrySTDinterwordspacing

\bibitem{healpix}
K.~M. {G{\'o}rski}, E.~{Hivon}, A.~J. {Banday}, B.~D. {Wandelt}, F.~K.
  {Hansen}, M.~{Reinecke}, and M.~{Bartelmann}, ``{HEALPix: A Framework for
  High-Resolution Discretization and Fast Analysis of Data Distributed on the
  Sphere},'' \emph{The Astrophysical Journal}, vol. 622, no.~2, pp. 759--771,
  Apr. 2005.

\bibitem{mitchell1992}
\BIBentryALTinterwordspacing
T.~P. Mitchell and J.~M. Wallace, ``The annual cycle in equatorial convection
  and sea surface tempera ture,'' \emph{Journal of Climate}, vol.~5, no.~10,
  pp. 1140 -- 1156, 1992. [Online]. Available:
  \url{https://journals.ametsoc.org/view/journals/clim/5/10/1520-0442_1992
  _005_1140_taciec_2_0_co_2.xml}
\BIBentrySTDinterwordspacing

\bibitem{okumura2004}
Y.~Okumura and S.-P. Xie, ``Interaction of the atlantic equatorial cold tongue
  and the african monsoon,'' \emph{Journal of Climate}, vol.~17, no.~18, pp.
  3589--3602, 2004.

\bibitem{pedlosky1990}
J.~Pedlosky, \emph{{Geophysical Fluid Dynamics}}, 2nd~ed.\hskip 1em plus 0.5em
  minus 0.4em\relax New York, NY, USA: Springer-Verlag, 1990.

\bibitem{hagen2005}
E.~Hagen, ``Zonal wavelengths of planetary rossby waves derived from
  hydrographic transects in the northeast atlantic ocean?'' \emph{J.
  Oceanography}, vol.~61, pp. 1039--1046, 2005.

\bibitem{kounta2018}
\BIBentryALTinterwordspacing
L.~Kounta, X.~Capet, J.~Jouanno, N.~Kolodziejczyk, B.~Sow, and A.~T. Gaye, ``A
  model perspective on the dynamics of the shadow zone of the eastern t ropical
  north atlantic -- part~1: the poleward slope currents along west africa,''
  \emph{Ocean Science}, vol.~14, no.~5, pp. 971--997, 2018. [Online].
  Available: \url{https://os.copernicus.org/articles/14/971/2018/}
\BIBentrySTDinterwordspacing

\bibitem{willett06}
\BIBentryALTinterwordspacing
C.~S. Willett, R.~R. Leben, and M.~F. Lavín, ``Eddies and tropical instability
  waves in the eastern tropical pacific: A review,'' \emph{Progress in
  Oceanography}, vol.~69, no.~2, pp. 218--238, 2006, a Review of Eastern
  Tropical Pacific Oceanography. [Online]. Available:
  \url{https://www.sciencedirect.com/science/article/pii/S0079661106000322}
\BIBentrySTDinterwordspacing

\bibitem{stramma2005}
\BIBentryALTinterwordspacing
L.~Stramma, S.~Hüttl, and J.~Schafstall, ``Water masses and currents in the
  upper tropical northeast atlantic off northwest africa,'' \emph{Journal of
  Geophysical Research: Oceans}, vol. 110, no. C12, 2005. [Online]. Available:
  \url{https://agupubs.onlinelibrary.wiley.com/doi/abs/10.1029/2005JC002939}
\BIBentrySTDinterwordspacing

\bibitem{combes2018}
\BIBentryALTinterwordspacing
V.~Combes and R.~P. Matano, ``The patagonian shelf circulation: Drivers and
  variability,'' \emph{Progress in Oceanography}, vol. 167, pp. 24--43, 2018.
  [Online]. Available:
  \url{https://www.sciencedirect.com/science/article/pii/S0079661117303130}
\BIBentrySTDinterwordspacing

\bibitem{painter2020}
\BIBentryALTinterwordspacing
S.~C. Painter, ``The biogeochemistry and oceanography of the east african
  coastal curren t,'' \emph{Progress in Oceanography}, vol. 186, p. 102374,
  2020. [Online]. Available:
  \url{https://www.sciencedirect.com/science/article/pii/S0079661120301130}
\BIBentrySTDinterwordspacing

\bibitem{chelton2011}
\BIBentryALTinterwordspacing
D.~B. Chelton, M.~G. Schlax, and R.~M. Samelson, ``Global observations of
  nonlinear mesoscale eddies,'' \emph{Progress in Oceanography}, vol.~91,
  no.~2, pp. 167 -- 216, 2011. [Online]. Available:
  \url{http://www.sciencedirect.com/science/article/pii/S0079661111000036}
\BIBentrySTDinterwordspacing

\bibitem{chatterjee2017dynamics}
A.~Chatterjee, D.~Shankar, J.~McCreary, P.~Vinayachandran, and A.~Mukherjee,
  ``Dynamics of andaman sea circulation and its role in connecting the
  equatorial indian ocean to the bay of bengal,'' \emph{Journal of Geophysical
  Research: Oceans}, vol. 122, no.~4, pp. 3200--3218, 2017.

\bibitem{huyer1983coastal}
A.~Huyer, ``Coastal upwelling in the california current system,''
  \emph{Progress in oceanography}, vol.~12, no.~3, pp. 259--284, 1983.

\bibitem{FreundMason1999}
Y.~Freund and M.~L., ``The alternating decision tree learning algorithm,'' in
  \emph{Proceedings of the 16$^{th}$ International Conference on Machine
  Learning}.\hskip 1em plus 0.5em minus 0.4em\relax San Francisco, CA, USA:
  Morgan Kaufmann Publishers Inc., 1999, pp. 124--133.

\bibitem{tanaka2019}
\BIBentryALTinterwordspacing
Y.~Tanaka and T.~Hibiya, ``Generation mechanism of tropical instability waves
  in the equatorial pacific ocean,'' \emph{Journal of Physical Oceanography},
  vol.~49, no.~11, pp. 2901 -- 2915, 2019. [Online]. Available:
  \url{https://journals.ametsoc.org/view/journals/phoc/49/11/jpo-d-19-0094.1.xml}
\BIBentrySTDinterwordspacing

\bibitem{chelton2001}
\BIBentryALTinterwordspacing
D.~B. Chelton, S.~K. Esbensen, M.~G.~S. a~nd Nicolai~Thum, M.~H. Freilich,
  F.~J. Wentz, C.~L.~G. nn, M.~J. McPhaden, and P.~S. Schopf, ``Observations of
  coupling between surface wind stress and sea surf ace temperature in the
  eastern tropical pacific,'' \emph{Journal of Climate}, vol.~14, no.~7, pp.
  1479 -- 1498, 2001. [Online]. Available:
  \url{https://journals.ametsoc.org/view/journals/clim/14/7/1520-0442_2001
  _014_1479_oocbsw_2.0.co_2.xml}
\BIBentrySTDinterwordspacing

\bibitem{ray2018}
\BIBentryALTinterwordspacing
S.~Ray, A.~T. Wittenberg, S.~M. Griffies, and F.~Zeng, ``Understanding the
  equatorial pacific cold tongue time-mean heat b udget. part i: Diagnostic
  framework,'' \emph{Journal of Climate}, vol.~31, no.~24, pp. 9965 -- 9985,
  2018. [Online]. Available:
  \url{https://journals.ametsoc.org/view/journals/clim/31/24/jcli-d-18-015
  2.1.xml}
\BIBentrySTDinterwordspacing

\bibitem{Rocha2016a}
C.~B. Rocha, T.~K. Chereskin, S.~T. Gille, and D.~Menemenlis, ``{Mesoscale to
  Submesoscale Wavenumber Spectra in Drake Passage},'' \emph{Journal of
  Physical Oceanography}, vol.~46, no.~2, pp. 601--620, feb 2016.

\bibitem{Rocha2016b}
C.~B. Rocha, S.~T. Gille, T.~K. Chereskin, and D.~Menemenlis, ``{Seasonality of
  submesoscale dynamics in the Kuroshio Extension},'' \emph{Geophysical
  Research Letters}, vol.~43, no.~21, pp. 11\,304--11\,311, nov 2016.

\bibitem{Arbic2018}
B.~K. Arbic, M.~H. Alford, J.~K. Ansong, M.~C. Buijsman, R.~B. Ciotti, J.~T.
  Farrar, R.~W. Hallberg, C.~E. Henze, C.~N. Hill, C.~A. Luecke, D.~Menemenlis,
  E.~J. Metzger, M.~M{\"{u}}eller, A.~D. Nelson, B.~C. Nelson, H.~E. Ngodock,
  R.~M. Ponte, J.~G. Richman, A.~C. Savage, R.~B. Scott, J.~F. Shriver, H.~L.
  Simmons, I.~Souopgui, P.~G. Timko, A.~J. Wallcraft, L.~Zamudio, and Z.~Zhao,
  ``{A Primer on Global Internal Tide and Internal Gravity Wave Continuum
  Modeling in HYCOM and MITgcm},'' in \emph{New Frontiers in Operational
  Oceanography}, E.~P. Chassignet, A.~Pascual, J.~Tintor{\'{e}}, and J.~Verron,
  Eds.\hskip 1em plus 0.5em minus 0.4em\relax GODAE OceanView, aug 2018,
  ch.~13, pp. 307--392.

\end{thebibliography}

%\begin{thebibliography}{1}
%\end{thebibliography}



%\vfill


\end{document}